\documentclass[namedreferences]{solarphysics}

\usepackage[optionalrh,solaromanenum]{spr-sola-addons} 
\usepackage{epsfig}     
\usepackage{graphicx}    
\usepackage{amssymb}    
\usepackage{color}      
\usepackage{url}       
\usepackage{breakurl}     

\renewcommand{\vec}[1]{{\mathbfit #1}}

\newcommand{\ion}[2]{#1\,{\sc #2}}

\DeclareMathAlphabet{\mathsc}{OT1}{cmr}{m}{sc}
\DeclareRobustCommand{\ion}[2]{\textup{#1\,\textsc{\lowercase{#2}}}}

\begin{document}
\begin{article}
%
\begin{opening}

\title{Non-Equilibrium Processes in the Solar Corona, Transition Region, Flares, and Solar Wind \textit{(Invited Review)}}
%

\author[addressref=aff1,corref,email={}]{\inits{J.}\fnm{Jaroslav}~\snm{Dud\'{i}k}}\sep
\author[addressref=aff1,email={}]{\inits{E.}\fnm{Elena}~\snm{Dzif\v{c}\'{a}kov\'{a}}}\sep
\author[addressref=aff2,email={}]{\inits{N.}\fnm{Nicole}~\snm{Meyer-Vernet}}\sep
\author[addressref=aff3,email={}]{\inits{G.}\fnm{Giulio}~\snm{Del Zanna}}\sep
\author[addressref={aff4,aff5,aff6},email={}]{\inits{P. R.}\fnm{Peter R.}~\snm{Young}}\sep
\author[addressref=aff7,email={}]{\inits{A.}\fnm{Alessandra}~\snm{Giunta}}\sep
\author[addressref=aff8,email={}]{\inits{B.}\fnm{Barbara}~\snm{Sylwester}}\sep
\author[addressref=aff8,email={}]{\inits{J.}\fnm{Janusz}~\snm{Sylwester}}\sep
\author[addressref=aff9,email={}]{\inits{M.}\fnm{Mitsuo}~\snm{Oka}}\sep
\author[addressref=aff3,email={}]{\inits{H. E.}\fnm{Helen E.}~\snm{Mason}}\sep
\author[addressref=aff10,email={}]{\inits{C.}\fnm{Christian}~\snm{Vocks}}\sep
\author[addressref=aff11,email={}]{\inits{L.}\fnm{Lorenzo}~\snm{Matteini}}\sep
\author[addressref=aff12,email={}]{\inits{S.}\fnm{S\"{a}m}~\snm{Krucker}}\sep
\author[addressref=aff13,email={}]{\inits{D. R.}\fnm{David R.}~\snm{Williams}}\sep
\author[addressref=aff14,email={}]{\inits{\v{S}.}\fnm{\v{S}imon}~\snm{Mackovjak}}
\address[id=aff1]{Astronomical Institute, Czech Academy of Sciences, Fri\v{c}ova 298, 25165 Ond\v{r}ejov}
\address[id=aff2]{CNRS, PSL, LESIA, Observatoire de Paris, 5 place Jules Janssen, F-92195 Meudon, France}
\address[id=aff3]{DAMTP, University of Cambridge, Wilberforce Road, Cambridge CB3 0WA, UK}
\address[id=aff4]{College of Science, George Mason University, Fairfax, VA 22030, USA}
\address[id=aff5]{NASA Goddard Space Flight Center, Code 671, Greenbelt, MD 20771, USA}
\address[id=aff6]{Northumbria University, Newcastle Upon Tyne, NE1 8ST, UK}
\address[id=aff7]{STFC Rutherford Appleton Laboratory, Chilton, Didcot, Oxon. OX11 0QX, UK}
\address[id=aff8]{Space Research Centre (CBK PAN), Warsaw, Bartycka 18A, Poland}
\address[id=aff9]{Space Sciences Laboratory, University of California, Berkeley, CA 94720, USA}
\address[id=aff10]{Leibniz-Institut f\"{u}r Astrophysik, An der Sternwarte 16, 14482 Potsdam, Germany}
\address[id=aff11]{Imperial College London, London SW7 2AZ, UK}
\address[id=aff12]{University of Applied Sciences and Arts Northwestern Switzerland, Bahnhofstrasse 6, 5210 Windisch, Switzerland}
\address[id=aff13]{ESAC, European Space Agency, Villanueva de la Ca\~{n}ada E-28692 Madrid, Spain}
\address[id=aff14]{Institute of Experimental Physics, SAS, Watsonova 47, 04001 Ko\v{s}ice, Slovak Republic}%

%

\runningauthor{Dud\'{i}k {\it et al.}}
\runningtitle{Non-thermal Distributions and Non-Equilibrium Processes}

\begin{abstract}
We review the presence and signatures of the non-equilibrium processes, both non-Maxwellian distributions and non-equilibrium ionization, in the solar transition region, corona, solar wind, and flares. Basic properties of the non-Maxwellian distributions are described together with their influence on the heat flux as well as on the rates of individual collisional processes and the resulting optically thin synthetic spectra. Constraints on the presence of high-energy electrons from observations are reviewed, including positive detection of non-Maxwellian distributions in the solar corona, transition region, flares, and wind. Occurrence of non-equilibrium ionization is reviewed as well, especially in connection to hydrodynamic and generalized collisional-radiative modelling. Predicted spectroscopic signatures of non-equilibrium ionization depending on the assumed plasma conditions are summarized. Finally, we discuss the future remote-sensing instrumentation that can be used for detection of these non-equilibrium phenomena in various spectral ranges.
\end{abstract}
\end{opening}

%
%
\section{Introduction}
\label{Sect:1}
Traditionally, optically thin astrophysical spectra are interpreted in terms of equilibrium. This means that the conditions in the emitting plasma are assumed to be time-independent and described by temperature, density, and emission measure. The very definition of temperature means that the distribution of particles is assumed to be Maxwellian. Alternatively, if the emission is time-dependent, it is assumed that the plasma evolves through a series of equilibria, and that its emission can be calculated at each instant from the instantaneous values of temperature, density, and emission measure.

When these assumptions are violated, non-equilibrium ensues. Processes creating the non-equilibrium can be either those leading to departures from a Maxwe\-llian, \textit{i.e.}, to non-Maxwellian distributions, or time-dependent non-equi\-librium ionization, or both. Non-equilibrium ionization refers to the ionic composition (charge state) of plasma being dependent on its evolutionary history. In the context of the upper solar atmosphere, transient ionization can arise due to dynamic phenomena involving changes in the local plasma properties faster than the ionization timescales, such as rapid heating or cooling \citep[\textit{e.g.},][]{Bradshaw03a,Bradshaw03b,Bradshaw06,Olluri13b,Martinez16}. Non-Maxwellians can occur due to a variety of processes, such as acceleration by electric fields \textit{e.g.} during magnetic reconnection \citep[\textit{e.g.},][]{Petkaki11,Zharkova11,Burge12,Burge14,Cargill12,Gordovskyy13,Gordovskyy14,Pinto16}, turbulence \citep{Hasegawa85,Laming07,Bian14}, shocks or interaction of various waves such as whistlers with particles \citep{Vocks08,Vocks16}, or long-range interactions inducing correlations among particles in the system \citep{Tsallis88,Tsallis09,Collier04,Leubner04a,Livadiotis09,Livadiotis10,Livadiotis11a,Livadiotis13}. Moreover, since the cross-section for the Coulomb collisions among the particles varies with kinetic energy as $E^{-2}$, the collison frequency scales as $E^{-3/2}$, and thus it takes longer for high-energy particles to equilibrate. This is the basic reason for occurrence of high-energy tails in astrophysical plasmas, as first pointed out by  \citet{Scudder79}. Furthermore, density and temperature gradients in the plasma can lead to non-Maxwellian tails \citep[\textit{e.g.},][]{Roussel-Dupre80a,Shoub83,Ljepojevic88a}. \citet{Scudder13} show that non-Maxwellians occur in regimes where the ratio of the electron mean free path to the pressure gradient (\textit{i.e.}, the electron Knudsen number) is larger than $10^{-2}$ at any point along a magnetic field line. Such points are argued to be common above 1.05\,$R_\odot$, \textit{i.e.}, in the solar corona, but could occur as low as in the solar transition region \citep[see also][]{Roussel-Dupre80a}.

The presence of non-Maxwellian distributions have significant consequences for the physics of the solar atmosphere. For example, any observational analysis done under the assumption of a Maxwellian will yield an ``observed'' temperature that is different from the one that corresponds to the second moment of the distribution \citep{Meyer-Vernet07,Nicholls12,Nicolaou16}, and it can also depend on the range of energies observed. In addition, any quantities derived from an integral over the particle distribution will be affected. These include not only the heat flux, which is the third moment of the distribution function, but also the rates of various spectroscopically important processes, such as excitation, ionization, and recombination \citep[\textit{e.g.},][]{Roussel-Dupre80b,Owocki83,Bryans06,Dzifcakova13a} that affect the emission line intensities. We note that non-Maxwellians also affect magneto-hydrodynamic (MHD) modelling, including the \citet{Chew56} approximation, since such models assume that the infinite hierarchy of moment equations can be truncated, \textit{e.g.}, by assuming an expression for the heat flux dependent only on local plasma properties \citep{Scudder92a}. Such assumptions are generally invalid if long-range interactions among particles are present \citep[\textit{cf.},][]{West08}.

The focus of the present review is to describe these non-equilibrium processes and their consequences in the solar corona and wind, including the coronal boundary in the transition region, and also in solar flares. The focus is on the mechanisms of optically thin emission in a collisionally dominated non-equilibrium plasma. We do not provide a review of the non-Maxwellians detected in other astrophysical plasmas, as these can be found elsewhere, \textit{e.g.}, in \citet{Pierrard10} and \citet{Bykov13}. An earlier review on the subject can be found in \citet{Bradshaw13}. A review on the kinetic effects in the solar wind, including non-Maxwellian distributions, can be found in \citet{Marsch06}. We also note that we do not treat the radio emission arising from non-Maxwellian plasmas, \textit{e.g.} due to beam or loss-cone instabilities. Information on such processes can be found in \citet{Krueger79}, \citet{Suzuki85}, \citet{Aschwanden05}, \citet{Karlicky09}, \citet{Benacek17}, and references therein. Similarly, we do not treat the non-equilibrium ionization of hydrogen and helium or other low ionization stages belonging to the solar chromosphere, such as \ion{Ca}{II}, as this would require the additional treatment of radiative transfer. The reader is instead referred \textit{e.g.}, to the works of \citet{Heinzel91}, \citet{Carlsson02}, \citet{Leenaarts07}, \citet{Kasparova09a}, \citet{Gudiksen11}, \citet{Wedemeyer11}, \citet{Golding14}, \citet{Allred15}, and \citet{Golding16}.

This review is structured as follows. Non-Maxwellian distributions and their consequences for the solar wind are described in Section \ref{Sect:2}, where the particular case of $\kappa$-distributions is treated in detail. This section also contains an overview of the processes that can lead to formation of the $\kappa$-distribution. The non-Maxwellian spectral synthesis is described in Section \ref{Sect:3}, which also includes a review of the detections of such non-Maxwellian distributions, or lack thereof. Solar flares are treated in Section \ref{Sect:4}, including the often present high-energy tails detected from bremsstrahlung emission, constraints on the particle distribution derived from imaging observations at lower energies, and also spectroscopic signatures of non-Maxwellians in the emission line spectra containing dielectronic satellite lines. Non-equilibrium ionization is treated in Section \ref{Sect:5}, where we discuss timescales for equilibration, their dependence on electron density and also on flows, and review the theoretical work on the importance of non-equilibrium ionization in the solar transition region, corona, and flares. Future instrumentation that may lead to advancement in the detection of non-equilibrium processes is described briefly in Section \ref{Sect:6}.

%
%
\section{$\kappa$-distributions and the solar corona and wind}
\label{Sect:2}

\begin{figure}[!t]
	\centering
	\includegraphics[width=0.5\textwidth]{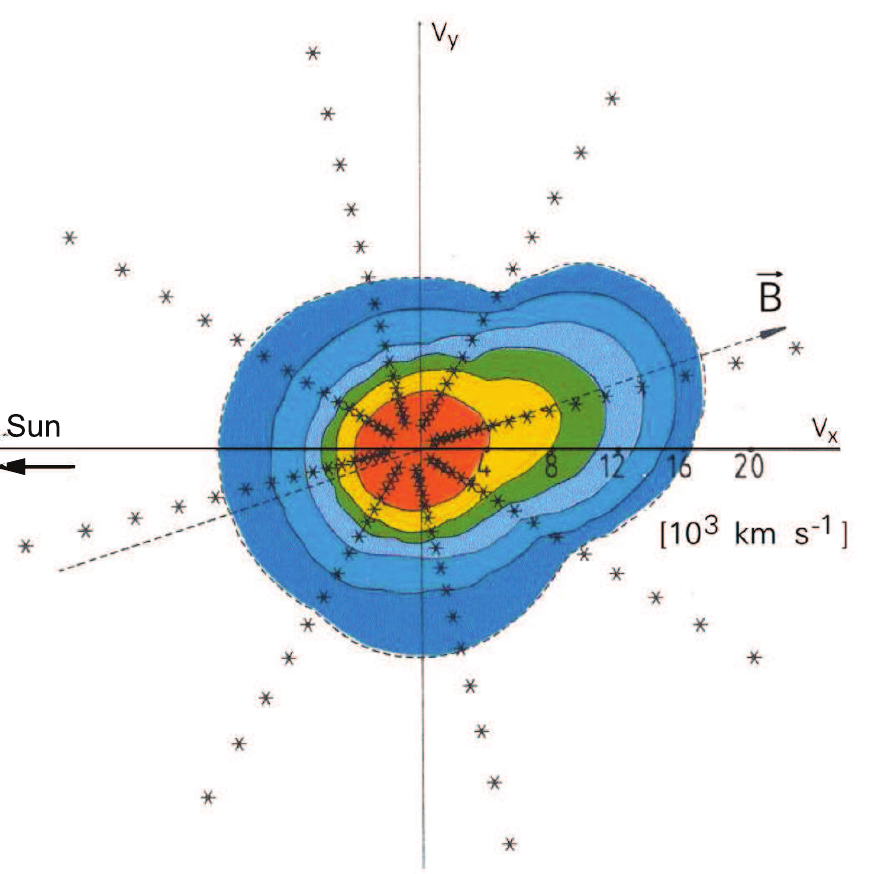}
	\caption{Electron velocity distribution in the fast solar wind measured by the plasma analyser on \textit{Helios} at 1 AU, showing three main components of the distribution: a cold core carrying bulk of the electrons, a hotter halo, and the strahl - a suprathermal component directed along the magnetic field. The particle density is denoted by logarithmically spaced contours, and the starred lines delineate the detector's resolution. The heat flux is carried by the strahl together with the halo, the hotter isotropic component which is slightly displaced with respect to the maximum of the core part indicated in red. Reproduced with permission from \citet{Pilipp87}, \textcircled{c} John Wiley and Sons.
\label{Fig:Felec}}
\end{figure}

As with most diffuse space and astrophysical plasmas, the outer solar corona and wind are not in equilibrium and their particle velocity distributions exhibit supra-thermal tails. This is illustrated in Figures \ref{Fig:Felec}, \ref{Fig:FKappa}, and \ref{Fig:LeChat2010}, as well as in Figure \ref{Fig:Core_beam}, which show classical examples of solar wind velocity distributions measured in situ.

Detailed analyses \citep[see][and references therein]{Tao16,Pierrard16} show that the solar wind electron velocity distributions are close to Maxwellians at low energies, whereas higher energy electrons (the so-called halo) have a power-law energy distribution which can be modelled by a $\kappa$-distribution (see Section \ref{Sect:2.1}), with an additional strahl component beamed along the magnetic field, as shown in Figure \ref{Fig:Felec}. Non-Maxwellian electron distributions in the quiet solar corona or coronal holes, and even in the high chromosphere have been proposed in order to explain these solar wind distributions \citep{Olbert81}, and to resolve inconsistencies between spectroscopic and radio solar observations \citep{Owocki99,Pinfield99,Esser00,Chiuderi04,Doyle04}.

%
%
This non-thermal character considerably complicates the modelling and diagnostics of the plasma, since instead of being characterised by the two independent parameters of the Maxwellian (particle number density $n$ and temperature $T$), the velocity distribution may in general have an infinite number of parameters.

\subsection{Kappa distributions}
\label{Sect:2.1}

\begin{figure}[!t]
	\centering
	\includegraphics[width=0.6\textwidth]{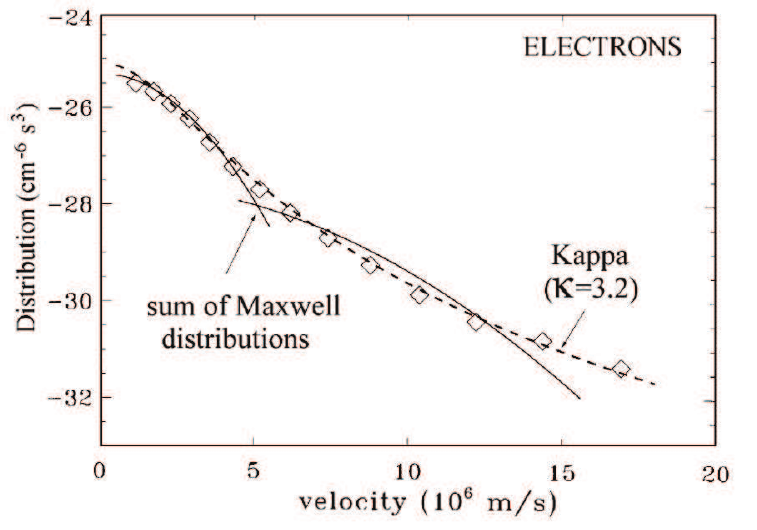}
	 \caption{An electron velocity distribution measured in the slow solar wind with the particle analyser on \textit{Ulysses}, and fitted with a sum of two Maxwellians (continuous lines) and with a $\kappa$-distribution (dashed). From \citet{Meyer-Vernet01}, adapted from \citet{Maksimovic97a}. Reproduced with permission, \textcircled{c} John Wiley and Sons.
\label{Fig:FKappa}}
\end{figure}

\begin{figure}[!t]
	\centering
	\includegraphics[width=1.\textwidth,clip=]{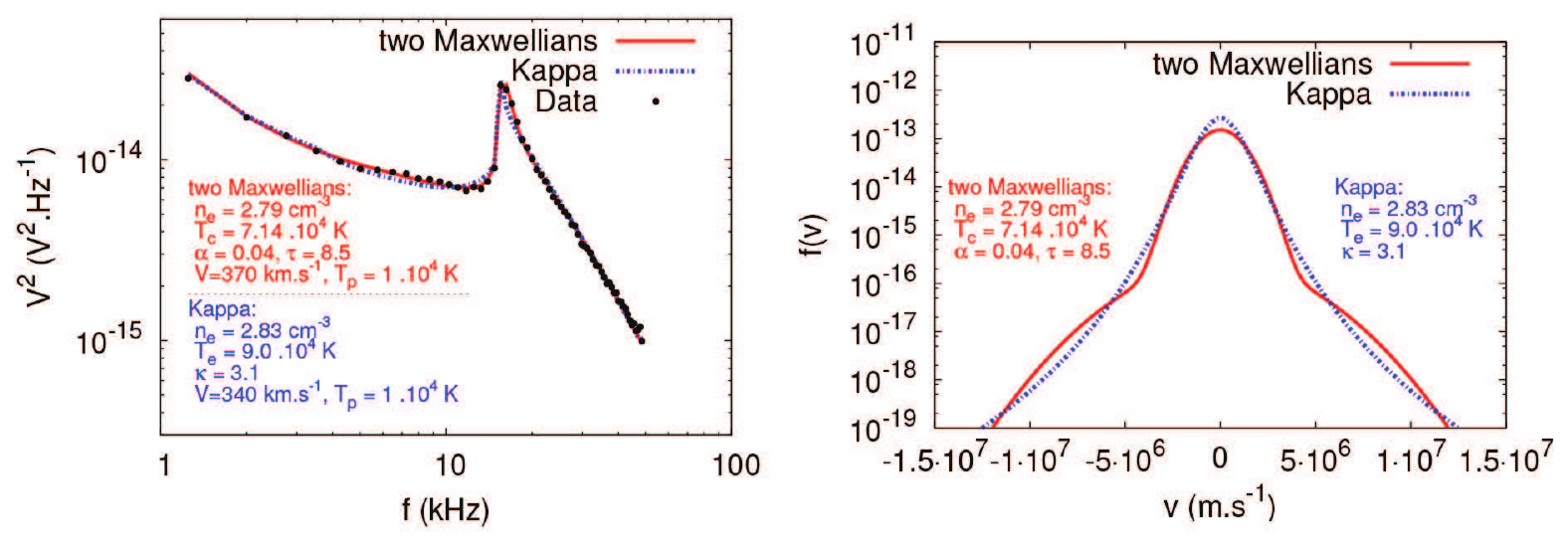}
	\caption{Plasma diagnostics with quasi-thermal noise spectroscopy. Left: Example of power spectrum measured in the solar wind with the URAP receiver on \textit{Ulysses}; the data are plotted as 64 heavy dots. The dash-dotted (solid) line shows the theoretical quasi-thermal noise with a $\kappa$-distribution and a sum of two Maxwellians that best fit the data; with the parameters shown. Right: Comparison of the $\kappa$ and sum of two Maxwellian distributions determined from the power spectrum plotted on the left. 
	Reproduced with permission from \citet{LeChat10}, \textcircled{c} AIP Publishing LLC.
\label{Fig:LeChat2010}}
\end{figure}

The $\kappa$-distribution of velocities, which may be written as
\begin{eqnarray}
	f_\kappa(v) & \propto & \left[1 + \frac{v^2}{\kappa v_\mathrm{th}^2}\right]^{-\kappa - 1}
	\label{Eq:Kappa}
\end{eqnarray}
has long been recognised \citep{Olbert68,Vasyliunas68a,Vasyliunas68b} as a very convenient tool to model distributions with supra-thermal tails (Figure \ref{Fig:FKappa}) because it is characterized by only three independent parameters only: $n$, $T$, and $\kappa$. $v_\mathrm{th}$ is the thermal speed, and the index $\kappa$ characterises the non-thermal properties. In the limit $\kappa \rightarrow \infty$ the distribution converges to a Maxwellian:
\begin{eqnarray}
	f_M(v) & \propto & \exp(-v^2/v_\mathrm{th}^2)\,.
	\label{Eq:Maxwellian}
\end{eqnarray}
Note that (\ref{Eq:Kappa}) and (\ref{Eq:Maxwellian}) represent isotropic velocity distributions, \textit{i.e.}, the probability for the velocity to have components lying in the ranges $[v_x, v_x+dv_x]$, $[v_y, v_y+dv_y]$, $[v_z, v_z+dv_z]$ is $f(v) dv_x dv_y dv_z$ with $v=(v_x^2+v_y^2+v_z^2)^{1/2}$; in other words, (\ref{Eq:Kappa}) and (\ref{Eq:Maxwellian}) represent cuts along any direction, and the probability for the speed to lie between $v$ and $v+dv$ is $f(v)$ times the corresponding volume in phase space, \textit{i.e.}, $4\pi v^2 dv$.

The distributions (\ref{Eq:Kappa}) and (\ref{Eq:Maxwellian}) must be normalised, and the temperature is defined from $m\langle v^2 \rangle/2=3k_\mathrm{B}T/2$, the average kinetic energy per particle of mass $m$, with $k_\mathrm{B}$ being the Boltzmann constant. The temperature $T$ defined in this way is a kinetic temperature, but it also coincides with the thermodynamic temperature defined from the nonextensive entropy \citep{Livadiotis09,Livadiotis15b}. For the $\kappa$-distribution  (\ref{Eq:Kappa}), this yields
\begin{equation}
T=\frac{\kappa}{\kappa -3/2} \frac{mv_\mathrm{th}^2}{2k_\mathrm{B}} \label{Eq:T}
\end{equation}
which requires $\kappa >3/2$.

%
%
%

\begin{figure*}[!t]
	\centering
	\includegraphics[width=0.49\textwidth]{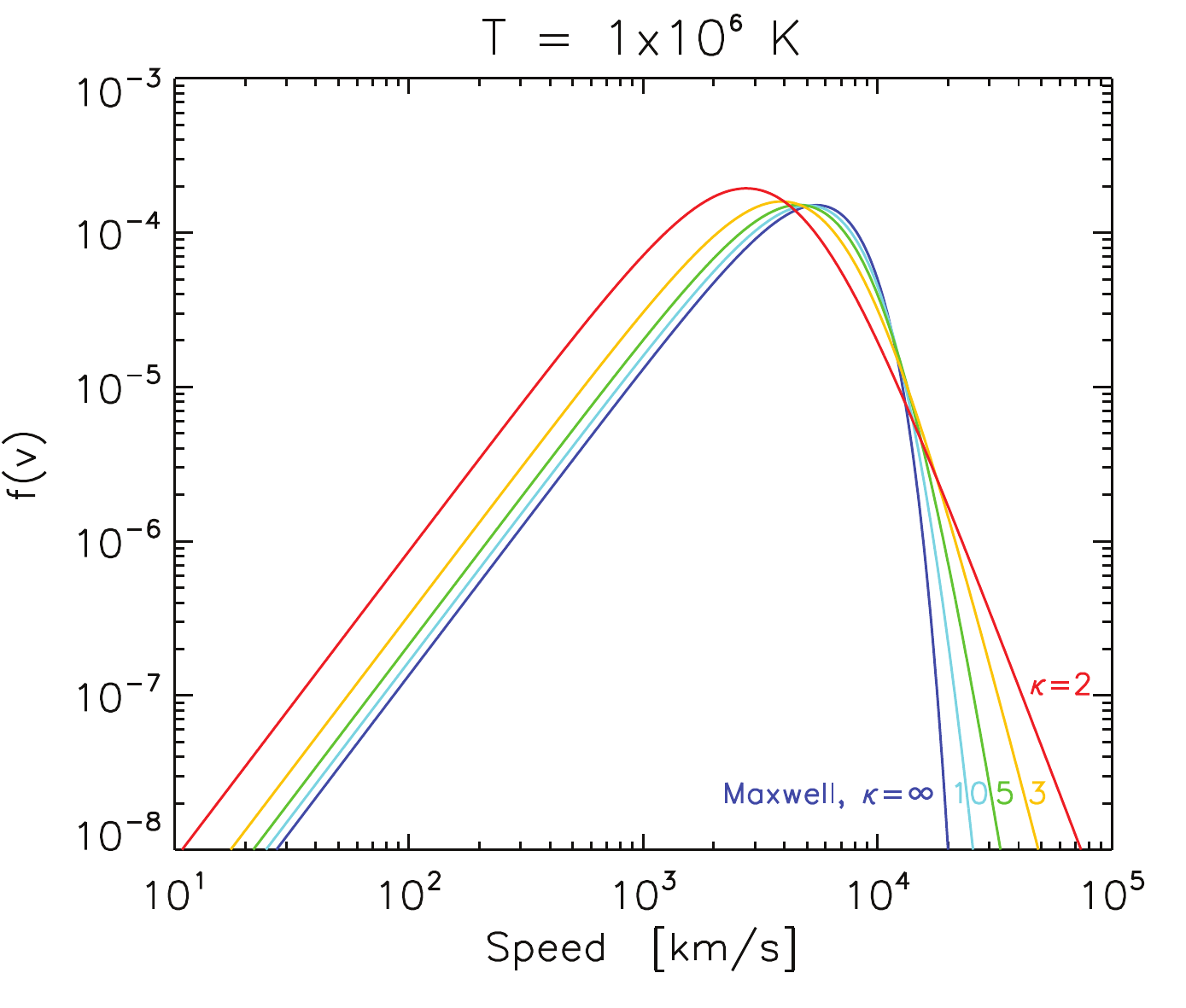}
	\includegraphics[width=0.49\textwidth]{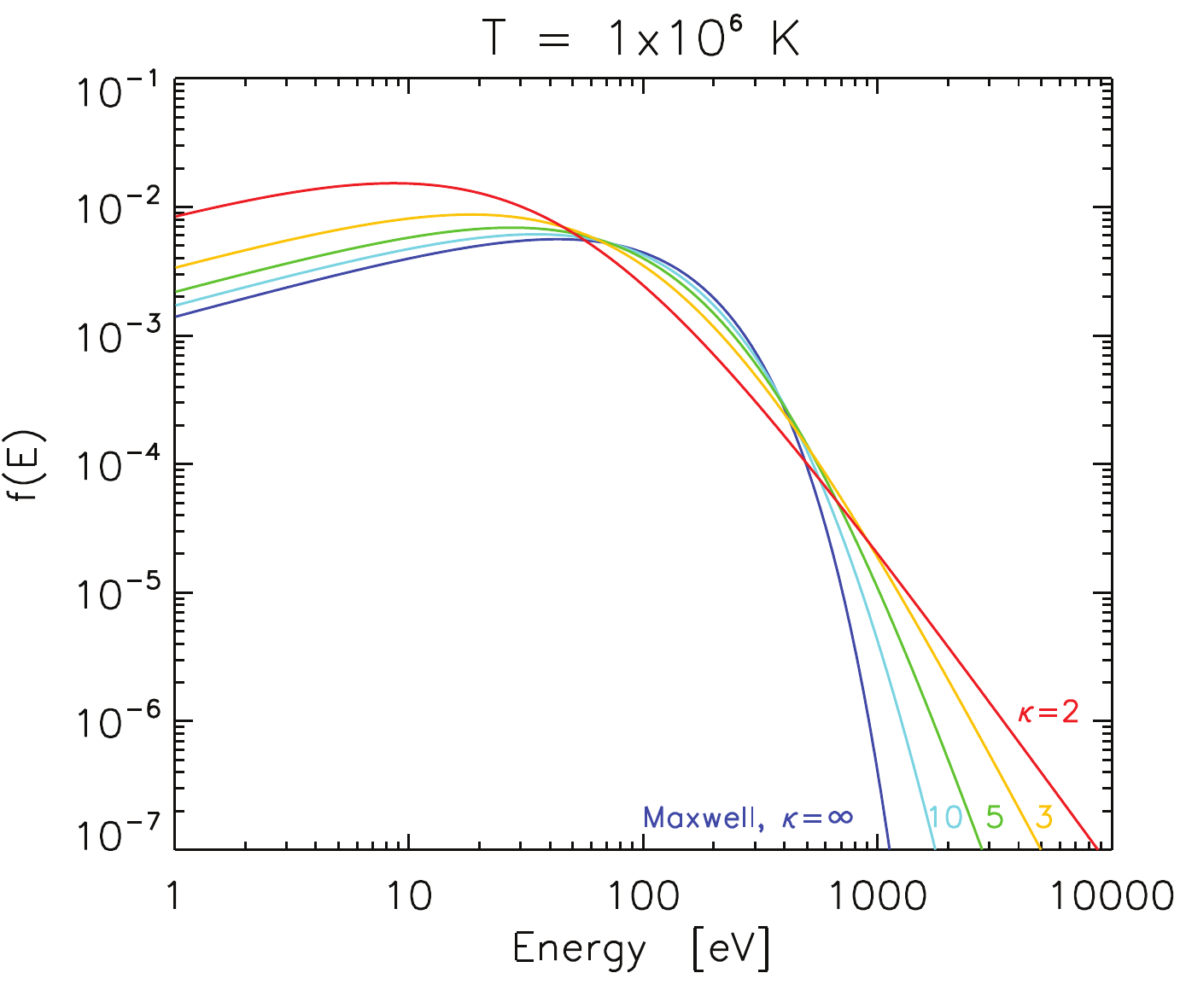}
	\caption{The speed ({\it left}) and energy ({\it right}) $\kappa$-distributions with $\kappa$\,=\,2, 3, 5, 10, 25 and the Maxwellian distribution with $T$\,=\,10$^{6}$\,K. Colors correspond to different values of $\kappa$.}
\label{Fig:Dst_k}
\end{figure*}

Using Equation (\ref{Eq:T}) for temperature and $mv^{2}/2 = E$, the $\kappa$-distribution can be expressed in terms of kinetic energy $E$ as \citep{Owocki83}
\begin{equation}
   f(E)\mathrm{d}E= A_{\kappa} \frac{2}{ \pi^{1/2}(k_\mathrm{B}T)^{3/2}} \frac{E^{1/2}\mathrm{d}E }
   { \left(1 +\frac { E}{( \kappa - 1.5) k_\mathrm{B}T}\right)^{ \kappa + 1 }}
   \label{Eq:Kappa_E}
\end{equation}
where
\begin{equation}
	A_{\kappa} = \frac {\Gamma ( \kappa + 1 )}{ \Gamma ( \kappa -0.5)
	( \kappa - 1.5 )^{3/2}}
	\label{Eq:A_kappa}
\end{equation}
is the normalization constant (Figure \ref{Fig:Dst_k}). As noted above, $T$ is the temperature.

%
%

It is important to note that the Debye length $L_D$, which is determined by the low-energy particles, is defined from $L_D^2 = [\epsilon _0 m/(ne^2)] / \langle v^{-2} \rangle$, with $\langle v^{-2} \rangle $ denoting the average of the inverse square speed \citep[see, \textit{e.g.},][]{Meyer-Vernet93}, so that \citep{Cha91} 
\begin{equation}
	L_D=\frac{1}{\omega_p}\left(\frac{k_\mathrm{B}T}{m}\frac{2\kappa -3}{2\kappa -1}\right) ^{1/2}
	\label{Eq:LD}
\end{equation}
with the (angular) plasma frequency $\omega _p=(ne^2/\epsilon _0 m)^{1/2}$, where $e$ is the electric charge and $\epsilon_0$ the permittivity of the vacuum.

Equations (\ref{Eq:T}) and (\ref{Eq:LD}) show that the smaller the index $\kappa$, the smaller the most probable speed and the Debye length for a given temperature. This highlights the fact that at low speeds, the $\kappa$-distribution can be fitted with a Maxwellian of temperature smaller than its actual kinetic temperature given by Equation\,(\ref{Eq:T}). In contrast, at high speeds, the $\kappa$-distribution has a power-law shape which could mimic, albeit in a very narrow velocity range, a Maxwellian of much higher temperature. We shall see an example of this application in Section\,\ref{Sect:2.5}; depending on the energy range used in the analysis, a different ``temperature" is revealed.
This behavior illustrates the two faces of the $\kappa$-distribution, discussed in detail by \citet{Nicholls12}. When properly taken into account, these two aspects enable one to resolve many apparent contradictions that can arise when Maxwellians are assumed to analyse observations. 

%
%
\subsection{Origin of supra-thermal tails in plasma velocity distributions}
\label{Sect:2.2}
Supra-thermal tails originate naturally from the plasma collisional properties, which are governed by Coulomb collisions. The cross-section for close Coulomb collisions is defined from the Landau radius, the distance at which the Coulomb interaction energy equals the kinetic energy. Hence the collisional cross-section decreases as the inverse square of the energy, which is the fundamental reason why velocity distributions have supra-thermal tails in astrophysical plasmas, which as a rule are both diffuse and inhomogeneous. Indeed, when the low-energy particles have a free-path (proportional to their inverse collisional cross-section) of the order of magnitude of the scale height, particles moving, say, 3 times faster, have a free-path greater by a factor of $3^4$, which amounts to two orders of magnitude, as illustrated with a numerical simulation by \citet{Beck08}. Hence, for the velocity distribution to be quasi-Maxwellian, and the plasma amenable to a fluid description, the Knudsen number, the ratio of the mean free-path to the scale height, should be much smaller than unity \citep{Scudder79,Scudder83}, namely, smaller than $10^{-3}$ \citep{Shoub83} or $10^{-2}$ \citep{Scudder13}, depending on the context. In practice, this means that in most space plasmas, including the solar transition region, corona and wind, velocity distributions should have supra-thermal tails, as is indeed observed in the solar wind \citep[see \textit{e.g.}][]{Pierrard10}.

A number of additional mechanisms have been proposed for producing supra-thermal tails in the corona and solar wind, such as turbulence, plasma instabilities, whistler waves, Langmuir fluctuations and nanoflares \citep[see][and references therein]{Meyer-Vernet01,Pierrard11,Vocks12,Che14,Yoon16}. The wave-particle interaction involving whistler waves is discussed in Section \ref{Sect:2.7}. For the generation of $\kappa$-distributions by turbulence with a diffusion coefficient proportional to $1/v$, we refer the reader to the works of \citet{Hasegawa85}, \citet{Laming07}, and \citet{Bian14}. In the latter, the $\kappa$-distributions are obtained analytically if the diffusive acceleration is balanced by Coulomb collisions in solar flare conditions.

Finally, non-extensive generalizations of thermodynamics involving long-range interactions in plasma have been shown to produce $\kappa$-distributions \citep[see \textit{e.g.}][]{Tsallis09,Livadiotis09,Livadiotis13,Livadiotis15a}.

%
%
\subsection{Effect of supra-thermal tails on the heat flux}
\label{Sect:2.3}
Since the heat flux is the third moment of the velocity distribution (in the frame where the mean velocity is zero), it is expected to be mainly carried by suprathermal electrons, which, as already noted, are generally non-collisional.

Indeed, as found already by \citet{Shoub83}, the heat flux in the solar transition region, corona, and wind is generally not given by the standard collisional \citet{Spitzer53} value, even though the Knudsen number is smaller than unity \citep{Scudder13}. This point has been confirmed recently by \citet{Landi14a}.

\begin{figure}[h!] 
	\centering
	\includegraphics[width=0.5\textwidth]{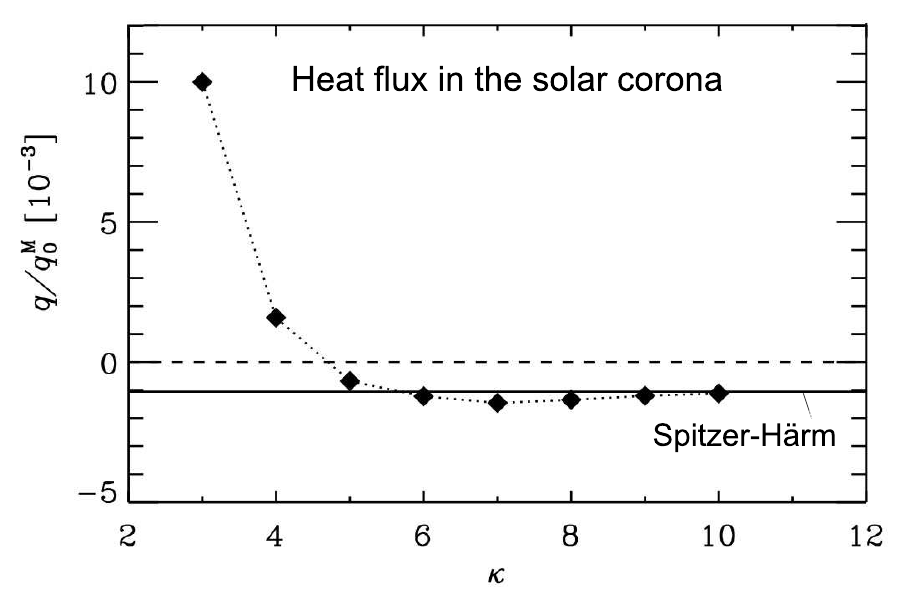}
	\caption{Heat flux $q$ in the solar corona (normalized, as explained in the text), from a numerical simulation with different values of $\kappa$ (dotted), compared to the classical Spitzer-H\"{a}rm heat flux (solid line). Credit: \citet{Landi01}, reproduced with permission \textcircled{c} ESO.
\label{Fig:LPheatflux}}
\end{figure}

Figure \ref{Fig:LPheatflux} shows how the heat flux in the solar corona changes when the velocity distributions differ from Maxwellians. The heat flux is calculated from a numerical simulation of \citet{Landi01}, taking collisions into account for different values of $\kappa$. It is normalised to $10^{-3} q_0^\mathrm{M}$\,=\,$10^{-3}m_e n_0 v_0^3$, \textit{i.e.} to 3 orders of magnitude less than the maximum heat flux that the plasma can presumably sustain; here, $n_0$ and $v_0=(2k_\mathrm{B}T_0/m_e)^{1/2}$ are the electron density and thermal speed at the base of the simulation. As expected, for $\kappa \rightarrow \infty$, the heat flux approaches the Spitzer-H\"{a}rm value. However, as the supra-thermal tail increases ($\kappa$ decreases), the heat flux changes, and when $\kappa <5$, the sign becomes opposite, with heat flowing upwards, from cold to hot regions. Furthermore, for $\kappa \lesssim 3$, the heat flux exceeds the classical Spitzer-H\"{a}rm value by one order of magnitude, in addition to flowing upwards \citep{Landi01}.

%

Recall that whereas the ions carry the plasma momentum, heat is mainly transported by the electrons, because of their small mass. In the solar wind, where the Knudsen number is much larger than in the corona, being of the order of magnitude of unity, the heat flux is expected to be much closer to the non-collisional heat flux than to the classical Spitzer-H\"{a}rm value. This non-collisional heat flux, introduced by \citet{Hollweg74}, is produced by the fast electrons escaping from the interplanetary electric potential, that skew the solar wind electron velocity distribution since in the absence of collisions, no electrons are coming from infinity. This non-collisional heat flux turns out to be of the same order of magnitude as the Spitzer-H\"{a}rm value in the solar wind at 1 AU \citep{Meyer-Vernet07}, but both vary very differently with plasma properties since the non-collisional heat flux is roughly proportional to the electron density and to the wind speed. It has been shown from data and simulations that the solar wind heat flux behaves indeed as the non-collisional heat flux when the Knudsen number exceeds 0.01 \citep{Landi14a}, as predicted by \citet{Scudder13}.

Since the non-collisional heat flux is produced by the tail of the electron velocity distribution, it increases when this tail exceeds the thermal value, which can have major consequences on energy transport, affecting the temperatures and the wind acceleration. Examples are outlined in the next section.

%
%
\subsection{Heating the solar corona and accelerating the solar wind with supra-thermal tails}
\label{Sect:2.4}
\citet{Scudder92a,Scudder92b} suggested that suprathermal particles at the base of the transition region could be at the origin of the temperature increase with altitude culminating at millions of degrees in the corona. This is because the particles around the Sun are subjected to an attracting potential, produced by gravity and by the electrostatic field that ensures plasma quasi-neutrality; this potential filters the particles, because it confines the low-energy particles at the base of the atmosphere since they do not have enough energy to climb up the potential well, whereas the more energetic particles are capable of reaching higher altitudes. Hence, the particles' kinetic temperatures increase with altitude. Since the final temperatures increase with this potential, which depends on the mass and charge state, being larger for heavy ions, minor ions can reach higher temperatures, as has been observed \citep[see \textit{e.g.}][]{Pierrard14}.


\begin{figure}[!t]
	\centering
	\includegraphics[width=0.5\textwidth,clip=]{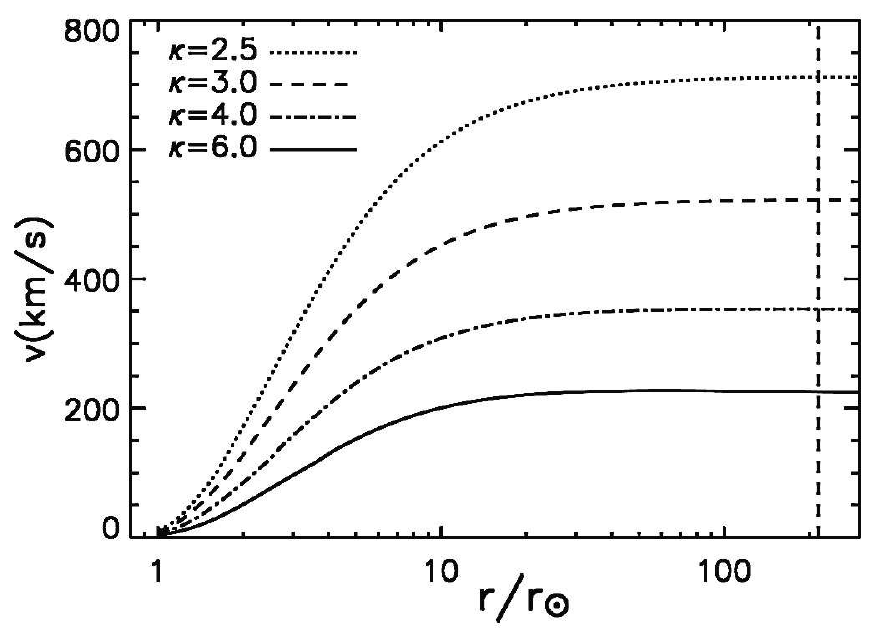}
	\caption{Solar wind speed as a function of heliospheric distance (normalised to the solar radius) from an exospheric model with different values of $\kappa$. From \citet{Zouganelis04}. \textcircled{c} AAS. Reproduced with permission.
	\label{Fig:Zouganelis}}
\end{figure}


Suprathermal electrons not only contribute to heating the solar corona, they should also be capable of accelerating the solar wind, as proposed by \citet{Olbert81}. This is easily understood by writing the energy equation between the base of the wind $r_0$ where the bulk speed is negligible, and a large distance where both the enthalpy and the gravitational energy are negligible compared to the values at $r_0$. This yields the terminal kinetic energy per wind unit mass \citep{Meyer-Vernet07}
\begin{equation}
	V^2/2=5k_\mathrm{B}T_0/m_p -M_{\astrosun}G/r_0 + q_0/ (\rho _0 V_0)
	\label{Eq:energy}
\end{equation}
Here, $V$ is the terminal wind speed, $m_p $ the proton mass, $G$ the gravitational constant, $M_{\astrosun}$ the solar mass, and $T_0$, $\rho _0 V_0$, and $q_0$ respectively the temperature, mass flux and heat flux at the base of the wind. Equation (\ref{Eq:energy}) shows that with typical values of the parameters, accelerating the fast wind requires a heat flux at the base $q_0 \geq 70$ W/m$^2$, which is much larger than the collisional heat flux \citep{Meyer-Vernet07}. This shows that not only does the heat flux play a crucial role for accelerating the wind, but it must be much larger than the classical collisional value.

The simplest models applicable in that case are the so-called exospheric models \citep{Lemaire10}. Such non-collisional models start with an assumed velocity distribution at the base of the wind, and calculate the distribution further away from the Sun by applying Jeans' theorem. Contrary to the fluid models \citep{Parker10}, the heat flux is not assumed a priori, but calculated everywhere from the third moment of the velocity distribution. The calculations, which are fully self-consistent since the electric potential is calculated at all distances from charge quasi-neutrality and zero charge flux, have been applied to $\kappa$-distributions \citep{Maksimovic97b}. The terminal wind speed and electron temperature at large distances can even be estimated analytically as a function of $\kappa$ \citep{Meyer-Vernet98i,Meyer-Vernet99}, suggesting that the high-speed wind can be produced with $\kappa \leq 3$. Figure \ref{Fig:Zouganelis} shows the wind speed computed as a function of heliocentric distance for different values of $\kappa$ \citep{Zouganelis04}. Interestingly, the wind speeds predicted by the exospheric models have been confirmed by numerical simulations taking collisions into account \citep{Zouganelis05}. However, exospheric models are unable to predict correctly the particle velocity distributions, which must be computed using full kinetic simulations taking collisions into account \citep{Landi12a}.

Such kinetic simulations, which take only into account the basic particle properties, \textit{i.e.}, expansion and collisions, can explain the solar wind observed electron heat flux driving the electron energetics, but the effect of direct heating from external sources as wave instabilities and turbulence is still in debate \citep[see, \textit{e.g.},][]{Cranmer09,Landi14a}.

%
%
\subsection{In situ diagnostics in the solar wind from thermal noise spectroscopy}
\label{Sect:2.5}
Plasma particle velocity distributions are traditionally measured with electrostatic particle analysers which separate and count the particles in different energy ranges. Such methods are often hampered by spacecraft charging effects; in the solar wind, this complicates the measurement of electrons of energy smaller than the (positive) spacecraft potential \citep{Maksimovic95}, which must be corrected for both this potential and the photoelectrons emitted by the spacecraft surfaces subjected to solar radiation, whose number density is much greater than that of the ambient solar wind electrons \citep{Garrett81,Salem01}. A complementary technique is thermal noise spectroscopy \citep{Meyer-Vernet98}, which is almost immune to these problems because it senses a much larger plasma volume since it measures the electrostatic fluctuations produced by the quasi-thermal motion of the plasma particles. The power spectrum at the ports of an electric antenna of length exceeding the Debye length has a peak at the plasma frequency $f_p$ (whose measurement reveals the electron density), of shape determined by the velocity distribution \citep{Meyer-Vernet89}. The high-frequency spectral power density is proportional to the electron kinetic temperature, the low-frequency level is mainly determined by the Debye length (thus by low-energy electrons), whereas the level of the peak reveals high energy electrons of speed close to the phase speed of Langmuir waves, which becomes very large when $f \rightarrow f_p$ \citep{Cha91}.

Figure \ref{Fig:LeChat2010} (left) shows an example of power spectral density measured by the \textit{Unified Radio and Plasma Wave Instrument} (URAP) on board the \textit{Ulysses} spacecraft \citep{Stone92} in the solar wind (dots) fitted with two models for the electron velocity distribution: a sum of 2 Maxwellians and a $\kappa$-distribution, with the plasma parameters shown. Both distributions are plotted on the right panel of Figure \ref{Fig:LeChat2010}. The fitting with 2 Maxwellians, labeled $c$, and $h$ (cold and hot), yields 4 electron parameters, which can be expressed as the total electron density $n_\mathrm{e} = n_c+n_h$, the temperature $T_c$ of the cold Maxwellian, and the ratios of the densities $\alpha = n_h/n_c$ and densities $\tau = T_h/T_c$. In contrast, the fitting with a $\kappa$-distribution yields 3 electron parameters: $n_\mathrm{e}$, $T_\mathrm{e}$, and $\kappa$ \citep{LeChat10}. Both fits yield a similar kinetic temperature $T_\mathrm{e}=T_c(1+\alpha \tau)$.

The technique of quasi-thermal noise spectroscopy has been used on a number of space probes for measuring electron properties in various solar system plasmas, including the solar wind near 1 AU and further away from the Sun, on board the \textit{ISEE3}, \textit{Ulysses}, \textit{WIND}, and \textit{STEREO} spacecraft. It will be implemented in the solar wind near 60 $R_\odot$ on \textit{Bepi-Colombo} \citep[with the instrument SORBET,][]{Moncuquet06}, on \textit{Solar Orbiter} \citep[with RPW,][]{Maksimovic05}, and on \textit{Solar Probe Plus} \citep[with the instrument FIELDS,][]{Bale16} near 10 $R_\odot$.

%
%
\begin{figure}
	\centering
	\includegraphics[width=0.55\textwidth,clip=]{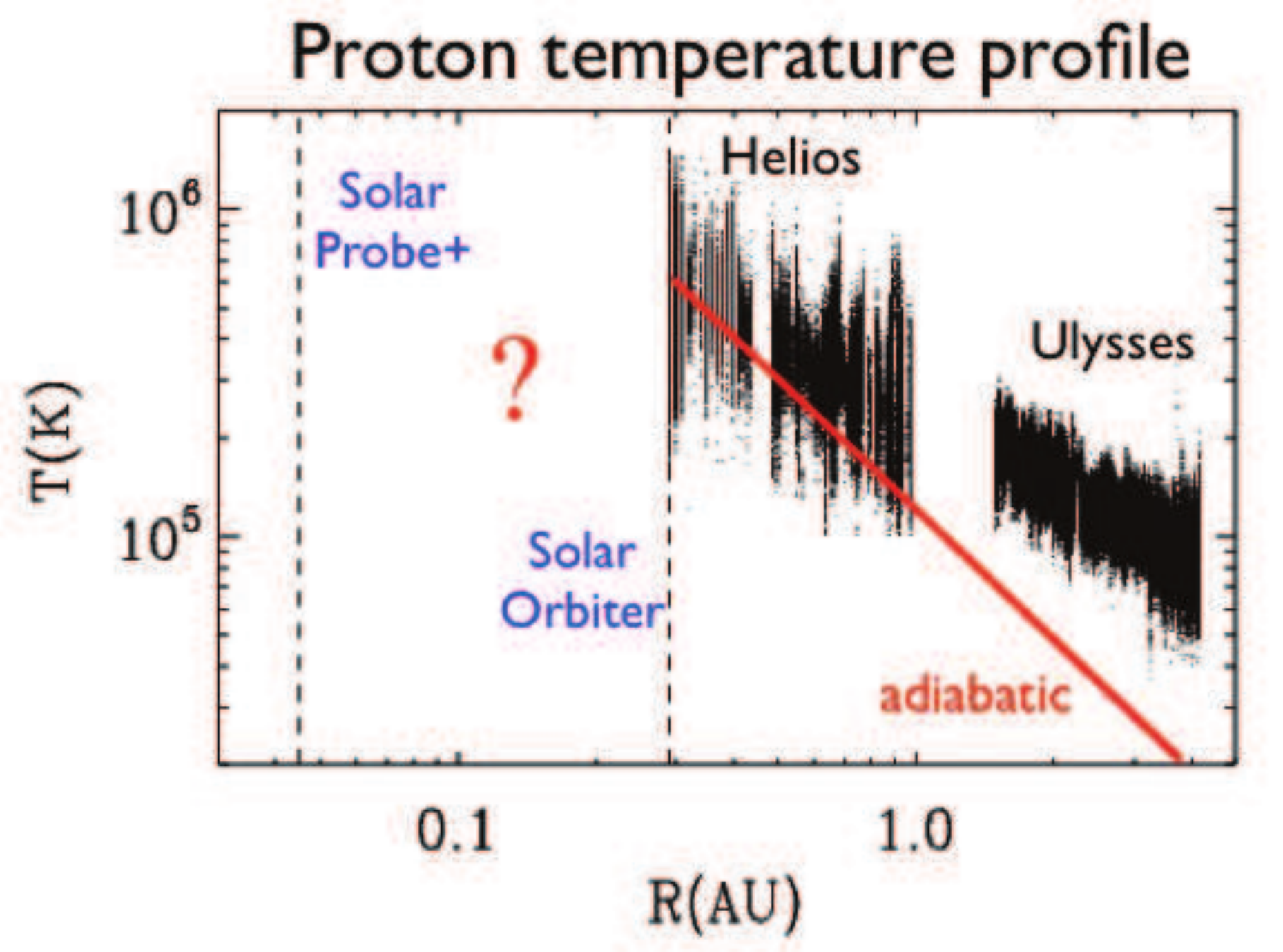}
	\caption{Proton temperature radial profile as measured by \textit{Helios} (0.3–1AU) and \textit{Ulysses} (1.5–5AU) in the fast solar wind. Adiabatic evolution for a collisional expanding gas is shown as the red line. Perihelions of the \textit{Solar Orbiter} and \textit{Solar Probe Plus} future missions are shown by dashed lines.}
	\label{Fig:Proton}
\end{figure}
%
\begin{figure}
	\centering
	\includegraphics[width=.9\textwidth,clip=]{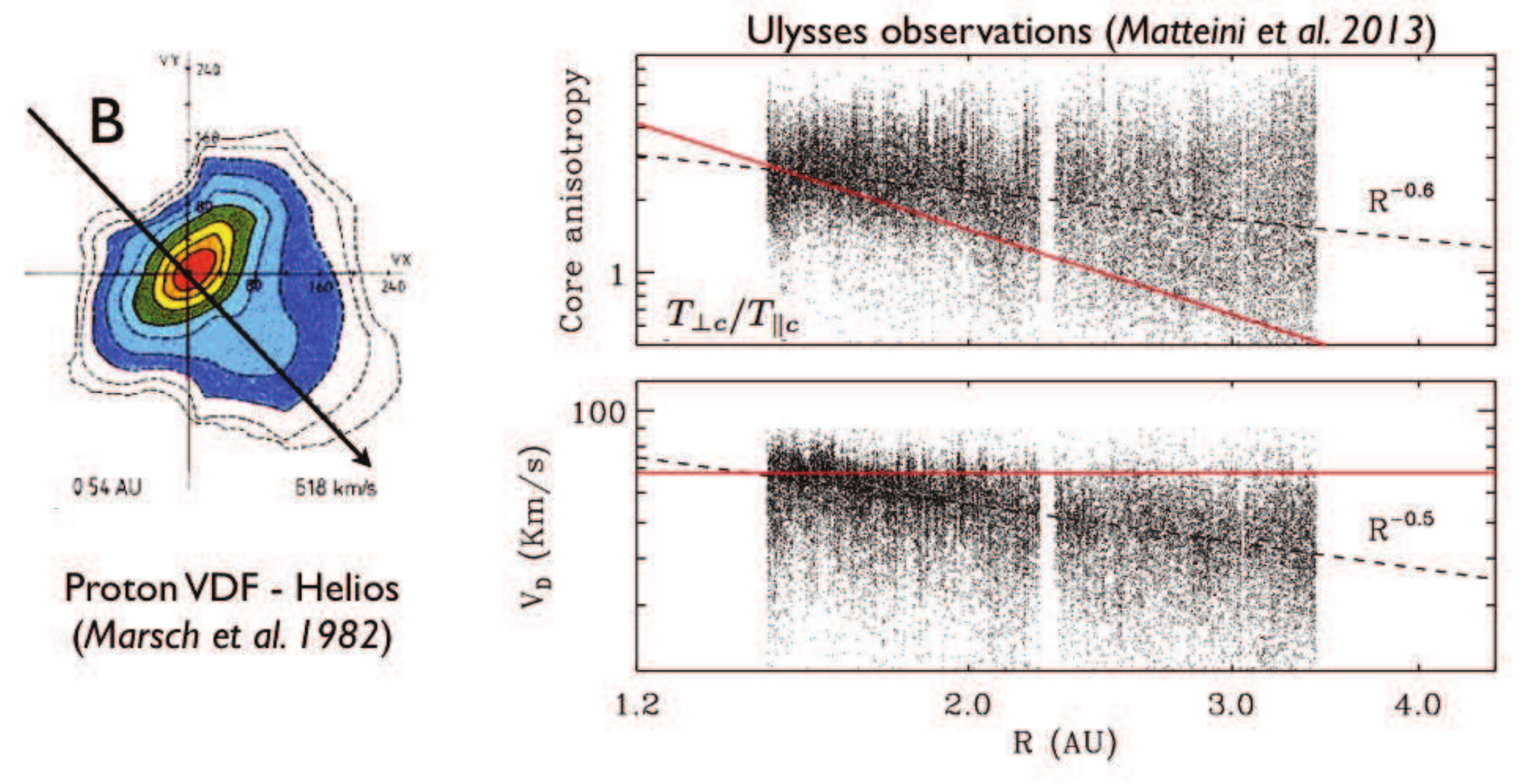}
	\caption{Left: typical proton velocity distribution measured in the fast solar wind by the \textit{Helios} spacecraft. Typical non-thermal features are observed: anisotropy with respect to the local direction of the magnetic field (solid black line) and relative drift between a thermal proton core and a secondary population (beam) streaming faster along the magnetic field.
	Right: Radial evolution of non-thermal properties core anisotropy (top right) and core-beam relative drift normalized to the local Alfv\'{e}n speed (bottom right). Red lines show the double-adiabatic expectations. Non-thermal properties are expected to be more enhanced closer to the Sun.
	Reproduced with permission from \citet{Marsch82b} and \citet{Matteini13}, \textcircled{c} John Wiley and Sons.
	\label{Fig:Core_beam}}
\end{figure}

%
%
\subsection{In situ observations of the solar wind ions}
\label{Sect:2.6}
As well as the electrons, ions in the solar wind, especially protons and $\alpha$ particles, also show interesting and unexplained characteristics. Figure \ref{Fig:Proton} shows a radial profile of the proton temperature as measured by \textit{Helios} and \textit{Ulysses} spacecraft, orbiting the Sun at 0.3--1 and 1.5--5 AU, respectively. During its expansion, the solar wind plasma cools down less than expected. The red solid line shows the adiabatic prediction $T\propto R^{-4/3}$ for the spherical expansion of a monoatomic gas. The physical reasons for this departure of the solar wind temperatures from the adiabatic profile are not fully known. Measurements at distances of about 0.3AU $\approx$ 60 $R_\odot$ are not yet available, but are planned with the next generation of space missions such as \textit{Solar Orbiter} and \textit{Solar Probe Plus} \footnote{\url{http://solarprobe.jhuapl.edu}}, which will explore the solar wind at regions as close as 10 $R_{\odot}$. (See Section \ref{Sect:6} for description of these spacecraft.)

Furthermore, like the electrons, the ion species in the solar wind are also characterized by non-thermal features \citep{Marsch82a,Marsch82b}. Distribution functions deviate from simple Maxwellians, but cannot be fully captured by a single $\kappa$ model. Figure~\ref{Fig:Core_beam} left shows an example of a typical proton distribution in the solar wind. The local magnetic field, indicated by the black arrow, introduces a preferential direction with respect to which an anisotropy exists in terms of the proton velocity distribution functions (hereafter, VDFs). This leads to the introduction of two different kinetic temperatures along ($T_\|$) and across ($T_\perp$) the main magnetic field. Moreover, a secondary population streaming along the magnetic field ahead of the main core population is also present. Its form is sometimes that of a distinct secondary beam, or that of an elongated shoulder. As a consequence, it is convenient to describe the proton distributions as a sum of two populations, core and beam, drifting along the magnetic field, each of them characterized by its own temperature anisotropy (usually assuming a bi-Maxwellian model). Detailed studies of such non-thermal properties suggests that wave-particle interactions driven by kinetic instabilities are at work in the solar wind and constrain the properties of proton distributions \citep[see \textit{e.g.}][and references therein]{Matteini13}. Similar characteristics are observed for $\alpha$ particles \citep{Maruca12,Matteini15b}.

Note that the presence of strongly non-thermal ion distributions as described above, which are the consequence of the low plasma collisionality, also makes questionable the comparison of the observations with the adiabatic temperature profile shown in Figure7, as this is only valid for a fully collisional Maxwellian gas. For a more consistent description of the evolution of a collisionless plasma, a different model has to be used. An example of such model is the double-adiabatic/CGL closure \citep[cfr.][]{Matteini12}, where the abbreviation stands for Chew-Golberger-Low closure \citep{Chew56}. This model separately considers the evolution of $T_\|, T_\perp$ and particle drifts. Following this approach, the right panels of Figure~\ref{Fig:Core_beam} show the radial evolution of the core anisotropy and the relative core-beam drift in the high-latitude polar wind. Both quantities are compared with the associated CGL predictions for a radial magnetic field shown as the red line in Figure~\ref{Fig:Core_beam}. Both profiles deviate from the prediction: the core anisotropy decreases less steeply than expected, suggesting some local heating sources for $T_\perp$, while the relative drift decreases with distance, which is a signature of slowing down of the beam with respect to the core. These results confirm that instabilities might act during the wind expansion and regulate its non-thermal properties. Detailed calculations of the proton energetics \citep{Hellinger11,Hellinger12} show that additional external sources are needed to explain the observed temperature profiles. These are likely related to complex processes such as turbulence, shocks, magnetic reconnection.

We note that local heating of ions along solar wind expansion cannot fully explain the presence of non-thermal features like those shown in Figure~\ref{Fig:Proton}. Moreover, the observed radial trends (dashed lines) suggest that the proton properties, such as the core anisotropy and beam drift, are expected to be even larger closer to the Sun. It is likely that these in situ signatures (anisotropy, beams) are generated in the corona and slowly relaxed during expansion. This implies that in situ measurements closer to the Sun should sample proton distributions with large departures from Maxwellian and significant contribution from supra-thermal plateaux and beams. How such ion properties affect remote sensing observations of the solar corona is an open question at present. 

%
%
\subsection{Formation and evolution of suprathermal electron distributions in the solar corona and transition region}
\label{Sect:2.7}
%
%
\subsubsection{Suprathermal electron production in the solar corona}
\label{Sect:2.7.1}
Solar wind observations of suprathermal particles, whose velocity distribution functions can be fitted by $\kappa$-functions in the case of electrons, have been discussed above. Since no direct in-situ observations are possible in the corona, we now discuss one possible mechanism for the formation of suprathermal electron tails in the solar corona and the transition region. \citet{Pierrard99} have shown that suprathermal solar wind electrons in the keV energy range can exist at altitudes of 4 $R_\odot$.

\citet{Vocks05} studied the formation of a suprathermal electron component from an initially Maxwellian distribution through resonant interaction with a given whistler wave spectrum. There, the wave spectrum was chosen as a power-law representing the high-frequency tail of wave spectra associated with coronal heating. This led to the formation of suprathermal tails in solar coronal and wind electron VDFs, showing that the suprathermal tail formation can be a by-product of the coronal heating mechanism.

To better understand the suprathermal tail formation process without the influence of open boundary conditions of the simulation box located both in the low corona and in the outer heliosphere, \citet{Vocks08} applied the same model on the closed volume of a coronal loop. These authors showed that power-law like electron VDFs do form inside the loop, with energies up to 10 keV. These models showed that the quiet solar corona is indeed capable of producing a substantial suprathermal electron population, as a by-product of coronal heating. We note that this result should be robust against the actual details of the coronal heating mechanism, as long as it is associated with some electron acceleration. The treatment of suprathermal electron propagation and thermalization in the model, which competes with the production mechanism, is independent of the latter.

%
%
\subsubsection{Propagation of suprathermal electrons in the transition region}
\label{Sect:2.7.2}
In the model of \citet{Vocks08}, the footpoints of the closed loop were located in the transition region. The interior of the simulation box was restricted to high temperatures above $5\,\times10^5\,{\mathrm K}$, and did not include the transition region and upper chromosphere. This was done to avoid the ranges of electron thermal speeds and phase space gradients inside the simulation box being too wide. However, suprathermal electrons from the corona do not stop at the transition region. Since the Coulomb collisional mean free-path scales as $v^4$, coronal electrons are capable of traversing the transition region, and entering the cooler and denser chromosphere. The strong temperature gradient and associated heat flux in the transition region readily suggests that transition region electron VDFs can substantially deviate from a Maxwellian.

To investigate the propagation and evolution of the suprathermal electron population in the strong temperature gradients of the transition region, \citet{Vocks16} extended the simulation box of \citet{Vocks08} into the upper chromosphere. This Vlasov kinetic model for electrons includes Coulomb collisions both with other electrons and the ion background. A Maxwellian distribution is assumed at the lower boundary in the chromosphere. For the upper boundary, the resulting VDF from the loop model of \citet{Vocks08}, with its suprathermal tail, is used. Since there are no heat sources or sinks in the model, a constant heat flux is assumed. The classic Spitzer $T^{5/2}$ law of thermal conductivity in a plasma then leads to a temperature profile with a steep temperature gradient, with a temperature change of the order of $10^4$ K on just a few meters.

%
%
\begin{figure}
	\centering
	\includegraphics[width=8cm]{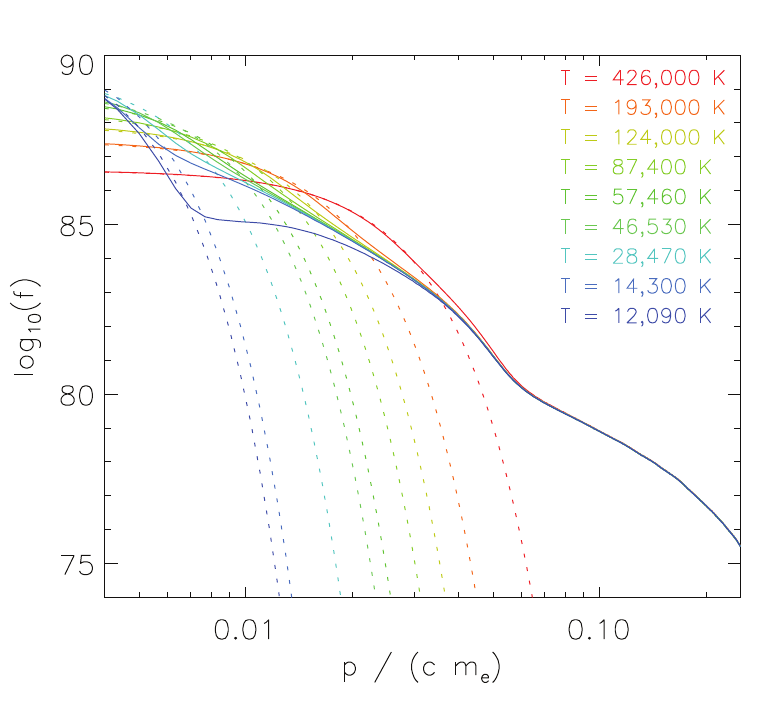}
	\caption{Pitch-angle averaged electron VDFs at different positions in the simulation box (solid lines), and Maxwellian VDFs with the same density and temperature (dotted lines). Adapted from \citet{Vocks16}.
	Credit: \citet{Vocks16}, reproduced with permission \textcircled{c} ESO.
\label{Fig:Vocks_VDFs}}
\end{figure}
%

Figure \ref{Fig:Vocks_VDFs} shows the resulting pitch-angle averaged electron VDFs for different temperature levels in the transition region. An electron momentum of $p = 0.1\,m_\mathrm{e}c$ corresponds to a kinetic energy of 2.5 keV. The energy range of the suprathermal electron shown here therefore covers a few tens of eV up to 10 keV. Maxwellian VDFs with the same densities and temperatures are plotted as dashed lines for reference. The power-law tail of suprathermal electrons is clearly visible in all plots. Note that the VDFs are not normalized. Instead, their numerical values are plotted in absolute units, \textit{i.e.}~$\mathrm{kg}^{-3}\,\mathrm{m}^{-6}\,\mathrm{s}^6$. In this way, it can readily be seen that the suprathermal electrons are hardly absorbed within the thin transition region and that their phase-space density stays nearly constant. Only the line for the lowest temperature, T = 12090 K, shows some deviation for momentum $p < 0.02\,m_\mathrm{e}c$, that corresponds to electron energies below 100 eV. These electrons are absorbed due to the high chromospheric density and the $v^4$ scaling of collisional mean free paths.

These simulation results show that transition region electron VDFs can be far from thermal equilibrium characterized by Maxwellian VDFs.

%
%
%
%
%
%
%
\section{Remote-sensing diagnostics of non-thermal distributions in the solar corona and transition region}
\label{Sect:3}

Remote-sensing observations of the solar corona are becoming increasingly detailed, and high-resolution spectroscopy is becoming more common. It has become clear that spectroscopic evidence of non-thermal electron distributions in the solar corona, other than during the impulsive phase of flares, and especially in the source regions of solar wind, requires not only very well calibrated observations, but also a wide range of accurate atomic data and modelling. In this section, we discuss the optically thin spectral synthesis and the behavior of the resulting synthetic spectrum if non-Maxwellian distributions are considered. Although we consider primarily the $\kappa$-distributions, the spectral synthesis for any non-Maxwellian distribution can be done in an analogous manner.

%
\subsection{Line intensities and atomic data}
\label{Sect:3.1}
The intensity of an optically thin spectral line at wavelength $\lambda_{ji}$ depends mainly on the abundance of the emitting ion $X^{+z}$ and the relative population of the upper emitting level $j$, $N(X^{+z}_j) / N(X^{+z})$ \citep[\textit{e.g.},][]{Mason94,Phillips08}
\begin{eqnarray}
	\nonumber I_{ji} &=& A_X \int \frac{hc}{\lambda_{ji}} A_{ji} N(X_j^{+z}) \mathrm{d}l = A_X \int \frac{hc}{\lambda_{ji}} \frac{A_{ji}}{n_\mathrm{e}} \frac{N(X_j^{+z})}{N(X^{+z})} \frac{N(X^{+z})}{N(X)} n_\mathrm{e} n_\mathrm{H} \mathrm{d}l \\
	&=& A_X \int G_{X,ji}(T,n_\mathrm{e},\kappa) n_\mathrm{e} n_\mathrm{H} \mathrm{d}l\,,
	\label{Eq:line_intensity}
\end{eqnarray}
where $hc/\lambda_{ji}$ is the photon energy, $\lambda_{ji}$ is the emission line wavelength, $A_{ji}$ is the Einstein coefficient for spontaneous emission, $N(X^{+z})/N(X)$ is the relative ion abundance of the $X^{+z}$ ion. The relative abundance of the element $X$ is denoted by $A_X$ and is usually assumed to be constant along the line of sight $l$. In the above expression, the $n_\mathrm{H}$ and $n_\mathrm{e}$ represent the hydrogen and electron number densities, respectively. The function $G_{X,ji}(T,n_\mathrm{e},\kappa)$ is called the line contribution function. It depends on $\kappa$ (or any other non-thermal parameter) due to the dependence of the individual underlying atomic processes resulting in the ionization, recombination, excitation, and de-excitation on $\kappa$ \citep[\textit{e.g.},][]{Dzifcakova92,Dzifcakova01,Dzifcakova02,Dzifcakova06,Wannawichian03,DzifcakovaMason08,Dzifcakova13a,Dudik14a,Dudik14b,Dzifcakova15}.

In case of isothermal plasma, the quantity EM\,=\,$\int n_\mathrm{e} n_\mathrm{H} \mathrm{d}l$ gives the total emission measure of the plasma. However, if the coronal plasma is not isothermal, but the emission along the line of sight arises in regions with different temperature, the intensity is given by
\begin{equation}
	I_{ji} = A_X \int G_{X,ji}(T,n_\mathrm{e},\kappa) \mathrm{DEM}_{\kappa}(T) \mathrm{d}T\,,
	\label{Eq:line_intensity_DEM}
\end{equation}
where the quantity DEM$_{\kappa}(T)$ =\,$n_\mathrm{e} n_\mathrm{H} \mathrm{d}l/\mathrm{d}T$ is the differential emission measure, generalized for $\kappa$-distributions by \citet{Mackovjak14}. 

The above expressions (\ref{Eq:line_intensity}--\ref{Eq:line_intensity_DEM}) assume a fundamental approximation that processes affecting the population of atomic states within an ion can be separated from those that affect the populations of different ions in the plasma. This is justified since the time-scales for ionization and recombination processes are typically longer than those for electron excitation. Departures from this situation, including non-equilibrium ionization, will be discussed in Section \ref{Sect:5}. In some cases, ionization and recombination can make significant modifications to the level populations of an ion, but in the case of ionization equilibrium these can be treated without the need to fully integrate all atomic processes in a single model that incorporates all ions of an element \citep[see][]{Landi06}.

Accurately solving the level balance equations for individual ions is the key goal of the CHIANTI atomic database \citep{Dere97,Landi13,DelZanna15b} and software. This requires large quantities of atomic data describing the ionization, recombination, excitation and de-excitation processes. These processes will be treated in the following subsections \ref{Sect:3.2}--\ref{Sect:3.4}. The dominant excitation process in the solar atmosphere is by electron collisions, but the low electron density means the dominant de-excitation process is spontaneous radiative decay and so the plasma is far from thermal equilibrium. This requires that the level balance equations are solved by including all relevant atomic processes between individual atomic states. CHIANTI ion models can contain hundreds of states, leading to tens of thousands of rates.

The key parameters for describing any electron collision process are the electron flux and the cross-section offered by the ion to change it from an excited atomic state $i$ to a state $j$. For an isotropic particle distribution, the flux of electrons of energy $E=mv^2/2$ is $n_\mathrm{e}v f(E){\rm d}E$, where $f(E)$ is the particle distribution function. The number of transitions occurring per unit volume per unit time is $n_i\sigma_{ij} n_\mathrm{e}v f(E){\rm d}E$, where $n_i$\,=\,$N(X_i^{+z})/N(X^{+z})$ is the number density of ions in a state $i$ and $\sigma_{ij}$ is the cross-section offered by the target ion for transitions to the state $j$. Integrating over energy leads to the following definition for the excitation rate coefficient
\begin{equation}\label{Eq:Rate_coeff}
C_{ij}=\int_{E_{ij}}^\infty \sigma_{ij} vf(E){\rm d}E
\end{equation}
where $E_{ij}=E_j-E_i$ is the energy separation of states $i$ and $j$, and $C_{ij}$ is usually given in units of cm$^3$~s$^{-1}$. For the ionization and recombination, the total rate coefficient is obtained analogously, with the state $j$ then belonging to the ionized ion $X^{+z+1}$, and for recombination it belongs to the recombined ion $X^{+z-1}$. From now on, we will use $C$, $q$ and $\alpha$ for the excitation, ionization and recombination rate coefficients, respectively.

%
%
\begin{figure}
	\centering
	\includegraphics[width=0.32\textwidth]{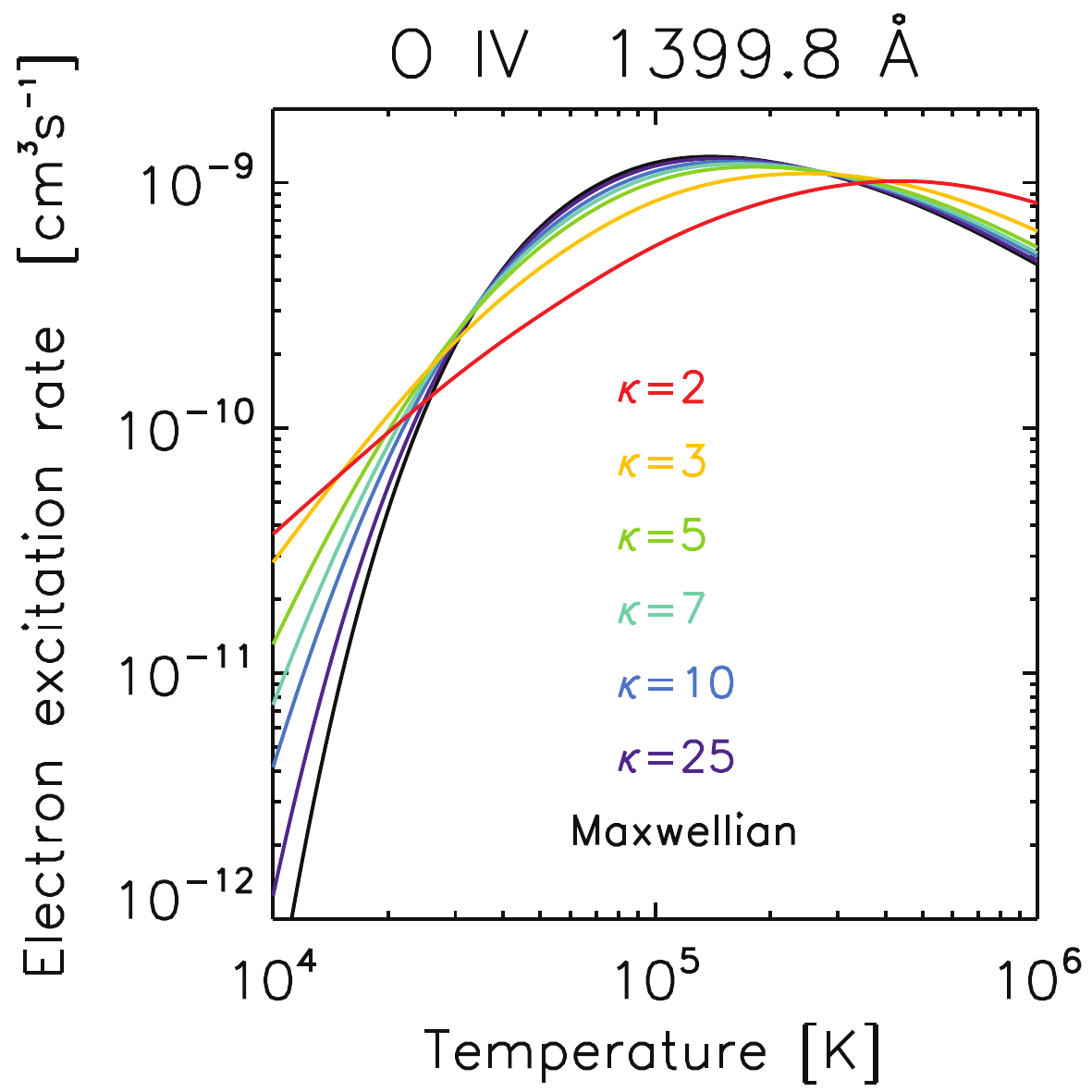}
	\includegraphics[width=0.32\textwidth]{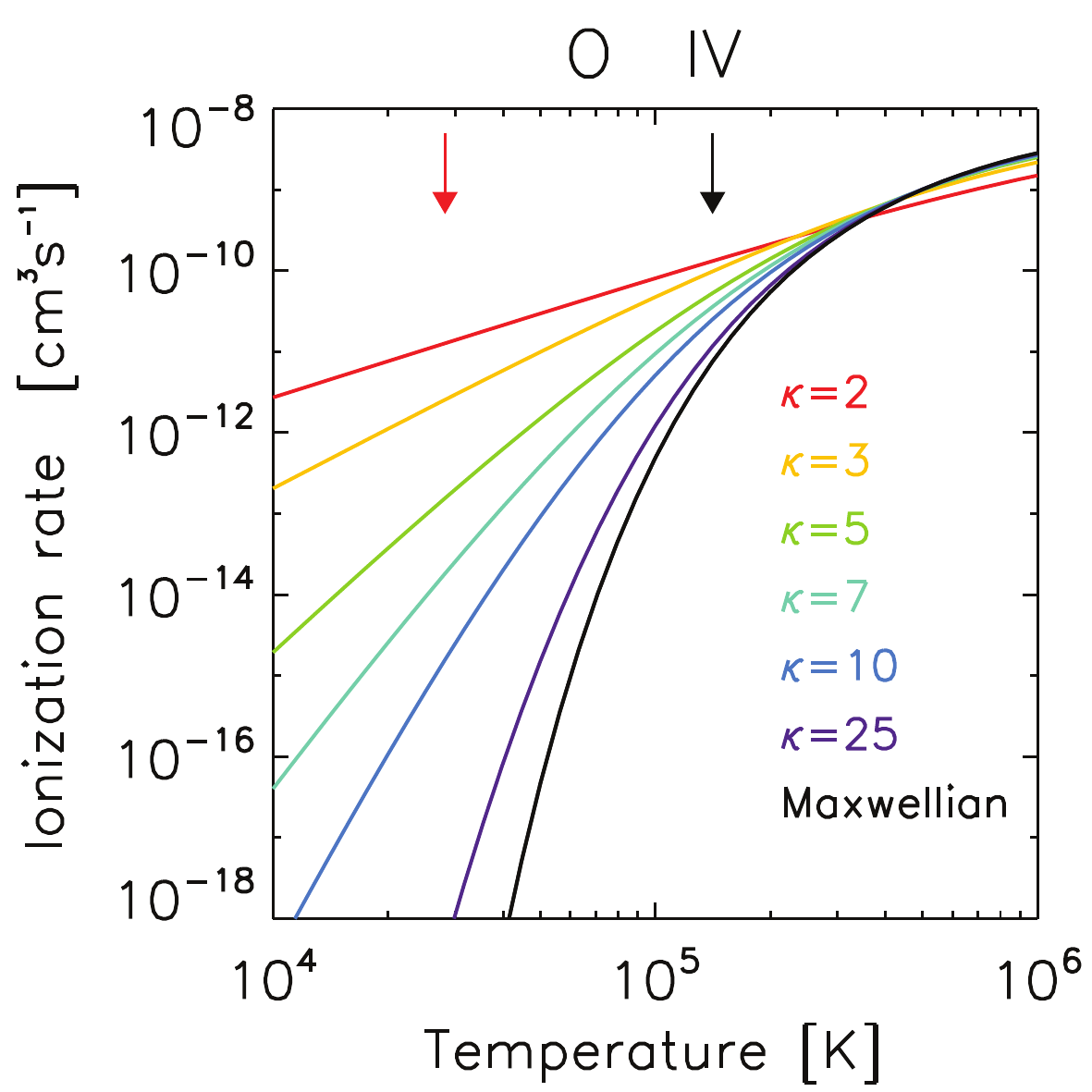}
	\includegraphics[width=0.32\textwidth]{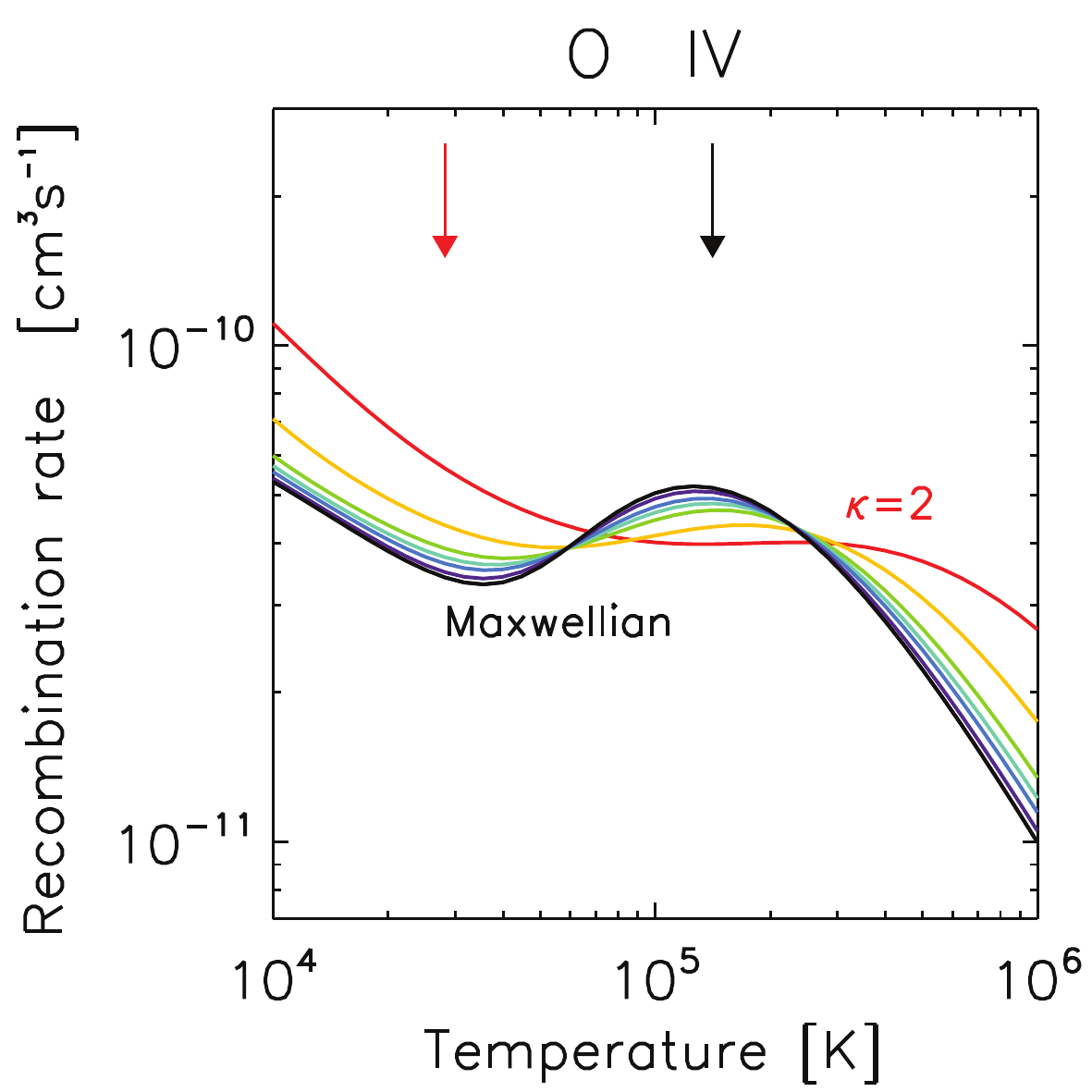}
	\caption{Collisional rates for \ion{O}{IV} and their behaviour for the $\kappa$-distributions: Electron excitation rate for the 1399.8\,\AA~line (left), ionization (middle) and recombination rate (right). Temperatures corresponding to the peak of the relative \ion{O}{IV} abundance for the Maxwellian and $\kappa$\,=\,2 distribution are shown by the black and red arrows, respectively.}
\label{Fig:Rates}
\end{figure}
%
%
\subsection{Electron excitation}
\label{Sect:3.2}
Accurate electron excitation cross-sections are the key data for computing the populations of upper emitting levels that enter into Equation~\ref{Eq:line_intensity}. Atomic physicists generally compute the \emph{collision strength} rather than the cross-section \citep[see, \textit{e.g.},][]{Phillips08}, and an integral over a Maxwellian distribution yields the \emph{effective collision strength}.

In general, atomic physicists publish only the effective collision strengths due to the large size of the collision strength data-sets, usually tens of thousands of data-points for tens of thousands of transitions, which prevents researchers from computing effective collision strengths for non-Maxwellian distributions. One solution is to express the non-Maxwellian as a superposition of Maxwellians, which was implemented in CHIANTI version 5 \citep{Landi06} and recently discussed in \citet{Hahn15b}, who provided an accurate way to decompose a $\kappa$-distribution into a linear series of Maxwellian distributions with different temperatures. Individual non-Maxwellian rates can then be calculated as a linear combination of the Maxwellian ones. Such decomposition is in principle possible for any non-Maxwellian distribution broader than a single Maxwellian. Summing over many Maxwellians however increases the computational time accordingly. Another way to model the non-Maxwellian spectra is to recover the (smooth) behavior of the cross-sections via a parametrisation method. This normally provides cross-sections that are accurate to within a few percent of the original values \citep[see, \textit{e.g.},][]{DzifcakovaMason08,Dzifcakova15}. Once the cross-sections are recovered, it is trivial to convolve them with non-Maxwellians to obtain the rate coefficients. For the $\kappa$-distributions, this method has been implemented using the CHIANTI database (version 7.1) and made available to the community via the KAPPA \footnote{\url{http://kappa.asu.cas.cz}} package \citep{Dzifcakova15}. An example of the electron excitation rate coefficient for the 1401.2\,\AA~line of \ion{O}{IV} and its behaviour for $\kappa$-distributions is shown in the left panel of Figure \ref{Fig:Rates}. We see that the excitation rate is enhanced and low and high temperatures, especially for low $\kappa$ values, and that its peak is decreased and shifted towards higher $T$.

The UK APAP network \footnote{\url{http://www.apap-network.org}} (led by N.R. Badnell, University of Strathclyde) has computed a large amount of electron excitation data for spectral modelling in recent years, and a database of collision strengths has been built that allows line intensities to be calculated in non-equilibrium conditions \citep{Dudik14a,Dudik14b}.

%
%
\subsection{Ionization and Recombination}
\label{Sect:3.3}
%
\subsubsection{Ionization}
\label{Sect:3.3.1}
Electron impact ionization can take place either directly or via excitation to an unstable state above the ionization threshold that then auto-ionizes. The cross-sections for both direct and excitation-autoionization can be expressed as analytic forms in terms of the electron energy or temperature \citep[\textit{e.g.},][]{Younger81,Arnaud92}, fitted to computational or laboratory data. These allow ionization rates to be calculated for arbitary electron distributions, including non-equilibrium distributions \citep{Dzifcakova13a}. An example of the total ionization rate and its behaviour with $\kappa$ is shown for \ion{O}{IV} in Figure \ref{Fig:Rates}. We see that the ionization rate increases by orders of magnitude especially at low $T$.

%
\subsubsection{Recombination}
\label{Sect:3.3.2}
Recombination is the capture of an electron into a stable excited level of an ion. The direct process is referred to as radiative recombination (RR), while capture into an unstable doubly-excited state that then radiatively decays to a stable state is called dielectronic recombination (DR). RR and DR rates are generally computed separately.

Atomic physicists compute RR rates from photoionization cross-sections as the two processes are related by the principle of detailed balance \citep[\textit{e.g.},][]{Phillips08}. Cross-sections for transitions from the ground and excited states of the recombined ion to the ground state of the recombining ion are needed, and the the rates are generally published as the parameters of analytical fit formulae that are functions of temperature \citep[\textit{e.g.},][]{Gu03,Badnell06a}. Non-Maxwellian rates would requires access to the original cross-sections, but \citet{Dzifcakova92} presented a method that allows $\kappa$-distribution rates to be computed directly from the Maxwellian fit parameters and this was applied by \citet{Dzifcakova13a} to all ions available in CHIANTI.

For DR, the inverse process to dielectronic capture is autoionization and modern calculations \citep[\textit{e.g.},][]{Badnell03} typically compute the DR rates from the autoionization rates and the radiative decay rates that stabilize the doubly-excited states. The DR rates are generally published as the parameters of a fitting formula that is a function of temperature \citep[\textit{e.g.},][]{Arnaud92}. \citet{Dzifcakova92} and \citet{Dzifcakova13a} provided a modified formula and fit parameters for the $\kappa$-distribution, while rates for other distributions can be derived numerically by integrating the dielectronic capture rates. An example of the total recombination rate for \ion{O}{IV} and its behaviour for $\kappa$-distributions is shown in Figure \ref{Fig:Rates}, right.

%
%
\begin{figure}
	\centering
	\includegraphics[width=1.00\textwidth]{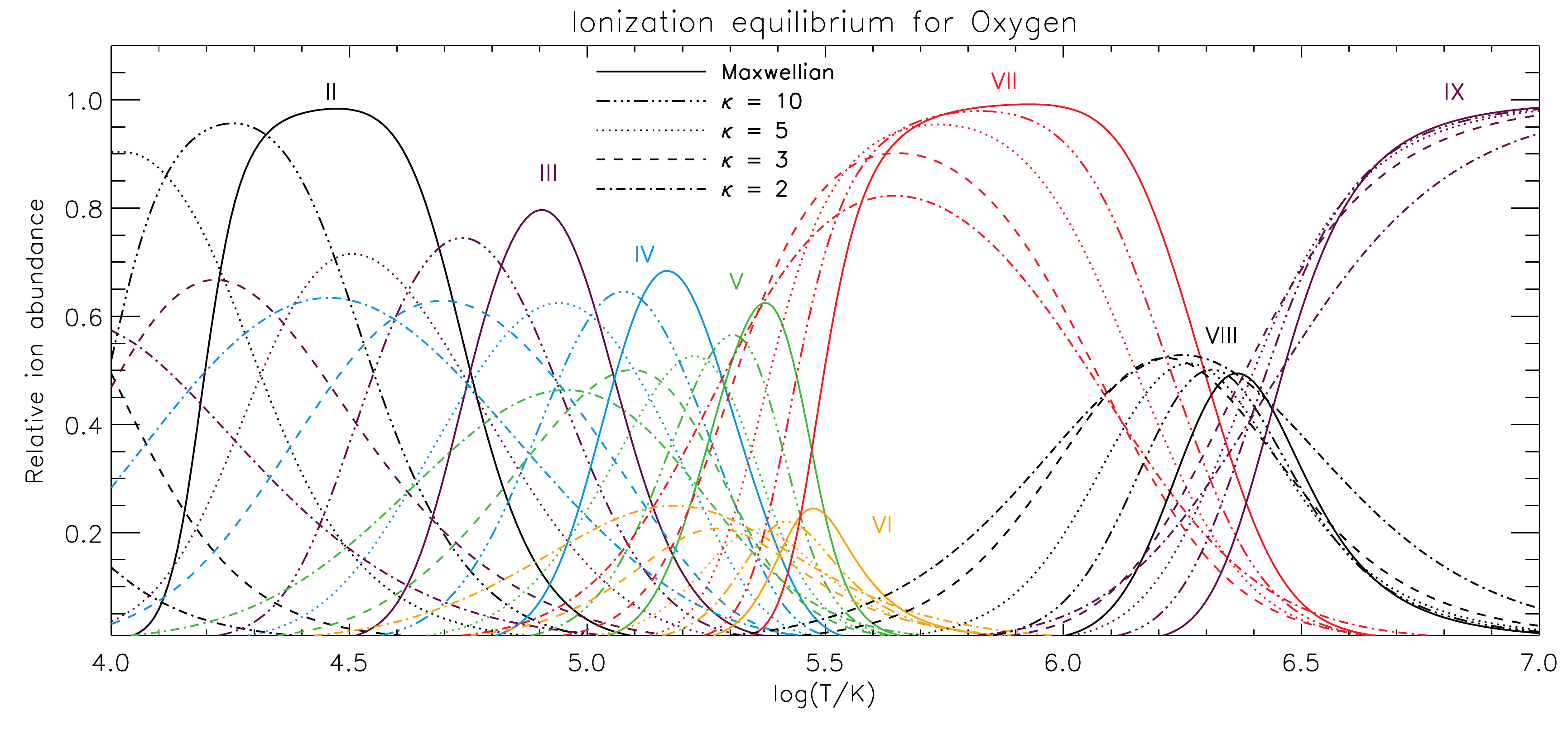}
	\caption{Behavior of the oxygen ionization equilibrium for the $\kappa$-distributions. Relative ion abundances for individual ions are shown in different colors, while different line-styles stand for different values of $\kappa$.}
\label{Fig:Ioneq_O}
\end{figure}
%
%
\subsubsection{Behavior of the ionization equilibrium for the $\kappa$-distributions}
\label{Sect:3.3.3}
Once the ionization and recombination rates are obtained, the ionization equilibrium can be calculated under an assumption that the total number of ionizing transitions is balanced by the total number of recombining transitions
\begin{equation}
	0 = n_\mathrm{e} [q^{(z-1)}N^{(z-1)} + (q^{(z)}+\alpha^{(z)})N^{(+z)} + \alpha^{(z+1)}N^{(z+1)}]
	\label{Eq:Ion_eq}
\end{equation}
where we used a shortened notation $N^{(+z)}$\,=\,$N(X^{+z})$. A number of such ionization equilibrium calculations have been published for different sets of cross-sections. For the Maxwellian distributions, older calculations of \citet{Arnaud85}, \citet{Arnaud92}, \citet{Mazzotta98} and \citet{Bryans09} were available in the CHIANTI database up to version 7.1 \citep{Landi13}. State-of-art calculations utilize the ionization rates of \citet{Dere07} and the dielectronic recombination rates calculated by \citet[][ and subsequent papers]{Badnell03}. These rates have been used by \citet{Dzifcakova13a} to calculate the corresponding ionization balances for the $\kappa$-distributions. These are available within the KAPPA package \footnote{\url{http://kappa.asu.cas.cz}}. 

For the $\kappa$-distributions, individual ions can typically exist in a wider range of temperatures than for the Maxwellian distribution. An example for oxygen is shown in Figure \ref{Fig:Ioneq_O}. The transition-region ions are furthermore strongly shifted towards lower temperatures as a consequence of rapidly increasing total ionization rate, by orders of magnitude \citep[see, \textit{e.g.}, Figure 2 in][]{Dzifcakova13a} dominated by the high-energy tail of the $\kappa$-distribution. In contrast, the recombination is dominated by the low-energy electrons. At coronal temperatures, the shift of the peak of the relative ion abundance can be either to higher or lower temperatures. For coronal Fe ions, the shift is typically towards higher log($T$). For oxygen, \ion{O}{viii} is shifted towards lower $T$, while the behavior is opposite for \ion{O}{ix} (Figure \ref{Fig:Ioneq_O}).

\subsubsection{Density-dependent effects on the ionization state}
\label{Sect:3.3.4}
The tabulations of ion charge state distributions that are widely used in the literature are computed in the zero-density approximation, \textit{i.e.}, recombination and ionization take place only out of the ground states of the ions. However, high electron densities can alter the ionization state of the plasma. A strong effect is the suppression of dielectronic recombination. This occurs because intermediate states below the ionization threshold that are populated through cascading as the stabilization process proceeds can be re-ionized by electron collisions. The suppression of dielectronic recombination have been estimated by Collisional-Radiative models and tabulated by \citep{Summers74a,Summers74b}. \citet{Nikolic13} used these tabulated rates to compute the density-dependent suppression factors for the zero-density DR rates of \citet{Badnell03} and subsequent papers.

Another effect of high density is that metastable states of the ions can be significantly populated so that the ionization and recombination from these states also need to be taken into account.
Both density effects are modelled in the Atomic Data and Analysis Structure (ADAS) \footnote{http://adas.ac.uk}, which uses the Generalized Collisional-Radiative (GCR) approach \citep{McWhirter84,Summers06} to simultaneously include all radiative and electron collision processes, and the influence of highly excited states onto the dominant revolved low-lying states. The calculation yields effective ionization and recombination rates as functions of temperature and density and these are made available through OPEN-ADAS \footnote{http://open.adas.ac.uk}. An example of the relative ion abundances of \ion{Si}{IV}, \ion{O}{IV}, and \ion{S}{IV} taking these effects into account can be found in Figure 13 of \citet{Polito16b}.



%
%
%
\begin{figure}
	\centering
	\includegraphics[width=0.49\textwidth]{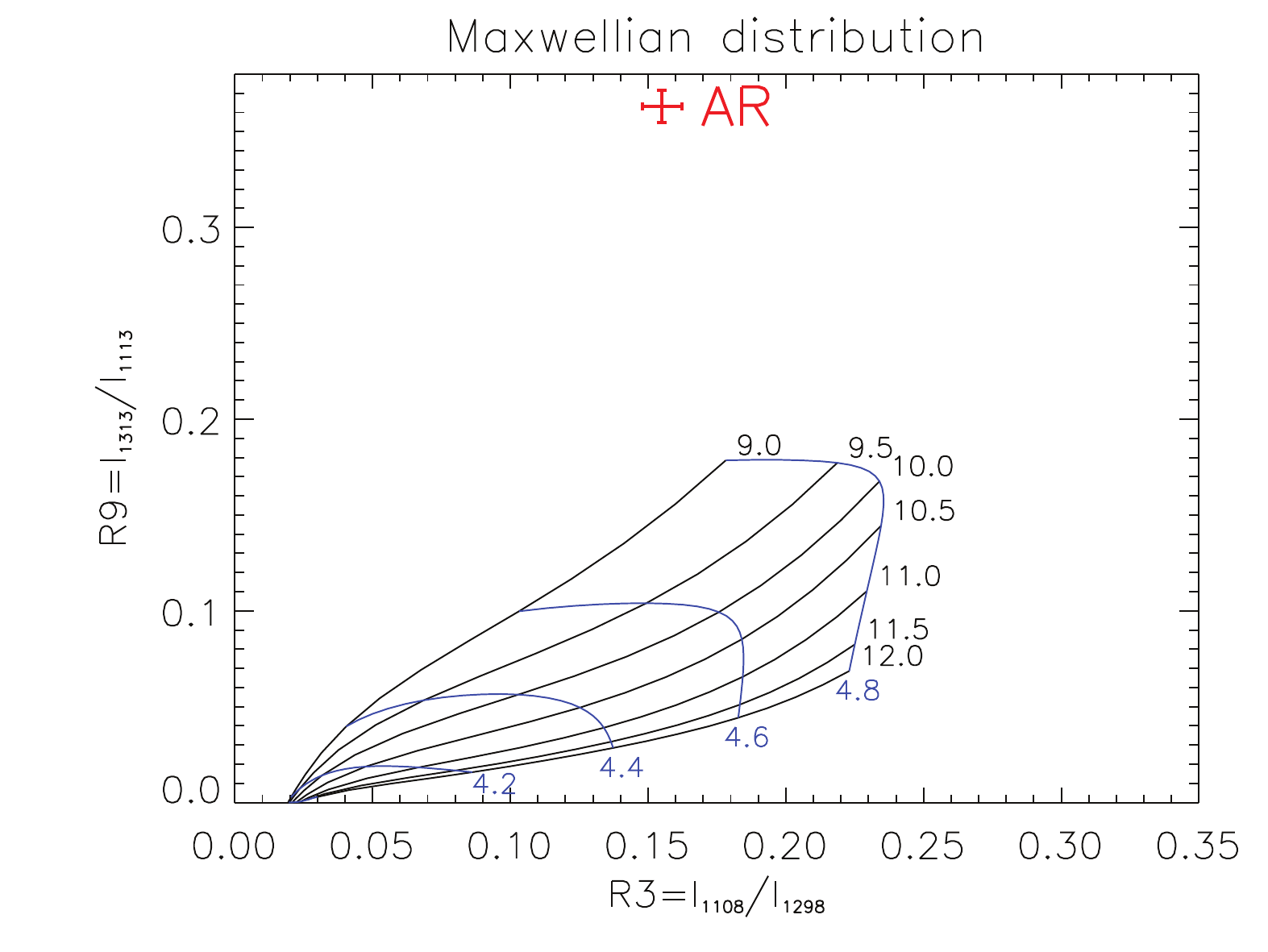}
	\includegraphics[width=0.49\textwidth]{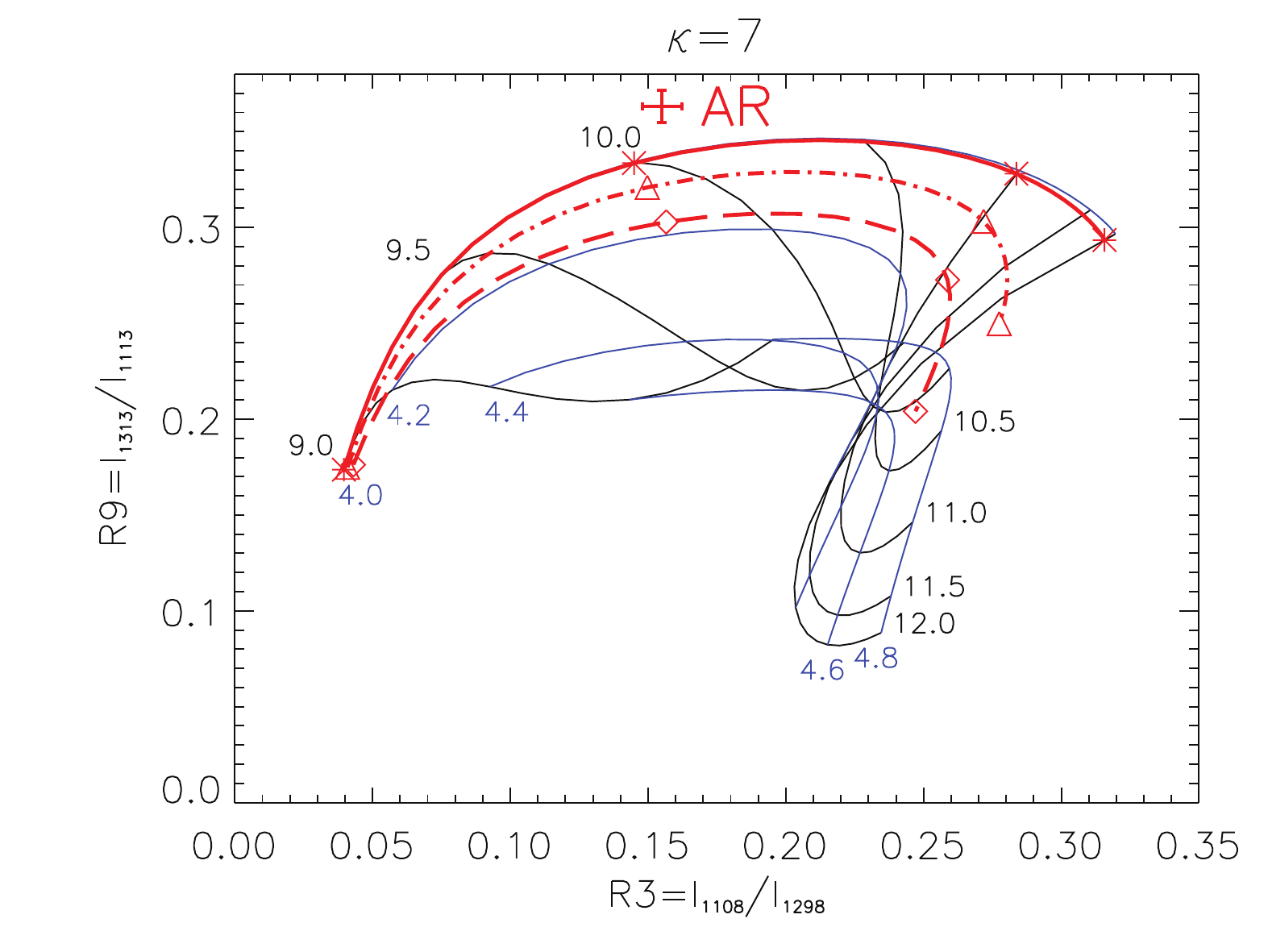}
	\caption{\ion{Si}{iii} line intensity ratios observed in the active region spectrum by \citet{Pinfield99} and analyzed for the $\kappa$-distributions by \citet{DzifcakovaKulinova11}. Left: Maxwellian-predicted ratios as a function of log($T$\,[K]) and log($n_\mathrm{e}$ [cm$^{-3}$]). Right: The same for the $\kappa$\,=\,7. The red lines show the predicted ratios for several DEMs from the CHIANTI database. 
	Credit: \citet{DzifcakovaKulinova11}, reproduced with permission \textcircled{c} ESO.
\label{Fig:Si3}}
\end{figure}
%
%
\begin{figure}[htbp]
	\centering 
	\includegraphics[width=0.6\textwidth,clip=]{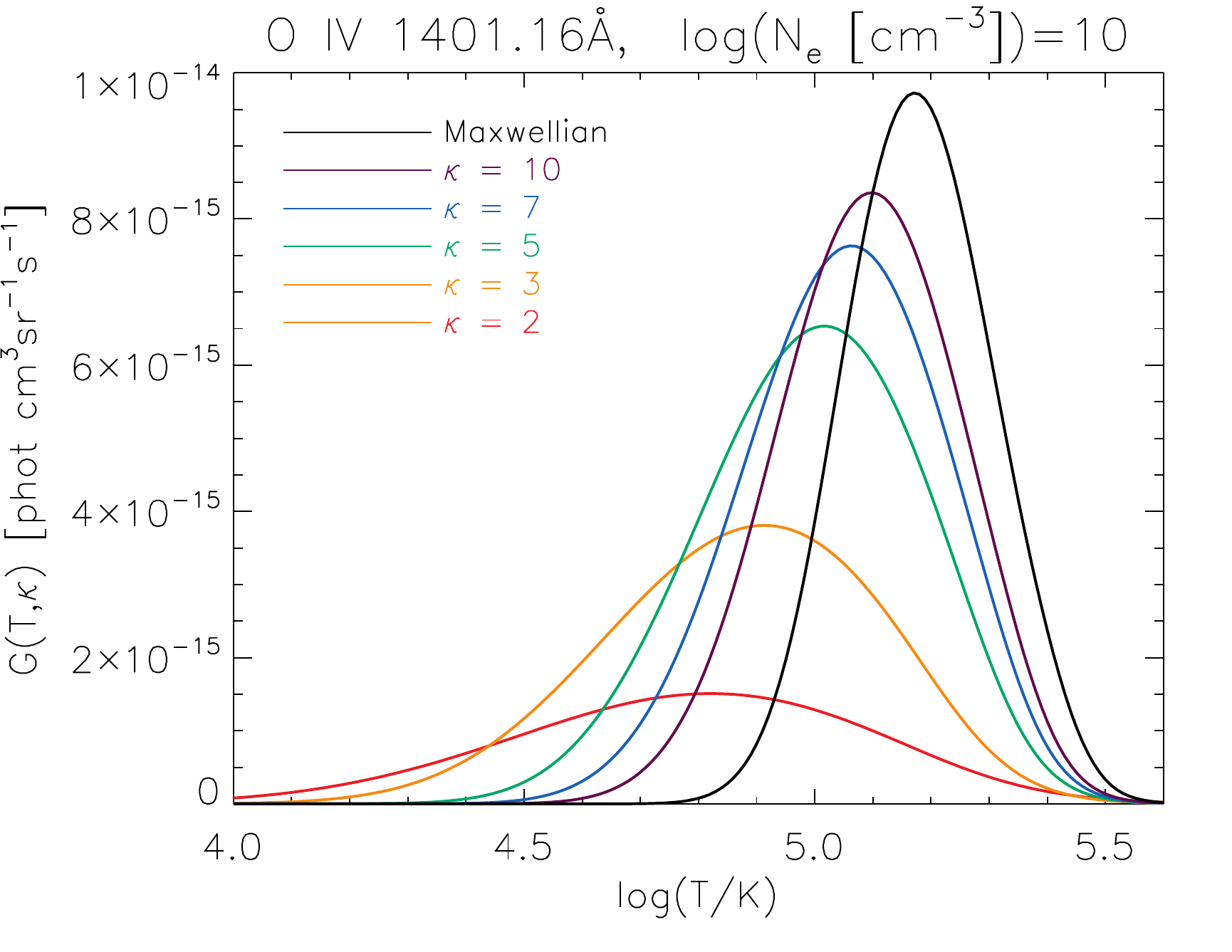}
	\includegraphics[width=0.6\textwidth,clip=]{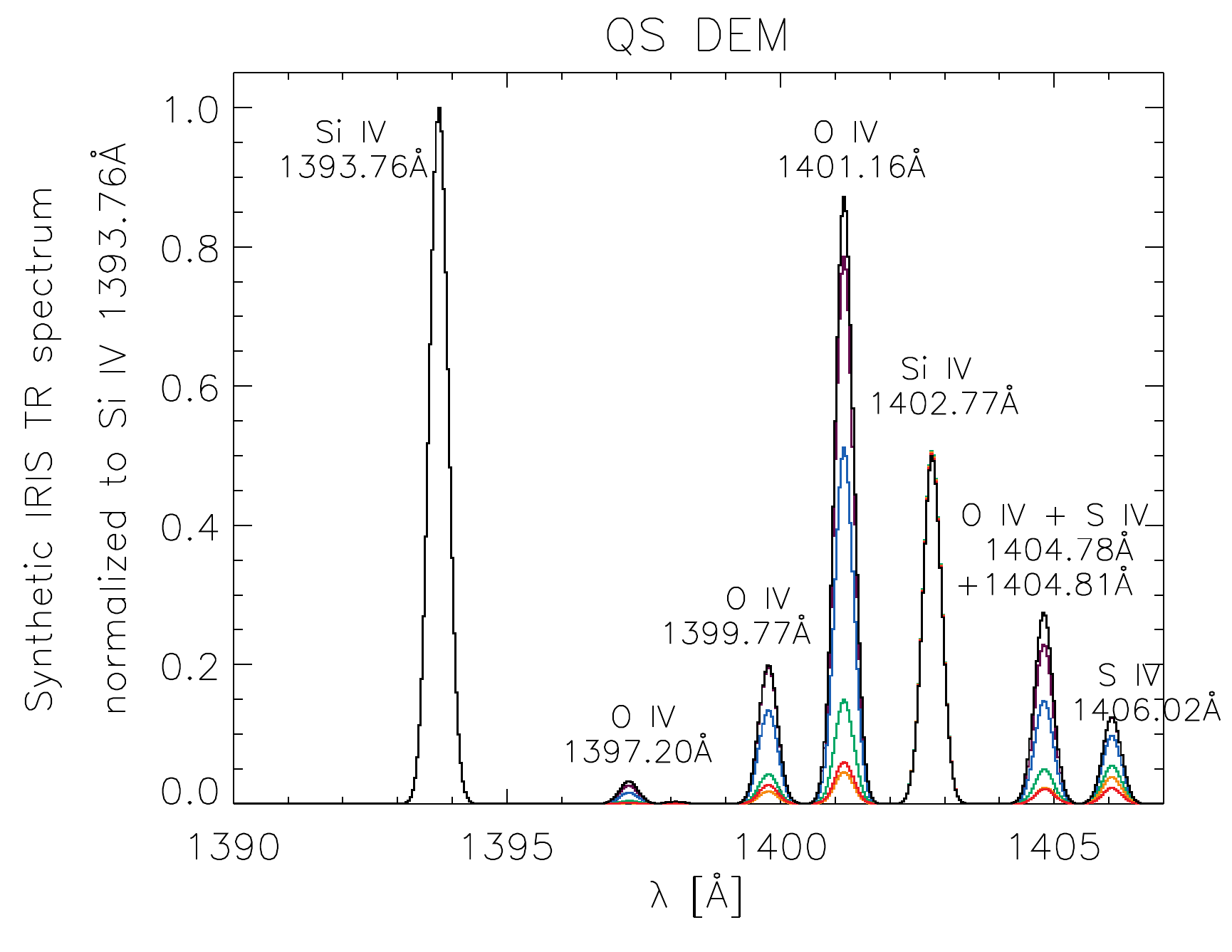}
	\caption{Behavior of the IRIS spectra for the $\kappa$-distributions. Top: contribution functions $G(T,n_\mathrm{e},\kappa)$ for the O IV 1401.16~\AA~line, calculated at fixed density of log($n_\mathrm{e}$ [cm$^{-3}$])\,=\,10 and for different $\kappa$. Bottom: Simulated IRIS \ion{Si}{IV}, \ion{O}{IV}, and \ion{S}{IV} line profiles in the IRIS FUV channel, normalised to the Si IV intensities. The calculations have been revised using the atomic data corresponding to CHIANTI 8. From \citet{Dudik14b}. \textcircled{c} AAS. Reproduced with permission.
\label{Fig:IRIS_TR_lines}}
\end{figure}
%
\subsection{Diagnostics from transition region lines}
\label{Sect:3.4}
The solar transition region is highly dynamic, with steep temperature gradients and strong flows, so non-equilibrium effects are expected. Section \ref{Sect:5} discusses how line intensities are affected by departures from ionization equilibrium. Here, we consider the effect of non-Maxwellian distributions.


Earlier suggestions of departures from Maxwellian distributions in the solar transition region were put forward by \citet{Dufton84} and subsequent authors \citep[\textit{e.g.},][]{Keenan89,Pinfield99}. \citet{Pinfield99} found anomalously high intensities in a high-excitation line from \ion{Si}{iii} observed by SOHO/SUMER \citep{Wilhelm95} and suggested that these discrepancies could have been caused by the presence of non-Maxwellian electron distributions with high energy tails. \citet{DzifcakovaKulinova11} analyzed these measurements and showed that the spectra are consistent with the presence of $\kappa$-distributions if the effect of photoexcitation is taken into account. The values of $\kappa$ found were rather high, $\kappa$\,=\,13 and 11 for the observed coronal hole and quiet Sun \ion{Si}{iii}, respectively, while $\kappa$\,=\,7 was found for the active region spectrum (Figure \ref{Fig:Si3}). However, \citet{DelZanna15a} showed, using new atomic data calculated using the R-Matrix method, that the observed \ion{Si}{iii} line intensities are actually consistent with Maxwellian electrons if one takes into account the temperature distributions of the plasma. The only exception is the active region spectrum. However the SUMER observations for all \ion{Si}{iii} lines were not simultaneous. The \citet{DelZanna15a} results clearly indicate the need for accurate atomic data and modelling before any non-Maxwellian effects can be established. 

Since 2013, the Interface Region Imaging Spectrometer \citep[IRIS][]{DePontieu14} has been observing the transition-region \ion{Si}{iv}, \ion{O}{iv}, and \ion{S}{iv} lines in the vicinity of 1400\,\AA. Previous work has demonstrated that some lines such as those from \ion{Si}{iv} can be enhanced relative to the \ion{O}{iv} lines by factors of five or more, compared to the expected ratios in equilibrium conditions \citep[\textit{e.g.},][]{Doyle84,Judge95,Curdt01,Yan15}, compared to other lines such as \ion{O}{iv} formed at similar temperatures, assuming equilibrium conditions. \citet{Dudik14b} showed that, using a Maxwellian distribution and typical transition-region DEMs, the \ion{O}{iv} 1401.16\,\AA~line should be stronger than the neighbouring \ion{Si}{iv} line at 1402.77\,\AA~if photospheric abundances are considered. Using coronal abundances would increase the predicted \ion{Si}{iv} line by a factor of 3--4, which is still not enough to explain the observed spectra. However, considering the $\kappa$-distributions, the \ion{Si}{iv} line strongly increases in intensity. This is mostly a consequence of the behavior of the ionization equilibrium for the $\kappa$-distributions with temperature. The ionization peaks for both \ion{Si}{iv} and \ion{O}{iv} are shifted to lower temperatures, but the shift for \ion{Si}{iv} is larger. In combination with the steeply increasing DEM with decreasing log($T$), a stronger increase of the \ion{Si}{iv} 1402.77\,\AA~line compared to the \ion{O}{iv} lines occurs, see Figure \ref{Fig:IRIS_TR_lines}.

Indications of non-Maxwellian distributions in the transition region were also obtained by \citet{Testa14}, who considered short-lived (tens of seconds) blue-shifts in IRIS transition-region lines during the chromospheric evaporation process at the footpoints of \ion{Fe}{xviii} coronal loops heated by coronal nanoflares. One-dimensional hydrodynamic modelling was unable to reproduce the observations without the inclusion of a power-law beam of non-thermal electrons. The authors have found that a low-energy cut-off of about $E_c$\,$\approx$\,10\,keV, with total event energies of up to 10$^{25}$\,ergs, was able to reproduce the observed blue-shifts in \ion{Si}{IV}.

%
\subsection{Diagnostics from coronal continuum and lines}
\label{Sect:3.5}
\subsubsection{Constraints from coronal X-ray continuum emission}
\label{Sect:3.5.1}

The relative number of high-energy electrons in the solar corona can be constrained by using the X-ray continuum bremsstrahlung (free-free) emission. This is because the high-energy tails strongly increase the bremsstrahlung emission in X-rays \citep[\textit{e.g.},][]{Brown71,Lin71,Holman03,Kontar11,Dudik12}. \citet{Hannah10} used quiet-Sun RHESSI \citep{Lin02} observations to obtain upper limits on the number of energetic particles at energies higher than 3\,keV that could be produced by coronal nanoflares \citep[see also][]{Klimchuk06,Klimchuk15}. In particular, it was found that the fraction of emission measure at temperatures above 5\,MK (representing the hot second Maxwellian) must be lower than about 10$^{-6}$ of quiet Sun corona, and that the possible power-law tails have to have an index higher than 5 for realistic low-energy cutoffs to be physically valid.

If the X-ray emission is instead interpreted as thin-target emission from isothermal corona with a $\kappa$-distribution (rather than with several Maxwellians), constraints for the high-energy tail that is characteristic of a $\kappa$-distribution \citep[see Fig 5 in ][]{Hannah10} can be obtained as a function of temperature. For example, at $T$\,=\,2\,MK, a $\kappa$\,=\,2 distribution requires the emission measure to be less than 10$^{43}$\,cm$^{-3}$, but for $\kappa$\,=\,4 the emission measure constraint is relaxed to about 10$^{45}$\,cm$^{-3}$. At higher $T$, the maximum emission measures decrease, with stronger decrease for larger $\kappa$ values.

Since the non-Maxwellian distributions can be approximated by a sum of two or several Maxwellians \citep[see][for the case of $\kappa$-distributions]{Hahn15b}, the multi-Maxwellian interpretation of X-ray observations can also yield constraints on the number of high-energy particles. Such studies have been done by \citet{Ishikawa14} and \citet{Hannah16}. \citet{Ishikawa14} used the FOXSI (\textit{Focusing Optics X-ray Solar Imager}) observations utilizing focusing X-ray optics mounted on a sounding rocket flight. The FOXSI instrument observed an active region for 6.5 minutes at energies greater than $4$\,keV. The observations were coordinated with \textit{Hinode}/EIS and \textit{Hinode}/XRT \citep{Golub07}. The XRT instrument observed in multiple X-ray filters, while EIS observed multiple emission lines up to \ion{Fe}{XXIV}, although only lines formed at temperatures of log$(T/$\,[K])\,$\lesssim$\,6.8 in equilibrium, including \ion{Ca}{XVII}, were detected. Using these coordinated observations, \citet{Ishikawa14} derived the DEM for the active region observed. The DEM peak was located at log$(T/$\,[K])\,$\approx$\,6.3--6.4, with FOXSI 6--7 and 7--8 keV bins providing strong constraints at temperatures of log$(T/$\,[K])\,$\gtrsim$\,6.8. Including these FOXSI data, the authors did not find evidence for a high-temperature component having an emission measure greater than 3\,$\times10^{44}$\,cm$^{-3}$, which would arise from the DEM inversion of EIS and XRT data only \citep[see also][]{Schmelz09}.

\citet{Hannah16} used the NuSTAR \citep{Harrison13} direct imaging observations of an active region at energies above 2\,keV to obtain constraints on the high-temperature component and its emission measure. It was found that the observed spectra are well-fitted with an isothermal component with temperatures of 3.1--4.4\,MK. Strong constraints were obtained on hotter sources, with their emission measure decreasing as $T^{-8}$. At 5\,MK, the upper limit was found to be 10$^{46}$\,cm$^{-3}$, while at 12\,MK, it was only 10$^{43}$\,cm$^{-3}$. No non-thermal contribution was detected, possibly due to insufficient effective exposure times. Exposure times of longer duration will be possible with diminishing solar activity as Cycle 24 approaches its minimum.

\subsubsection{Diagnostics from coronal lines}
\label{Sect:3.5.2}
Signatures of non-Maxwellian distributions in coronal lines have been sought from spectrometric observations in the UV, EUV and X-rays. \citet{Feldman07} studied the active region UV spectra of $1s2s {}^{3}S - 1s2p {}^{3}P$ lines of multiple He-like ions observed by SOHO/SUMER \citep{Wilhelm95}. These authors searched for signatures of a hot electron population that was modeled by a second Maxwellian distribution. The ions formed at higher temperatures, such as \ion{Ne}{IX}, \ion{Mg}{XI}, and especially \ion{Si}{XIII}, provided tight constraints on the number of high-energy electrons. In particular, the \ion{Si}{XIII} lines were not detected. The authors thus found no indications that a second Maxwellian with temperature log($T$\,[K])\,=\,7.0 would be required to explain the spectra. Instead, the spectra were consistent with multi-thermal plasma where the temperature of the second Maxwellian is at most log($T$\,[K])\,=\,6.5 \citep[][see Table 4 therein]{Feldman07}.

\citet{Ralchenko07} investigated the quiet Sun UV spectra of \ion{Si}{VIII}--\ion{Si}{XII}, \ion{Ar}{XI}--\ion{Ar}{XIII} and \ion{Ca}{XIII}--\ion{Ca}{XV} ions observed by SUMER. These authors showed that these spectra are consistent with a bi-Maxwellian distribution, where the hot-temperature Maxwellian contains only a few per cent of particles. This fraction depended on the assumed temperature of this hot Maxwellian. It was at most 5--7\% for the temperature of 300 eV (2.3\,MK) and decreased down to 1\% for 1 keV (7.7\,MK).

\begin{figure}[htbp]
	\centering 
	\includegraphics[width=0.69\textwidth,clip=]{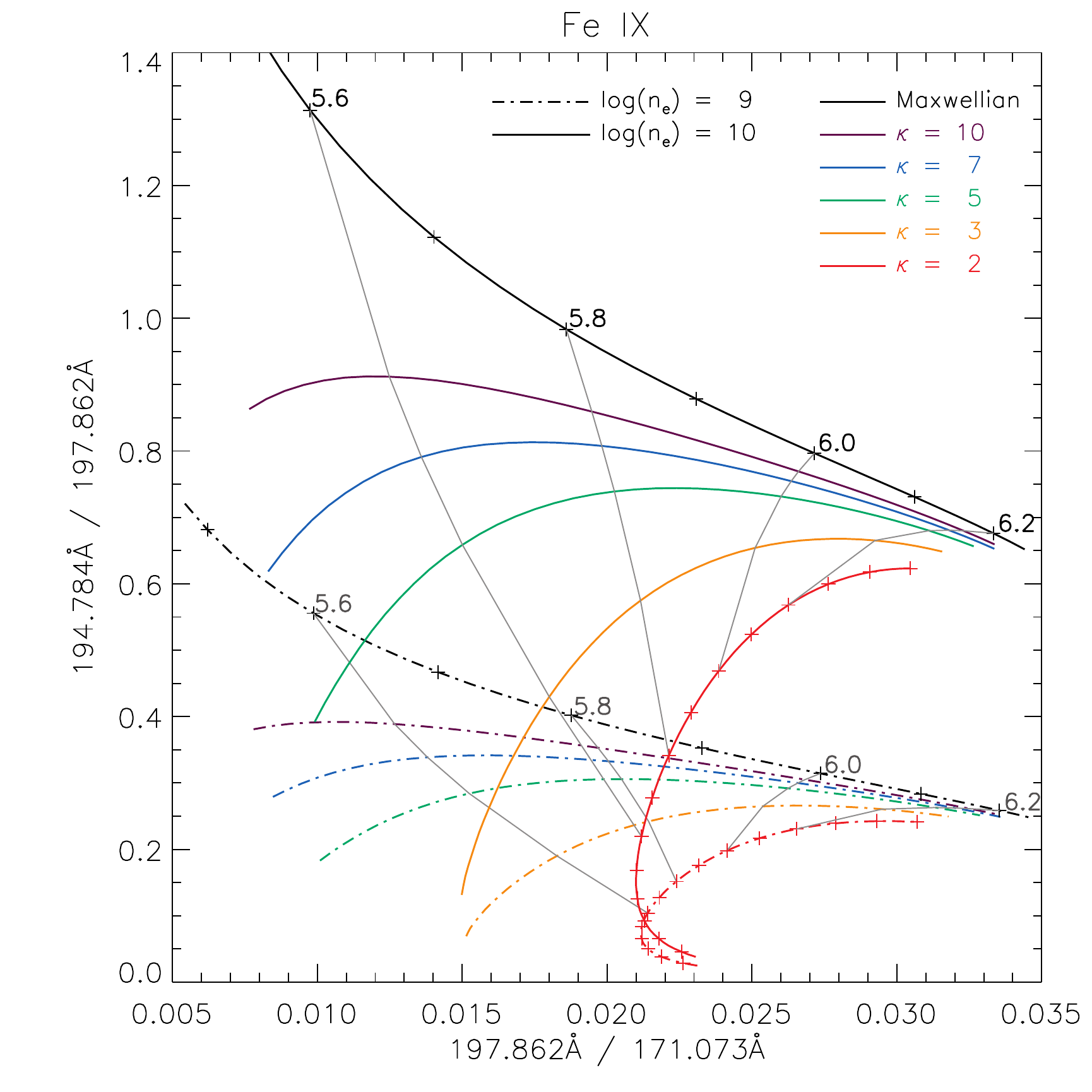}
	\caption{Theoretical ratio-ratio diagram for simultaneous diagnostics of $T$ and $\kappa$ using \ion{Fe}{IX} EUV observations. Individual colors stand for different $\kappa$, while line-styles denote $n_\mathrm{e}$. The thin gray lines connect points with the same log($T$\,[K]). Wavelengths of the lines constituting the intensity ratios are indicated at each axis. 
	Credit: \citet{Dudik14a}, reproduced with permission \textcircled{c} ESO.
\label{Fig:Diag_tk_fe9}}
\end{figure}

\begin{figure}[htbp]
	\centering 
	\includegraphics[width=0.69\textwidth,clip=]{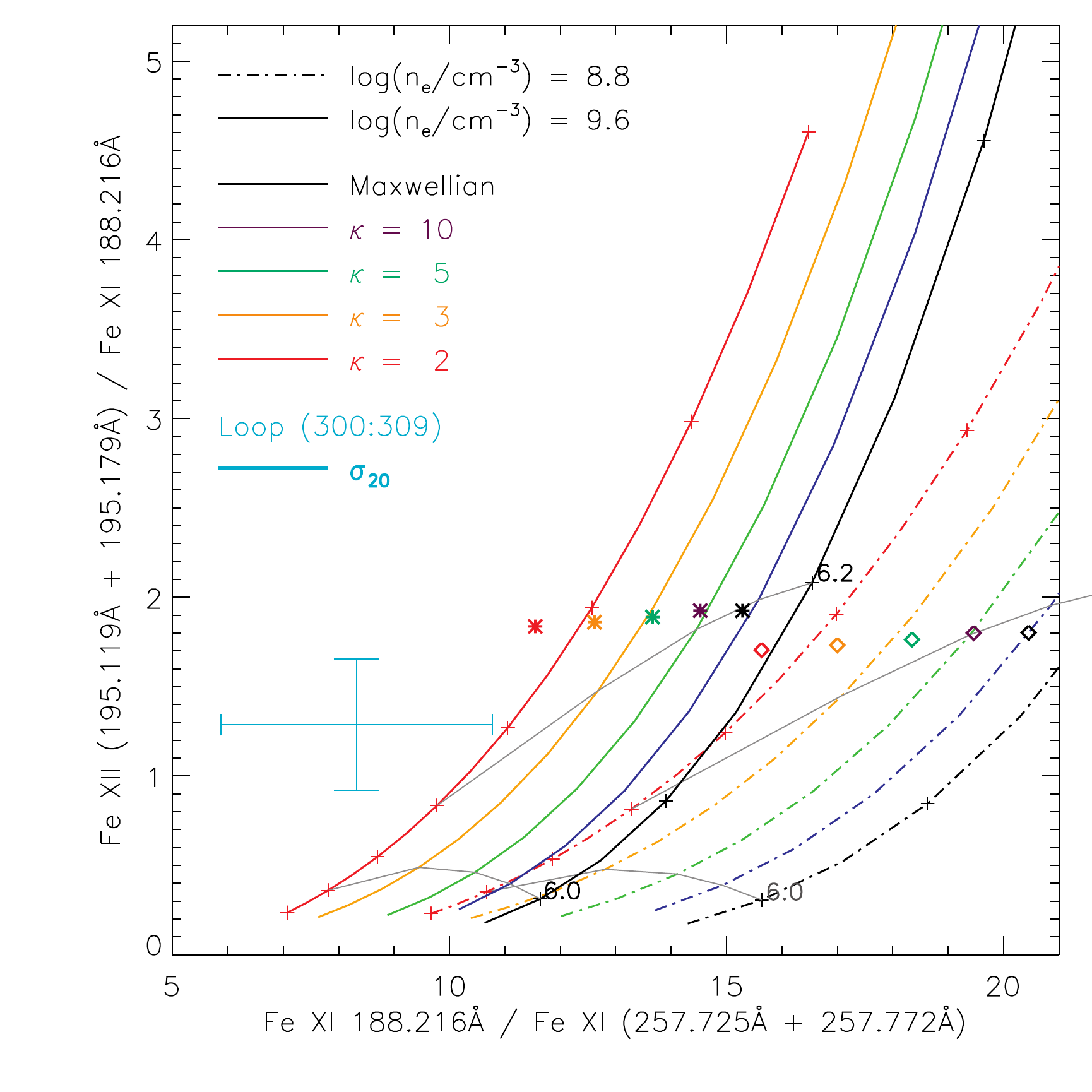}
	\caption{Diagnostics of a $\kappa$-distribution in a transient loop observed by \citet{Dudik15}. The observed ratios of the EIS lines are indicated by the azure cross, while the theoretical ratios for individual $\kappa$ are indicated by the different colors. The theoretical ratios are shown for the two densities of log($n_\mathrm{e}$ [cm$^{-3}$])\,=\,8.8 (dot-dashed) and 9.6 (full lines), respectively. Diamonds and asterisks represent the DEM-predicted line ratios. From \citet{Dudik15}. \textcircled{c} AAS. Reproduced with permission.
\label{Fig:Diag_tk_loop}}
\end{figure}

Systematic search of signatures of the $\kappa$-distributions in the \textit{Hinode}/EIS \citep{Culhane07} spectra were performed by \citet{Dzifcakova10} for Fe ions, and \citet{Mackovjak13} for other elements. A key finding was that diagnostics of $\kappa$ are always coupled to diagnostics of $T$, so at least two line ratio pairs (\textit{i.e.}, at least three lines) are needed for each ion – see Figures \ref{Fig:Diag_tk_fe9} and \ref{Fig:Diag_tk_loop}. For the Fe ions, the diagnostics are complicated further by the strong density sensitivity of many of the Fe emission lines. Although it was found that many of the standard EIS density diagnostics \citep[\textit{e.g.},][]{Young09,Watanabe09,DelZanna12a} do not depend strongly on $\kappa$ or $T$, the uncertainties on the density measurements are passed on to the $\kappa$ determinations, leading to greater uncertainty in derived values. Fe XVII was identified as the most useful Fe ion as the line ratios are insensitive to density, but the diagnostics were hampered by uncertainties in the atomic data.

\citet{Mackovjak13} determined that only Ca, Ni, S and O of the non-Fe elements provide diagnostics of $\kappa$ within the EIS wavelength range. Diagnostics involving lines from neighbouring ionization stages were found to be the most sensitive. A key single-ion diagnostic was identified for \ion{O}{IV} involving lines at 207.2 and 279.9\,\AA, but the former was found to be blended in EIS spectra. A $\kappa$-diagnostic belonging to \ion{S}{X} was identified and applied to an off-limb data-set, but results were not conclusive due to the weakness of the lines and the contributions of multi-thermal structures.

%

An ongoing effort is underway to capitalize on state-of-art atomic data produced by the UK APAP team. \citet{Dudik14a} used these data to identify a number of $\kappa$-diagnostics of \ion{Fe}{IX} to \ion{Fe}{XIII} over a wide wavelength region from the visible to soft X-rays. The authors stressed that accurate radiometric calibration and temperature measurements are critical for application of the diagnostics. Two particular diagnostics were highlighted: ratios formed from EUV allowed lines and visible forbidden lines, and \ion{Fe}{IX} ratios involving the 197.86\,\AA~emission line. For the former, the forbidden lines become stronger relative to the EUV lines as $\kappa$ decreases, with \ion{Fe}{X} 6378.26\,\AA~(the coronal red line) highlighted as an excellent line. \ion{Fe}{IX} 197.86\,\AA~is unusual in that it is a transition between two energy levels that are much higher in energy than other nearby \ion{Fe}{IX} transitions, and examples of theoretical sensitivity curves as functions of $\kappa$, $T$ and $n_{\rm e}$ are shown in Figure \ref{Fig:Diag_tk_fe9}. Although these lines are observed by EIS, the instrument sensitivity at 171.07\,\AA~is extremely low, but one example for which the lines were well-observed was presented by \citet{DelZanna14}. They found that \ion{Fe}{IX} intensities for a coronal loop leg anchored in a sunspot were consistent with the assumption of a Maxwellian electron distribution.

%

Since the single-ion Fe spectra typically do not have strong sensitivity to $\kappa$, lines from neighbouring ionization stages have to be used. This has the advantage of increased sensitivity to $\kappa$ since the ratio of two lines formed in neighbouring ionization stages is dependent on $\kappa$ through ionization equilibrium (Section \ref{Sect:3.3.3}). A disadvantage however is that such ratios can also become sensitive to departures from ionization equilibrium (see Section \ref{Sect:5}). Nevertheless, \citet{Dudik15} used the \ion{Fe}{XI}--\ion{Fe}{XII} observations from \textit{Hinode}/EIS to diagnose a $\kappa$\,$\lesssim$\,2 distribution in a transient coronal loop. Many combinations of different \ion{Fe}{XI} and \ion{Fe}{XII} lines were used, all yielding consistent results. An example is shown in Figure \ref{Fig:Diag_tk_loop}. The revision of the EIS radiometric calibration performed earlier by \citet{DelZanna13a} was instrumental to these diagnostics. This is because the long-wavelength detector of EIS degraded differently from the short-wavelength one, a fact noted and quantified by \citet{DelZanna13a}. The $\kappa$ diagnostics of \citet{Dudik15} relied on using pairs of \ion{Fe}{XI} lines from two different EIS detectors, such as the \ion{Fe}{XI} 188.2\,\AA\,/\,\ion{Fe}{XI} 257.7\,\AA~ratio (Figure \ref{Fig:Diag_tk_loop}). A significant problem in such diagnostics remains the high absolute calibration uncertainty, which is typically 20\% for the coronal spectrometers \citep{Lang00,Lang06,Culhane07}. Such uncertainty can be comparable to the spread of the individual diagnostic curves (see Figure \ref{Fig:Diag_tk_loop}), which hampers the diagnostics of moderate $\kappa$ values.

\begin{figure}[htbp]
	\centering 
	\includegraphics[width=0.49\textwidth,clip=]{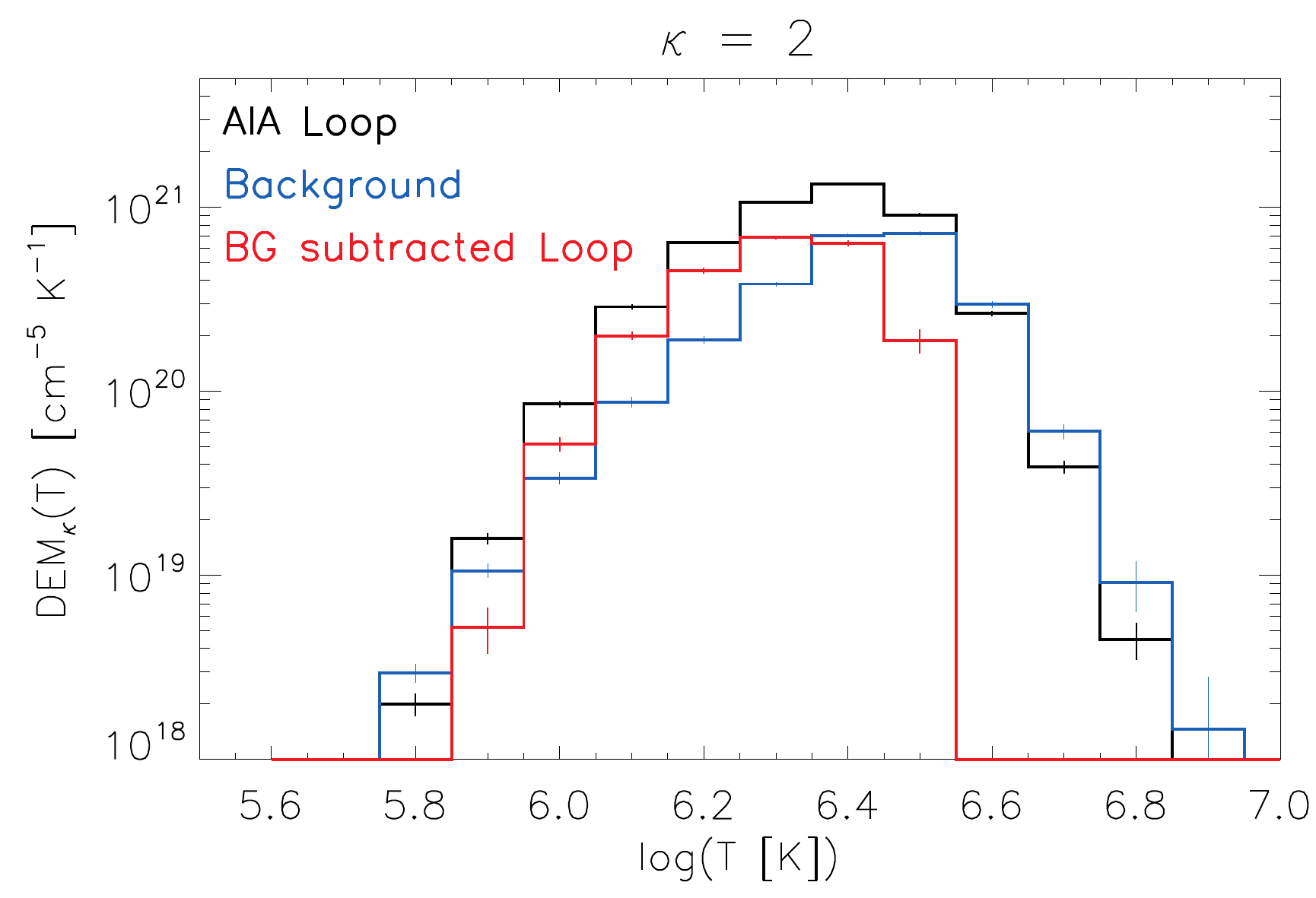}
	\includegraphics[width=0.49\textwidth,clip=]{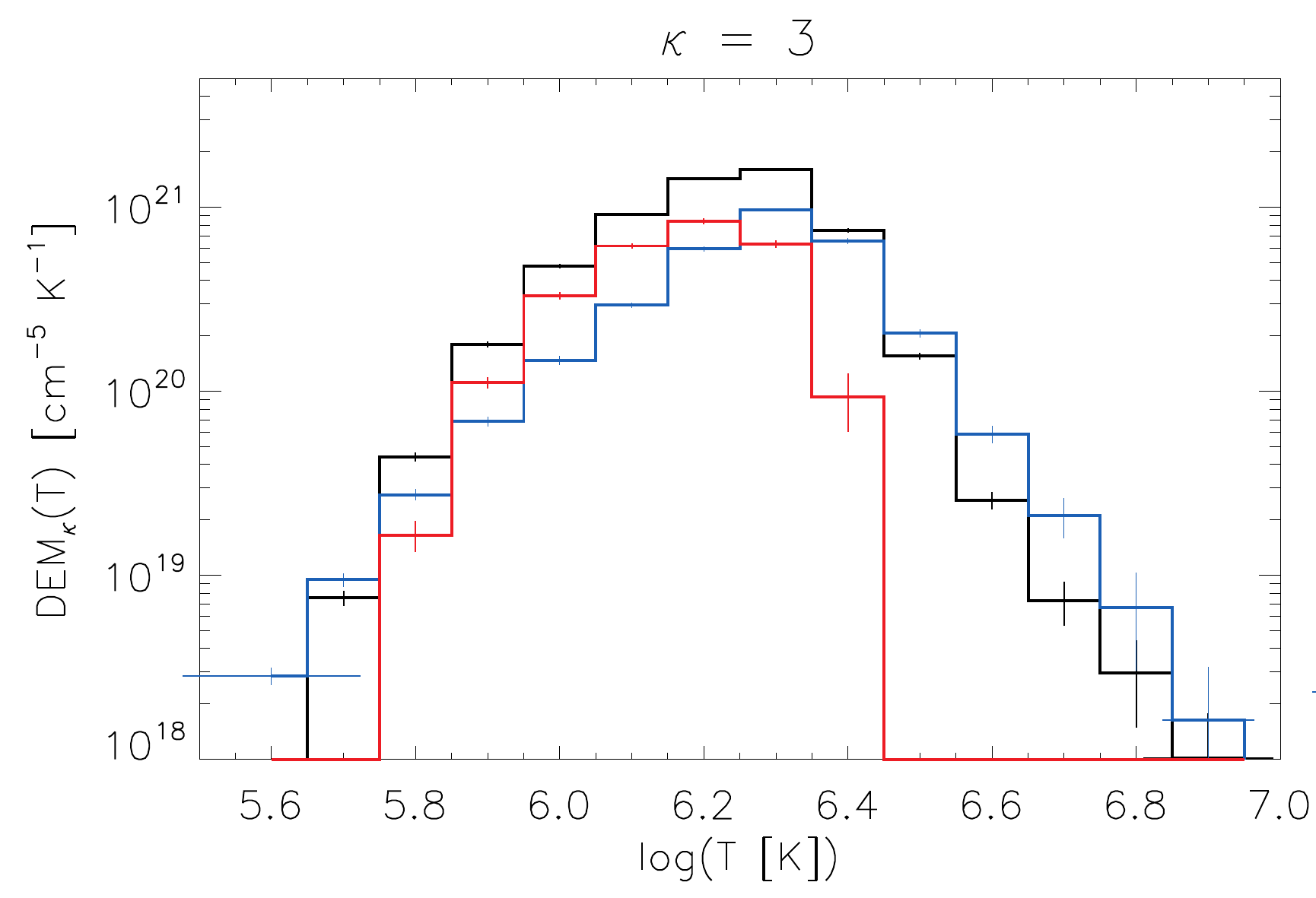}
	\includegraphics[width=0.49\textwidth,clip=]{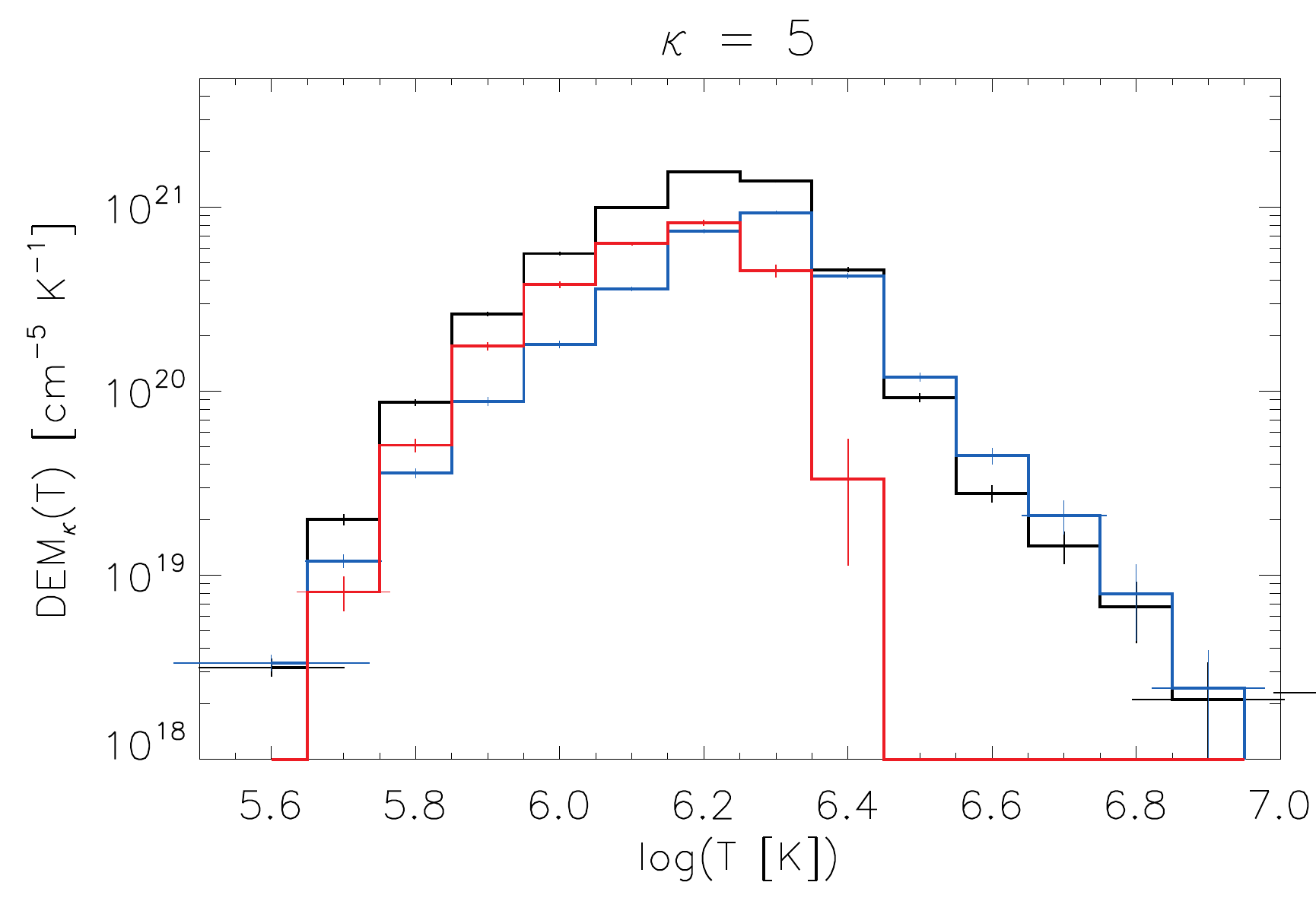}
	\includegraphics[width=0.49\textwidth,clip=]{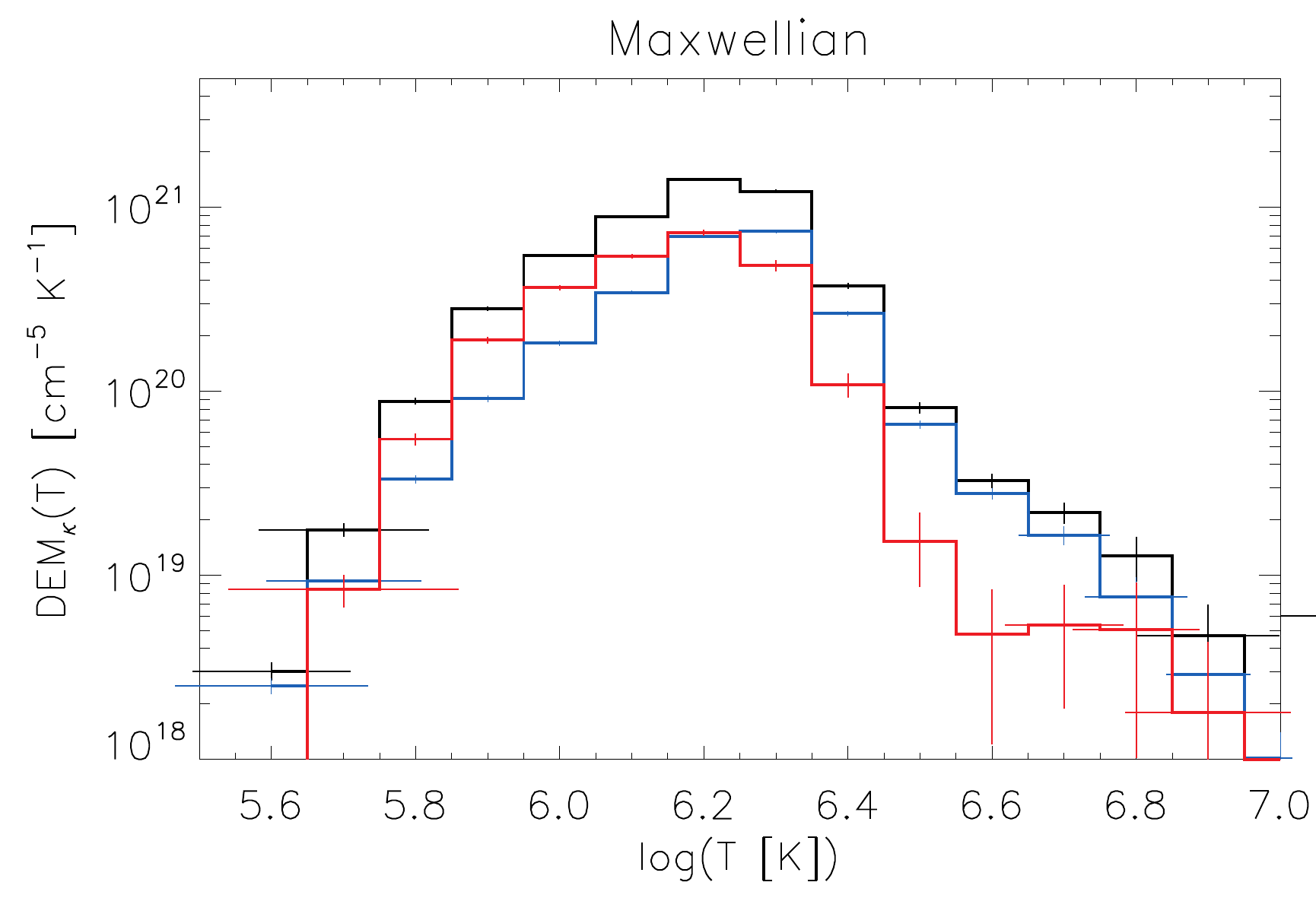}
	\caption{Behaviour of the DEM$_\kappa(T)$ for different $\kappa$\,=\,2, 3, 5, and the Maxwellian distribution. The DEMs are derived from AIA data. Black lines represent the original AIA data at the position of the transient loop. Blue lines stand for the neighbouring background, while the red denotes the background-subtracted DEMs for the transient loop. Horizontal and vertical error bars are shown at each temperature bin. From \citet{Dudik15}. \textcircled{c} AAS. Reproduced with permission.
\label{Fig:DEM_loop}}
\end{figure}

Furthermore, \citet{Dudik15} found that the transient coronal loop studied was not isothermal. Data from the \textit{Atmospheric Imaging Assembly} \citep[AIA][]{Lemen12,Boerner12} on-board the \textit{Solar Dynamics Observatory} were used to derive the corresponding DEMs as a function of $\kappa$ (Figure \ref{Fig:DEM_loop}, see also Equation \ref{Eq:line_intensity_DEM}). When folded with the line contribution functions, it was found that the DEM-predicted ratios approach the observed ones with decreasing $\kappa$ (asterisks and diamonds in Figure \ref{Fig:Diag_tk_loop}). This confirmed the result of $\kappa$\,$\lesssim$\,2 for this coronal loop.

\begin{figure}[htbp]
	\centering 
	\includegraphics[width=0.69\textwidth,clip=]{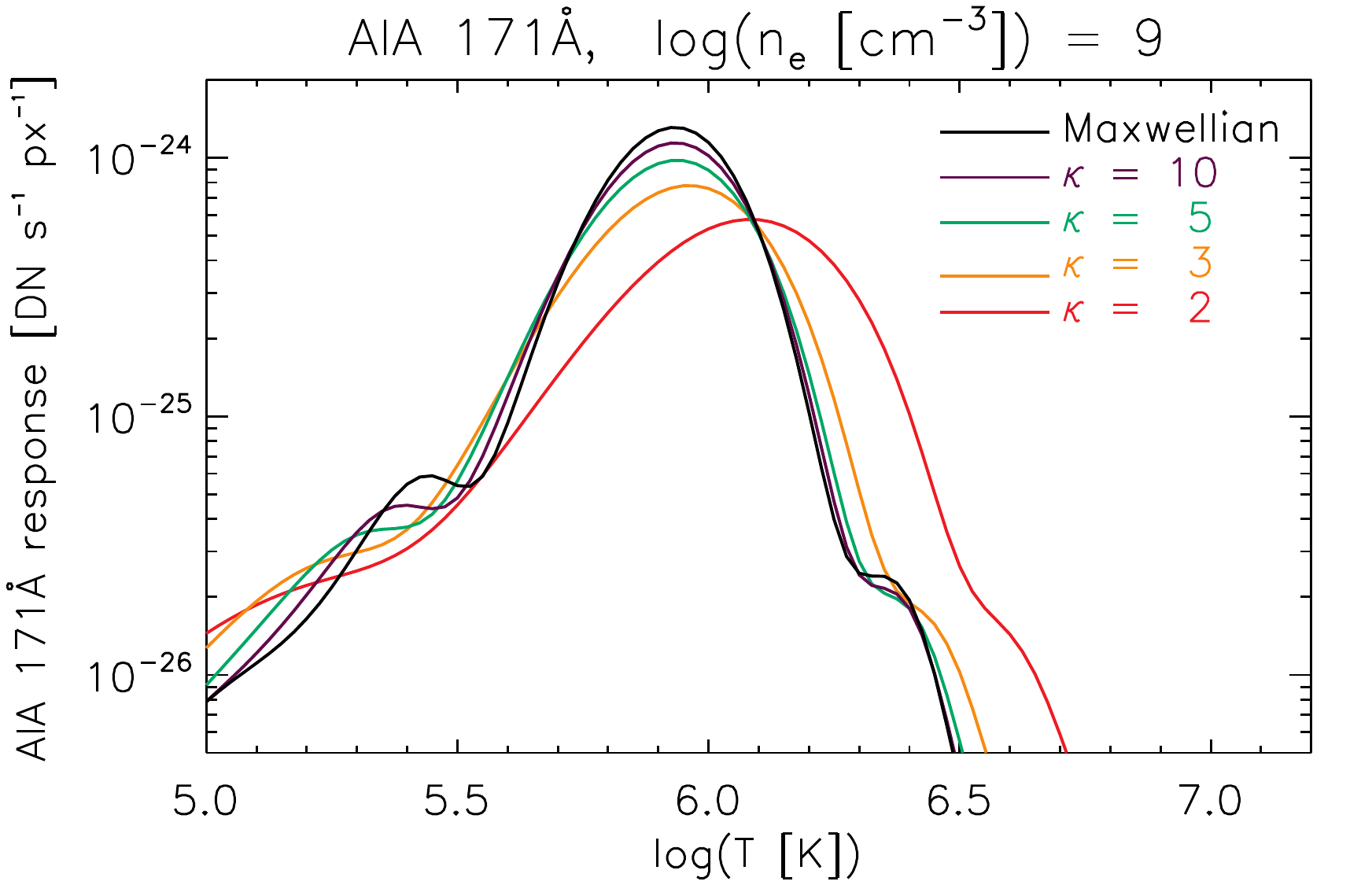}
	\caption{Example of the behavior of the AIA temperature responses with $\kappa$. The AIA 171\AA~response is shifted towards higher temperatures with decreasing $\kappa$. From \citet{Dzifcakova15}. \textcircled{c} AAS. Reproduced with permission.
\label{Fig:AIA_resp}}
\end{figure}

We note that the low value of $\kappa$ found could have significant consequences for coronal physics. For example, the total radiative losses from non-thermal plasmas can be modified by up to a factor of 2 compared to Maxwellian plasmas, with details depending on temperature and $\kappa$ \citep{Dudik11}. Similarly, the nature of the AIA temperature responses changes with decreasing $\kappa$. Generally, the dominant changes in the AIA responses are given by the behavior of the ionization equilibrium. The coronal peaks of the AIA responses are typically broadened and shifted towards higher temperatures for low $\kappa$ (Figure \ref{Fig:AIA_resp}).

%
%
%
%
\subsection{Differential emission measures}
\label{Sect:3.6}

Since the non-Maxwellian distributions can be approximated by a sum of multiple Maxwellians, analyses of coronal observations involving calculation of DEM (Equation \ref{Eq:line_intensity_DEM}), which consists of contributions from several Maxwellians, could in principle yield different results if non-Maxwellian distributions are used instead. Namely, the resulting DEM can depend on the value of $\kappa$, which can be either assumed or independently diagnosed (see Section \ref{Sect:3.5}). \citet{Mackovjak14} studied quiet Sun \ion{Fe}{VIII}--\ion{Fe}{XV} spectra as well as active region spectra including several ions of Fe, Si, S, Ca, and Ar. These authors showed that in the case of quiet Sun, the EM-loci plots \citep[consisting of multiple $I_{ji}/G_{X,ji}(T,n_\mathrm{e},\kappa)$ curves for each observed line $ji$, see \textit{e.g.}][]{Strong78,Veck84,DelZanna03} appear less multithermal, as the corresponding $I_{ji}/G_{X,ji}(T,n_\mathrm{e},\kappa)$ curves move towards a common isothermal crossing-point for $\kappa$\,=\,2. Since these curves constitute an upper bound on the corresponding DEM$_\kappa$, the resulting DEM$_\kappa$ are also less multithermal. We note that the term ``multithermal'' for the case of $\kappa$-distributions refers to DEM$_\kappa(T)$ having contributions from more than one temperature bin \citep{Mackovjak14}.

In contrast to that, the degree of multithermality in the cores of active regions was found to be the same for all $\kappa$. The DEMs were only shifted towards higher $T$ as a result of the behavior of the ionization equilibrium. 
This is of importance since in active region cores, the DEM slope at temperatures lower than its peak are thought to provide constraints on the recurrence frequency of the coronal nanoflares \citep[\textit{e.g.},][]{Viall11,Bradshaw12,Warren12,Winebarger12a,Cargill14,DelZanna15c}. Since the low-temperature slope of the DEM$_\kappa$ does not change with $\kappa$, the constraints on nanoflare timescales derived from observations do not change with respect to the relative number of accelerated particles (\textit{i.e.}, different $\kappa$) present in the active region corona.

Examples of the effect of different $\kappa$ values on the shapes of derived DEM$_\kappa$ curves \citep[from][]{Dudik15} are shown in Figure \ref{Fig:DEM_loop}. It can be seen that the Maxwellian DEMs peak at log($T$\,[K])\,=\,6.2, with the peaks shifting to higher temperatures with higher $\kappa$ value: the $\kappa$\,=\,2 DEM peaking at log($T$\,[K])\,=\,6.4.

%
%
%
%
%
\section{Flare X-Ray Emission}
\label{Sect:4}

Solar flares are characterized by a rapid increase of the X-ray emission over short timescales. It is thought that this occurs as a consequence of magnetic energy release involving magnetic reconnection \citep[\textit{e.g.},][]{Priest00,Zweibel09} accelerating particles and heating the plasma. Soft X-ray emission is mostly emitted by flare loops during the peak and gradual phases, while hard X-ray emission is present during the impulsive phase. The presence of accelerated particles during the impulsive phase is routinely detected from hard X-ray bremsstrahlung (free-free) emission (Section \ref{Sect:4.1}). The presence of these accelerated particles should therefore have an effect also on the EUV and X-ray line emission originating in the flaring corona (Sections \ref{Sect:4.2} and \ref{Sect:4.3}). 

%
\subsection{Hard X-ray continuum}
\label{Sect:4.1}
We start with a review of hard X-ray (HXR) signatures that indicate non-equilibrium processes in the solar corona.
Although much was already known from previous observations, we focus here on the latest HXR mission, 
the \textit{Reuven Ramaty High-Energy Solar Spectroscopic Imager} \citep[RHESSI][]{Lin02}, which
achieved spatial and spectral resolutions much higher than those of earlier solar hard X-ray missions 
\footnote{\url{http://hesperia.gsfc.nasa.gov/rhessi3/news-and-resources/results/top_rhessi_accomp.html}}. Various new features of solar flare X-rays have been discovered by RHESSI and these have already been reviewed elsewhere \citep[\textit{e.g.}][]{Benz08, Krucker08a, Fletcher11, Hannah11, Kontar11, Lin11, Vilmer11, White11, Zharkova11}. 

We note that strong HXR emission is normally found at the footpoints of flare loops. These footpoint sources are much more intense than those in the corona \citep[\textit{e.g.}][]{Duijveman82, Sakao94, Saint08}. However, the coronal sources have received considerable attention recently because of their proximity to the possible reconnection region. In the following subsections, we briefly review observations and interpretations of coronal hard X-ray sources.

\subsubsection{Overview of hard X-ray coronal observations}

Solar flares produce non-thermal electrons with energies up to tens of MeV. While magnetic reconnection has been considered as a key mechanism of energy-release during solar flares \citep[\textit{e.g.}][]{Masuda94, Tsuneta96, Sui03}, it still remains unclear how the requisite production of non-thermal electrons can be explained by magnetic reconnection. \textit{Yohkoh'}s discovery of a coronal hard X-ray (HXR) source that appeared to float above the tops of flaring loops indicates that electron energization (\textit{i.e.}, heating and/or production of non-thermal component) takes place in the corona \citep[\textit{e.g.}][]{Masuda94}. Since then, various theoretical ideas have been proposed to explain energetic electrons associated with the ‘above-the-looptop’ HXR source. Some of the key theories proposed are the fast-mode termination shocks \citep[\textit{e.g.}][]{Tsuneta98, Guo12, Li13, Nishizuka13}, turbulence \citep[\textit{e.g.}][]{Miller97, Bian14}, collapsing magnetic fields \citep[\textit{e.g.}][]{Somov97, Karlicky04, Bogachev05, Giuliani05} and magnetic islands \citep[\textit{e.g.}][]{Drake06, Oka10, Bian13}. Many of the recent theories use more than one of these mechanisms. For more details, readers are referred to more comprehensive reviews by, for example, \citet{Miller97}, and \citet{Zharkova11}.

RHESSI established that, during flares, a non-thermal power-law tail often exists in the energy spectrum of a coronal HXR source \citep[\textit{e.g.}][]{Lin03,Veronig04, Battaglia06, Krucker08b, Kasparova09b, Simoes13}, in particular in the above-the-looptop source \citep[\textit{e.g.}][]{Krucker10, Ishikawa11, Chen12, Oka13, Oka15, Krucker14}. This clearly indicates that time-dependent and non-equilibrium processes exist in the solar corona during flares. The measured spectral slopes of the high-energy tail as well as the inferred energy partition between thermal and non-thermal electrons are expected to provide key information on the possible non-equilibrium processes and to constrain theories of electron acceleration. However, the non-thermal and thermal emissions generally originate from different sources, \textit{i.e.} above-the-loop-top sources and coronal flare loops, and therefore the combined diagnostic capability can be applied for the rare cases where both sets of emissions originate from the same volume.

%
%
\begin{figure} 
	\centering
	\includegraphics[width=1.0\textwidth]{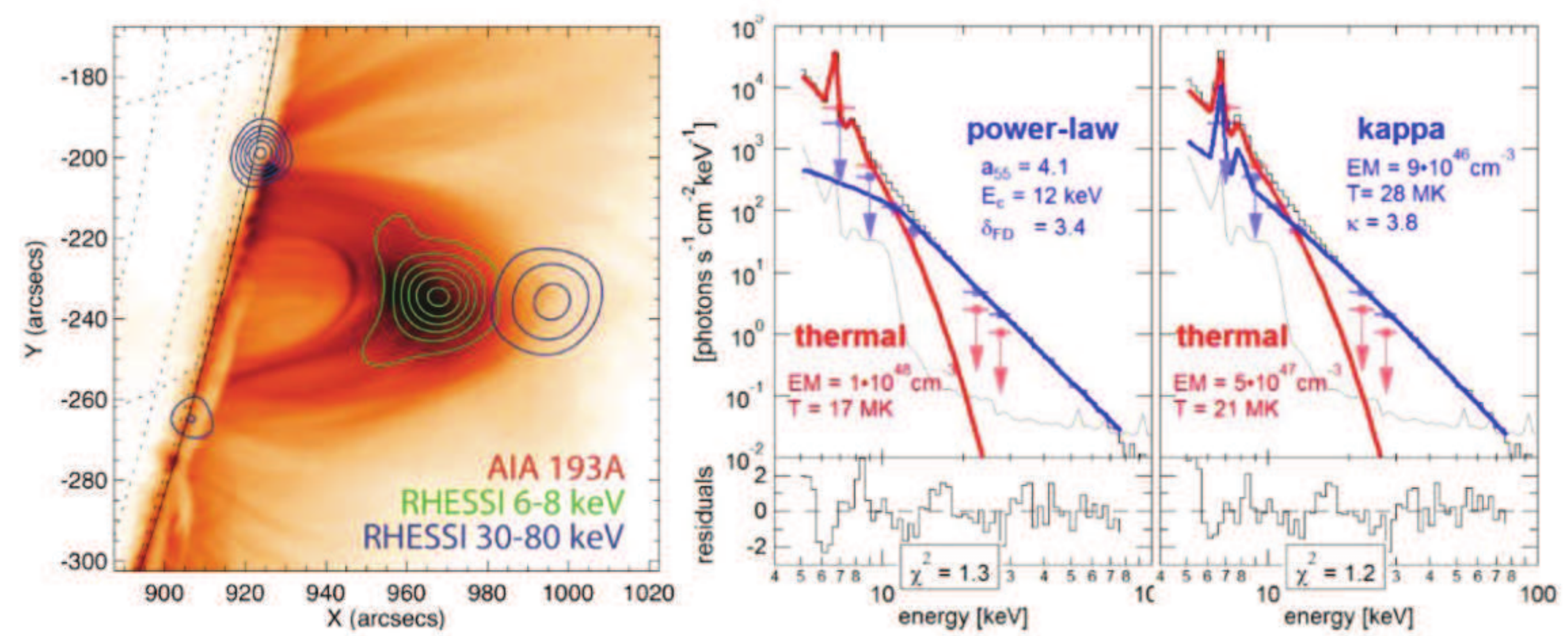}
	 \caption{Left: X-ray (contours) and EUV (background) images of the limb flare of SOL2012-07-19T05:58. Lower energy thermal X-rays come from the looptop region (green contours) whereas higher energy non-thermal X-rays come from the footpoint as well as `above-the-looptop' regions (blue contours). From \citet{Krucker14}). Center: The combined data from both the looptop and above-the-looptop regions of another flare SOL2007-12-31T01:11 (histogram) fitted with a thermal and a power-law distributions (red and blue curves, respectively). Right: The same, combined data are now fitted with a thermal and a $\kappa$-distribution. From \citet{Oka13}. \textcircled{c} AAS. Reproduced with permission.
	 \label{Fig:RHESSI}}
\end{figure}
%
\subsubsection{Electron energy partition and the $\kappa$-distributions}

In two particular flares with very intense above-the-loop-top sources, it was shown that all electrons are accelerated in a bulk energization process to form a power-law with no thermal pre-flare plasma left \citep{Krucker10, Krucker14}. This requires an extremely efficient process for the formation of the non-thermal electron distribution. However, the non-thermal component is significantly weaker than this case in most flares, and for some events the non-thermal component is even undetected. Some studies have reported a super-hot ($\gtrsim$30MK) thermal source in addition to the hot ($\lesssim$30 MK) loop top source \citep{Caspi10, Caspi14, Longcope10}, suggesting a need for strong heating in addition to acceleration to non-thermal energies. 

A caveat is that the main thermal flare loop is generally so bright that instrumental limitations are masking the above-the-loop-top spectrum in the lower-energy ($\lesssim$15 keV) range. Furthermore, the above-the-loop-top source is generally faint and its spectrum reaches the detection limits at higher energies (\textit{e.g.}, $\gtrsim$80 keV in the case of the fortuitously bright above-the-loop-top source, but more generally only up to $\approx$30 keV). Hence, the above-the-loop-top source is only observed over a rather limited energy range (\textit{i.e.}, 15--80\,keV) and the observations therefore only provide limited diagnostic power to distinguish different interpretations. 

Recently, $\kappa$-distributions have been proposed as yet another way of interpreting the spectral form in the above-the-loop-top region \citep{Kasparova09b,Oka13,Oka15}. In \citet{Oka13,Oka15} it was argued that there should always be a thermal core component due to thermalization (in the lower energy range) by Coulomb collisions and/or wave-particle interactions \citep[see also][]{Holman11}. By varying the index $\kappa$, the $\kappa$-distribution can, in fact, represent not only cases with negligible non-thermal component (including super-hot cases) but also cases with saturated non-thermal component (with no spectral break between the thermal and non-thermal components). Note that the spatial and temporal scales of kinetic processes are much smaller than those of {\it RHESSI} observations. Thus, it was argued that it is reasonable to assume a simple functional form for the lower energy ($\lesssim$15 keV) range rather than a complex, non-thermal or non-stationary spectral forms such as flat-top, bi-Maxwellian, bump, beam, \textit{etc.} The thermal Maxwellian and $\kappa$-distributions are simple in the sense that they represent the state of maximum Boltzmann-Gibbs and Tsallis entropies, respectively \citep[\textit{e.g.}][]{Livadiotis09}. Based on the analysis of 5 events using the $\kappa$-distribution model, it was suggested that the smallest value of $\kappa$ is $\sim$4 and that the corresponding non-thermal fractions of electron density and energy are $\sim$20\% and $\sim$50\%, respectively. The observations can be considered as bulk energization with all electrons being energized as the thermal part of the $\kappa$-distribution is much hotter than the pre-flare plasma. 

\subsubsection{Event-to-event variations of the non-thermal tail}

Despite the difference in the interpretation (or extrapolated form) of the lower-energy range, at higher-energy the non-thermal tail in the coronal X-ray source seems to be a generic feature. From a statistical analysis of 55 coronal source events during flares, it was reported that the power-law indices of photon spectra fall between $\sim$4 and $\sim$6 \citep{Krucker08a}. Note that coronal sources are often interpreted by the thin-target model and the range corresponds to the power-law index $\delta$ = 3 -- 5 of electron flux distribution under the assumption of the thin-target model\footnote{A power-law index $\gamma_{\rm thin}$ of a thin-target photon spectrum is related to a power-law index $\delta$ of electron flux distribution and a power-law index $\delta'$ of electron density distribution by $\gamma_{\rm thin} = \delta +1$ and $\gamma_{\rm thin} = \delta' + 0.5$, respectively \citep[\textit{e.g.}][]{Brown71,Tandberg88}}.

On the other hand, flare events with a non-thermal coronal X-ray source but no clear emission from footpoints have been reported \citep{Veronig04, Lee13}. It has been argued that the column densities of the non-thermal coronal source were high and that energetic electrons must have lost their energies before reaching the footpoints. Then, the thick-target model can be applied to the coronal source of these events. The derived spectral index $\delta$ was between $\sim$6 and $\sim$7, indicating very soft (steep) electron spectra\footnote{Here, a power-law index $\gamma_{\rm thick}$ of a uniform thick-target photon spectrum is related to a power-law index $\delta$ of electron flux distribution and a power-law index $\delta'$ of electron density distribution by $\gamma_{\rm thick} = \delta - 1$ and $\gamma_{\rm thick} = \delta' - 1.5$, respectively \citep[\textit{e.g.}][]{Holman11}}.

A variation of the power-law index was also discussed for the above-the-looptop source events analyzed with the $\kappa$-distribution \citep{Oka15}. It was reported that the $\kappa$ values obtained were between $\sim$4 and $\sim$14 (or $\delta$ between $\sim$4 and $\sim$14)
\footnote{Because the phase space density of the $\kappa$-distribution has the form $f(v)\propto v^{-2(\kappa+1)}$, the corresponding power indices of electron flux distributions are $\delta$ $\sim$4 and $\sim$14 (or $\sim$4.5 and $\sim$14.5 for the density distribution)} indicating a very wide range of power-law spectral index. A larger value of $\kappa$ indicates that the spectrum is closer to Maxwellian. In fact, \citet{Galloway10} showed that the non-thermal, high-energy tail of a $\kappa$-like distribution can be thermalized into a Maxwellian over a timescale of 100 - 1000 $\tau_{\rm coll}$, where $\tau_{\rm coll}$ is the electron-electron collision time of the core particles and it can be $\sim$0.1s in a plasma with 30 MK temperature and 10$^9$ cm$^{-3}$ density. Thus, it was speculated that, while electron acceleration may be achieved primarily by collisionless magnetic reconnection, the above-the-looptop source density determines the electron energy partition and the associated spectral slope of the source spectra. 

\subsubsection{Confinement of electrons}

A standard scenario for footpoint emission is that electrons accelerated in the corona precipitate along flare loops and lose the bulk of their energy through electron-ion bremsstrahlung X-ray emission in the lower solar atmosphere. It has been reported that measured differences between the coronal (thin-target) and footpoint (thick-target) power-law indices are consistent with this scenario \citep[\textit{e.g.},][]{Battaglia06, Ishikawa11}.

A more recent quantitative analysis indicated that there is a sufficient number of accelerated electrons in the coronal source to suggest not only precipitation into footpoint sources but also a fraction of the population being trapped within the source \citep{Simoes13}. In fact, the observations of the above-the-looptop source over a time period longer than the electron transit time across the source already indicates that electrons are trapped within the source region. 

From a theoretical point of view, \citet{Minoshima11} developed and solved a drift-kinetic Fokker-Planck model to discuss quantitatively the importance of pitch angle scattering for trapping of electrons within the above-the-looptop region. \citet{Kontar14} also used a Fokker-Planck approach to deduce mean free-paths of electrons trapped in a coronal source. On the other hand, particle-in-cell (PIC) simulations of energetic electron transport showed that a localized electrostatic potential or ‘double layer’ is formed and that it inhibits further transport of energetic electrons \citep{Litc12, Litc13, Litc14}.

Interestingly, \citet{Bian14} proposed that a local stochastic acceleration of electrons within the X-ray source is still possible because a collisional loss can be taken into account instead of escape from the source region to explain the power-law formation. They have further derived the $\kappa$-distribution analytically based on the balance between the stochastic acceleration and the collisional loss.

To understand how the above-mentioned physics (scattering, double-layer, collisional loss, \textit{etc.}) are related to each other and how they are related to the dynamics of flares including magnetic reconnection, it is essential to have larger scale, kinetic simulations. It is also desirable to have more detailed observations of the above-the-looptop source.


%
\subsection{Mean electron fluxes from RHESSI and AIA observations}
\label{Sect:4.2}

Since 2010, a number of solar flares have been observed by both RHESSI and the \textit{Atmospheric Imaging Assembly} \citep[AIA,][]{Lemen12,Boerner12} onboard the \textit{Solar Dynamics Observatory} \citep[SDO,][]{Pesnell12}. While RHESSI provides direct information on the high-energy part of the electron distribution function above $\approx$3\,keV, the low-energy part up to about 1--1.5\,keV can be constrained from AIA observations following the approach of \citet{Battaglia13}. These authors consider the electron mean flux spectrum $\left<n_\mathrm{e}VF(E)\right>$
\begin{eqnarray}
	\nonumber \left< n_\mathrm{e} V F(E)\right> &=& \int_{V} n_\mathrm{e}(r) F(E,r) \mathrm{d}V = \int_{T} n_\mathrm{e}^2(r) f(E,r) \frac{\mathrm{d}V}{\mathrm{d}T} \mathrm{d}T\, \\
	&=& \int_{T} \xi_{f}(T) f(E,r) \mathrm{d}T\,,
	\label{Eq:NVF}
\end{eqnarray}
where $V$\,=\,$A \int dl$ is the emitting volume, $A$ the emitting plane-of-sky area on the Sun, $F(E) = n_\mathrm{e}f(E)$, $\xi_f(T)$ = $\int n_\mathrm{e}^2 \mathrm{d}V / \mathrm{d}T$ = $A C \int n_\mathrm{e} n_\mathrm{H} \mathrm{d}l / \mathrm{d}T$ = $AC$ DEM$_{f}(T)$ is the volumetric differential emission measure (\textit{cf.}, Sections \ref{Sect:3.1} and \ref{Sect:3.6}), and $C$\,=\,$n_\mathrm{e}/n_\mathrm{H}$\,=\,1/0.83 for coronal and flare conditions. 

\citet{Battaglia13} considered $f(E)$ to be a Maxwellian and/or power-law distribution. They used the Maxwellian AIA responses (\textit{cf.}, Section \ref{Sect:3.5}) to calculate the $\xi_f(T)$ functions and the corresponding mean electron flux $\left<n_\mathrm{e}VF\right>$ for the loop source from RHESSI and AIA separately. This was done for two events, a C4.1-class event SOL2010-08-14T10:05 and a well-known M7.7-class SOL2012-07-19T05:58 \citep[see also, \textit{e.g.}, Figure \ref{Fig:RHESSI} \textit{left} and][]{Krucker14,Sun14,Oka15}. Thermal and power-law fits to the RHESSI data were performed and extrapolated to lower energies (up to 2\,keV) corresponding to AIA observations. For both flares the RHESSI fits over-predicted the AIA observations by a factor of at least 2 for all regions studied. For the case of the above-the-loop-top source in the M7.7 flare, the over-prediction was by more than an order of magnitude. Conversely, the AIA DEMs, derived using the regularized inversion method of \citet{Hannah12,Hannah13}, and the corresponding $\left<n_\mathrm{e}VF\right>$ values were consistent with a hot isothermal plasma, but only up to energies of 1 and 0.5\,keV for the C4.1 and M7.7 flares, respectively. Above these energies, excess flux above the isothermal $\left<n_\mathrm{e}VF\right>$ level was present. The authors suggested that the derived mean electron flux spectrum could be described by a combination of a Maxwellian core, secondary halo, and a non-thermal tail similar to the situation observed in the solar wind.

The analysis was applied for several above-the-loop-top sources by \citet{Oka15}. These authors fitted the observed RHESSI high-energy tails using several different components, including a power-law, thermal and power-law, and a $\kappa$-distribution. The power-law fits over-predicted the AIA $\left<n_\mathrm{e}VF\right>$ values by orders of magnitude \citep[Figure 5]{Oka15}, while the thermal and power-law fits under-predicted them by at least an order of magnitude. A caveat is that for the above-the-loop-top sources, the RHESSI thermal components can be outside the temperature sensitivity of AIA. Thus, the AIA data represent only an upper limit because of the possible contribution from unresolved cooler plasma at temperatures of $\approx$\,10\,MK or lower.
Therefore, the thermal+power-law fit to the RHESSI data, which under-predicts the AIA $\left<n_\mathrm{e}VF\right>$ is not ruled out by the constraints derived from AIA. \citet{Oka15} however pointed out that such fits returned unreasonable fit parameters due to an artifact of a sharp lower-energy cut-off. In contrast to that, the third class of fit used, the $\kappa$-distribution, fitted well the high-energy tail observed by RHESSI. In addition, the $\kappa$-distribution is not ruled out by the mean electron flux $\left<n_\mathrm{e}VF\right>$ derived from AIA \citep[Figure 5]{Oka15} and dispenses with the need of an artificial low-energy cut-off. 

%
\begin{figure}
	\centering
	\includegraphics[width=0.49\textwidth]{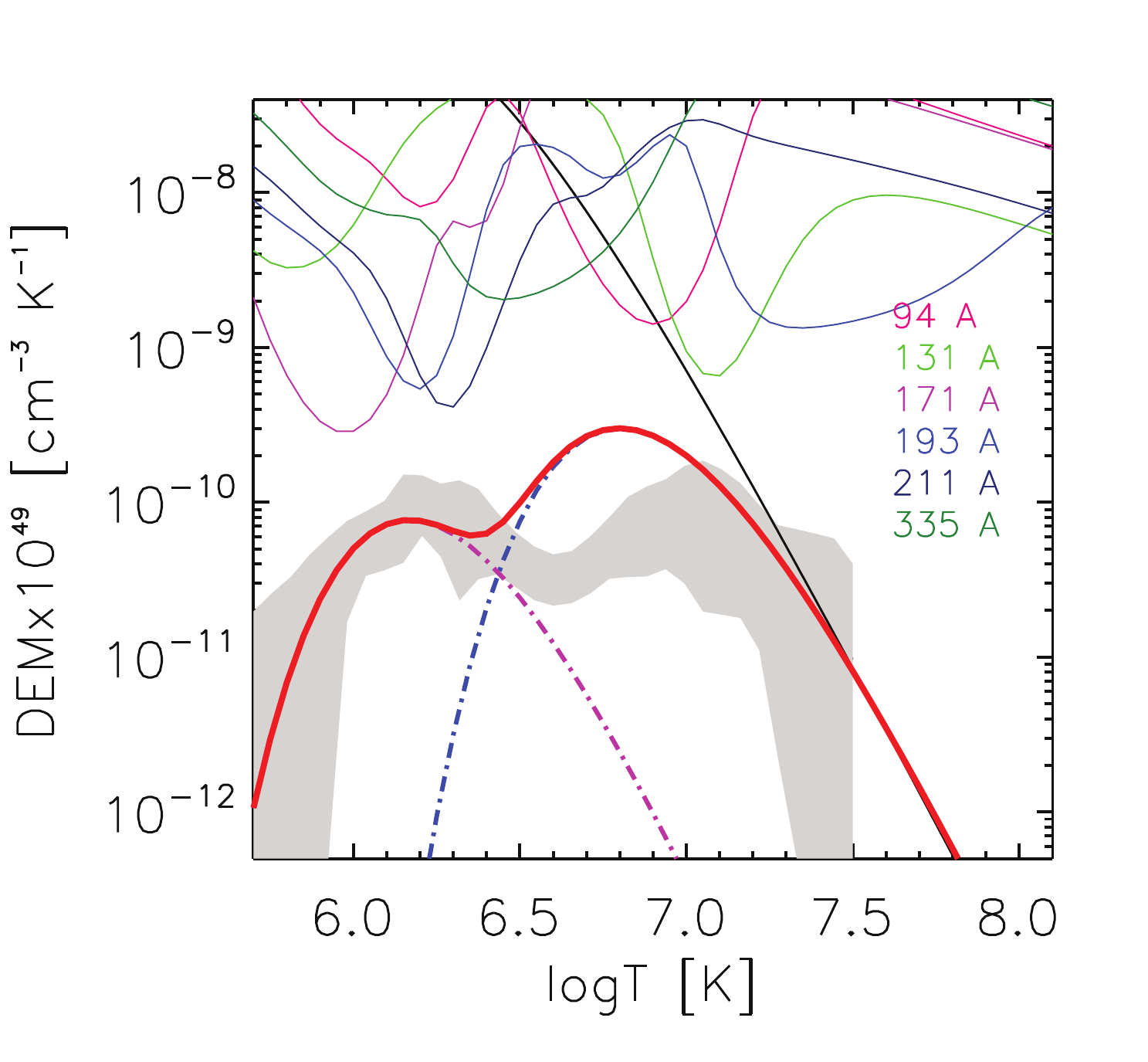}
	\includegraphics[width=0.49\textwidth]{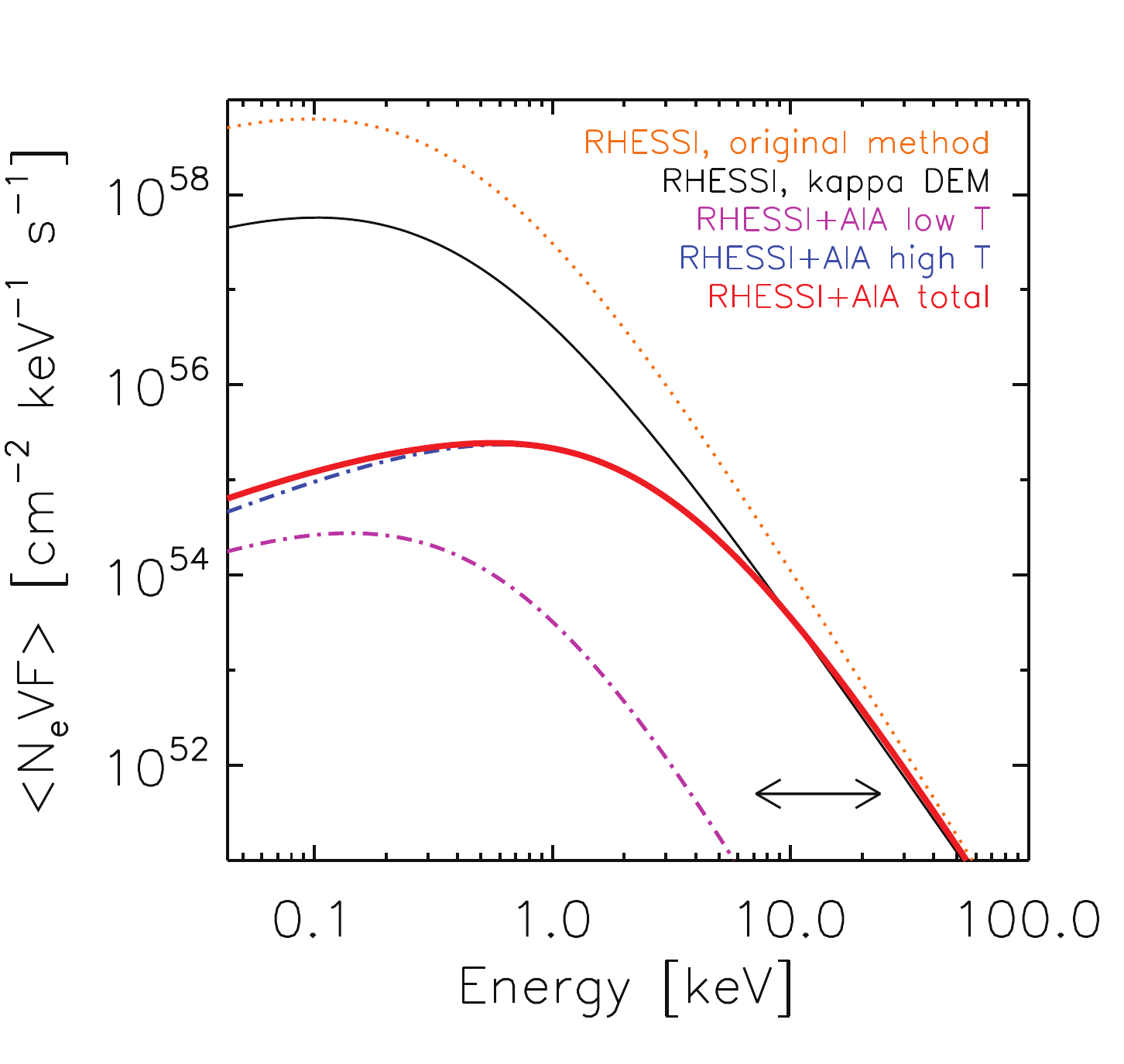}
	\caption{Comparison of the DEMs (\textit{left}) and the electron mean flux spectrum (\textit{right}) from analysis of RHESSI and AIA data. The $\kappa$-fit to the RHESSI data is shown in thin orange line, while the $\xi_\kappa(T)$ and the corresponding $\left<n_\mathrm{e}VF\right>$ is shown by thin black line. The $\xi_\mathrm{AIA}(T)$ is denoted by the gray shaded area in the \textit{left} image. The combined RHESSI and AIA fit is shown by the thick red line. It consist of two components, $\xi^\mathrm{hot}(T)$ and $\xi^\mathrm{cold}(T)$, shown in blue and purple, respectively. From \citet{Battaglia15}, \textcircled{c} AAS. Reproduced with permission. 
	\label{Fig:RHESSI_AIA}}
\end{figure}

We note here that in these papers, the $\left<n_\mathrm{e}VF\right>$ values were derived under the assumption that $f(E)$ is a Maxwellian distribution. \citet{Battaglia15} remedied this by using a $\kappa$-distribution instead (see Section 2 therein). The RHESSI 7--24\,keV data for the loop source in the confined SOL2010-08-14T10:05 C4.1 event \citep[studied previously by][]{Battaglia13} were fitted with a $\kappa$-distribution yielding $\kappa$\,=\,4.1 $\pm$0.1, and also with a $\xi_\kappa(T)$, yielding $\kappa$\,=\,3.6 $\pm$0.1. This $\xi_\kappa(T)$ function is shown by a full black line in Figure \ref{Fig:RHESSI_AIA}, while the fitting with a $\kappa$-distribution is shown by orange dotted line. The fitting interval of 7--24\,keV is shown by the horizontal arrow in the right panel of Figure \ref{Fig:RHESSI_AIA}. The regularized inversion DEMs obtained from the AIA data alone, $\xi_\mathrm{AIA}(T)$, are shown by the gray area in the left panel, denoting the uncertainties in the resulting DEMs. It contains a dip at log($T$\,[K])\,$\approx$\,6.6, a consequence of the lower AIA sensitivities at these temperatures. It is again seen that the $\xi_\kappa(T)$ derived from RHESSI alone over-predicts the $\xi_\mathrm{AIA}(T)$ at temperatures log($T$\,[K])\,$<$\,7.5 and the $\left<n_\mathrm{e}VF\right>$ fluxes at energies below about 10\,keV (Figure \ref{Fig:RHESSI_AIA}).

In addition to that, \citet{Battaglia15} found that fitting the combined RHESSI and AIA data together can be done by using a two-component $\xi_\kappa(T)$ function, consisting of $\xi_\kappa^\mathrm{hot}(T)$ and $\xi_\kappa^\mathrm{cold}(T)$. These two components are shown in Figure \ref{Fig:RHESSI_AIA} by purple and blue dash-dotted curves. Since the peak of $\xi_\kappa^\mathrm{cold}(T)$ function is located at log($T$\,[K])\,=\,6.2, it probably represents the foreground/background emission from the non-flaring corona. Therefore, the flare emission is likely well-represented by a single $\xi_\kappa^\mathrm{hot}(T)$ function peaking at about 11\,MK. The values of $\kappa$ obtained were $\kappa^\mathrm{hot}$\,=\,3.9$^{+0.2}_{-0.1}$ and $\kappa^\mathrm{cold}$\,=\,4.5$^{+4}_{-1}$, in broad agreement with those derived from RHESSI fitting alone.

These analyses show that taking account of the larger observed energy range provides tighter constraints on the mean electron flux spectrum present in the flare plasma. Additionally, approximating only a limited energy range (\textit{e.g.}, only RHESSI observations above several tens of keV) could in principle lead to misleading results on the nature of the non-thermal component present in the plasma. We note that these analyses used the AIA responses calculated for the Maxwellian distribution and that the $\kappa$-distributions can influence these (Section \ref{Sect:3.5}). This may be of potential importance mainly for low $\kappa$ since the corresponding DEMs derived from AIA under the assumption of a $\kappa$-distribution can differ from the Maxwellian one Section \ref{Sect:3.6}.

Finally, we note that independent indication of the presence of $\kappa$-distributions in solar flares were obtained by \citet{Jeffrey16,Jeffrey17} by analyzing of the \textit{Hinode}/EIS \ion{Fe}{XVI} and \ion{Fe}{XXIII} line profiles observed during flares. After accounting for the instrumental profile (via convolution), the line profiles were found to be consistent with $\kappa$ values as low as 2, although larger values of 3--6 were found for \ion{Fe}{XXIII}.

%
%
\subsection{X-ray line spectra}
\label{Sect:4.3}

In this section, we discuss the non-Maxwellian effects on the flare X-ray line spectra, especially those formed in the 1--7\,\AA~range. Since these wavelengths correspond to $\approx$12.4 and 1.8\,keV, respectively, and the lines are excited by electrons of these or higher energies, their intensities can be directly influenced by the presence of high-energy tails observed using bremsstrahlung emission (Section \ref{Sect:4.1}), but also other non-Maxwellian effects (Section \ref{Sect:4.3.2}).

%
\begin{figure}
	\centering
	\includegraphics[width=0.95\textwidth,clip=,angle=0]{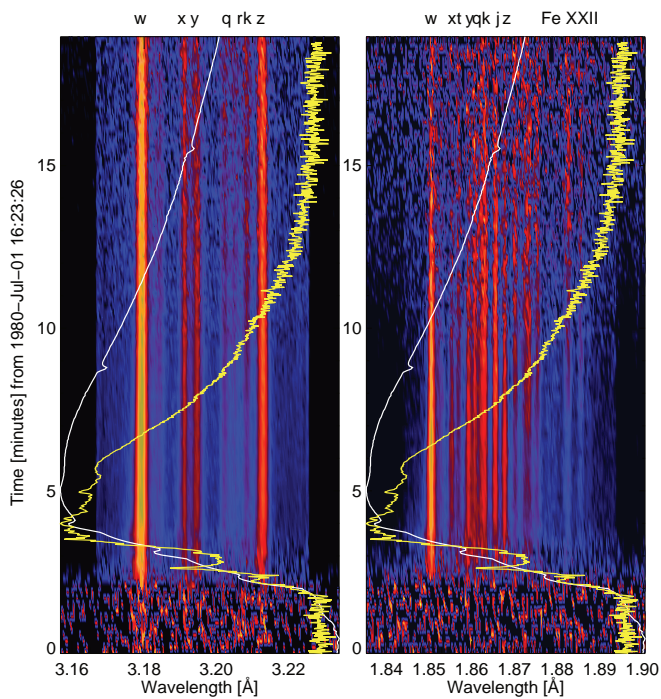}
	\caption{Time stack of normalized spectra for the flare SOL1980-07-01T16:29 observed by the BCS, a part of XRP on SMM. Left: Spectral region of \ion{Ca}{XIX} He--like triplet (Channel~1). In the upper part, the line designations are presented after \citet{Gabriel72}. Right: Spectral region of \ion{Fe}{XXV} He--like triplet including lines formed in lower ionization stages (see above the stack). On both panels, the lightcurve of the hard X-ray emission as observed by HXRBS full Sun spectrometer in the range 20-500 keV is drawn (yellow) as well as the soft X-ray lightcurve measured by GOES in the 1~--~8~\AA\ band (white). Both Ca and Fe He-like lines are pronounced from the very beginning of the impulsive flare for this event. More than 100 flare spectral cubes are available for the analysis from the SMM BCS data archive.}
	\label{Fig:XRP}
\end{figure}
%
\begin{figure}
	\centering
	\includegraphics[width=7.5cm, angle=90]{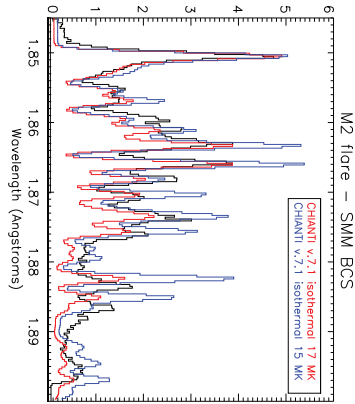}
	\caption{An SMM/BCS spectrum during a solar flare (black) with superimposed CHIANTI simulated spectra at two slightly different temperatures (blue and red).}
\label{Fig:bcs_sp}
\end{figure}
%
%
%
\subsubsection{Overview of the X-ray line observations at 1--7\,\AA}
\label{Sect:4.3.1}

The diagnostic power and scientific value of high-resolution X-ray line spectra emitted by hot, 1--30\,MK astrophysical plasmas have been well established over the past half-century. Here, we briefly discuss the spectra observed in the 1--7\,\AA~range during solar flares. These spectra are characterized by a multitude of emission lines, including (1) allowed lines of H-like and He-like ions from Si to Ni, (2) dielectronic satellite lines located longward of the respective allowed lines of the parent ion (Sections \ref{Sect:4.3.1} and \ref{Sect:4.3.2}), (3) inner-shell excitation lines, (4) lines formed from excitation levels corresponding to high principal quantum number (Section \ref{Sect:4.3.3}), and (5) free-free and free-bound continua \citep[see, \textit{e.g.},][Chapter 6.2 therein]{Phillips08}. A brief description of important instruments that observed or will be observing the lines in these short wavelength bands is discussed below and in Section \ref{Sect:6}.
We also briefly describe below two of the main non-equilibrium diagnostics provided in this wavelength range, \textit{i.e.} the possibility to search for departures from Maxwellian distributions (Section \ref{Sect:4.3.2}) and from ionization equilibrium (Section \ref{Sect:5.4}). Both cases rely on observations of the satellite lines, formed by electrons dielectronically captured by ions (see Section \ref{Sect:3.3.2} and \citet{Gabriel72}).

Although there is no X-ray spectrometer observing the 1--7\,\AA~solar flare emission lines at present, several of these instruments have been flown in the past. The first of these, flown in the 1960s and 1970s, observed the prominent spectral lines due to the H-like and He-like ions. The next generation of spectrometers in the late 1970s and early 1980s was flown on the U.S. Navy \emph{P78-1} \citep[SOLFLEX, SOLEX;][]{Doschek79}, the NASA \emph{Solar Maximum Mission} \citep[X-ray Polychromator (XRP)][\textit{cf.} Figure \ref{Fig:XRP}]{Acton80}), and the Japanese \emph{Hinotori} spacecraft \citep[SOX1, SOX2;][]{Tanaka82}. These allowed the time evolution of temperatures, emission measures, flows and turbulent velocities to be studied and related to current theories of solar flares. This was later extended with the \textit{Bragg Crystal Spectrometer} (BCS) \citep{Culhane91} aboard the Japanese \textit{Yohkoh} spacecraft. The BCS had a collimator field of view of about 6'$\times$6' and obtained spectra with eight proportional counters in the 1.7--3.2\,\AA~range, with a resolving power of about 10$^4$. Figure~\ref{Fig:bcs_sp} shows as an example a BCS spectrum obtained in the \ion{Fe}{xxv} channel during the peak phase of a solar flare, with superimposed theoretical spectra obtained from CHIANTI v.7.1 \citep{Landi13} assuming two isothermal distributions with temperatures of $T$\,=\,15 and 17\,MK, respectively. The figure shows how sensitive the intensities of individual lines are to the electron temperature. The spectral lines near the \ion{Fe}{xxv} line are also sensitive to departures from Maxwellian distributions and ionization equilibrium. 

%
%
\begin{figure}
	\centering
	\includegraphics[width=0.85\textwidth,clip=]{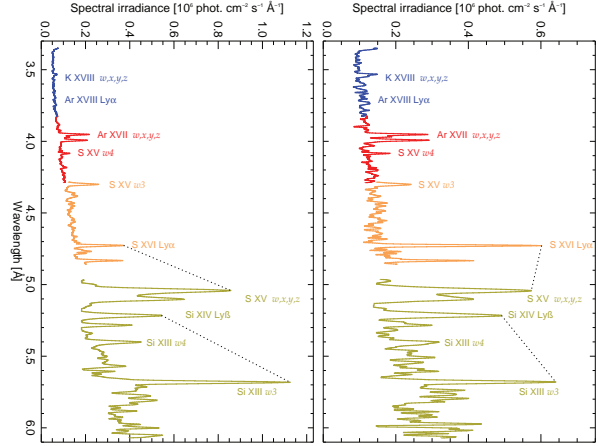}
	\caption{RESIK spectra for the rise and decay phase of C5.8 SOL2003-02-22T09:29 flare, obtained during 09:27:30--09:29:00\,UT (top) and 09:30:00--09:34:00\,UT (bottom). Colors denote individual spectral bands. Identification of prominent spectral features is given. Dotted lines connect maxima of H- and He resonance transitions in S and Si.}
	\label{Fig:bjs1}
\end{figure}


At longer wavelengths above 3.4\,\AA, the X-ray spectra were obtained during 2001--2003 by the RESIK bent-crystal spectrometer \citep[Rentgenovskij Spectrometer s Izognutymi Kristalami,][]{Sylwester05} on-board the CORONAS-F satellite. This spectrometer collected spectra of more than 200 flares in four bands covering nearly continuously the 3.4--6.1~\AA~range\footnote{Reduced level~2 spectra containing absolute fluxes for 101 flares and active region plasma is accessible at www.cbk.pan.wroc.pl/experiments/resik/RESIK\_Level2/index.html}. The RESIK spectra include a number of strong emission lines due to transitions $1s^2 - 1s(np)$ and $1s - np$ in He-like and H-like ions, denoted by letters \emph{w} and Ly, respectively. Sample RESIK spectra in four energy channels are shown in Figure \ref{Fig:bjs1} together with the identification of the main spectral features for the rise and decay phase of the C5.8 flare SOL2003-02-22T09:29. RESIK channels 1, 2, and 3 are dominated by \ion{K}{XVIII}, \ion{Ar}{XVII} triplets, and \ion{S}{XVI} Ly$\alpha$ lines, respectively. Channel 4 contains strong \ion{S}{XV} and \ion{Si}{XIII}--\ion{Si}{XIV} allowed lines, as well as a clearly resolved number of dielectronic satellite lines of \ion{Si}{XII}.

%
\begin{figure}
	\centering
	\includegraphics[width=0.9\textwidth,clip=]{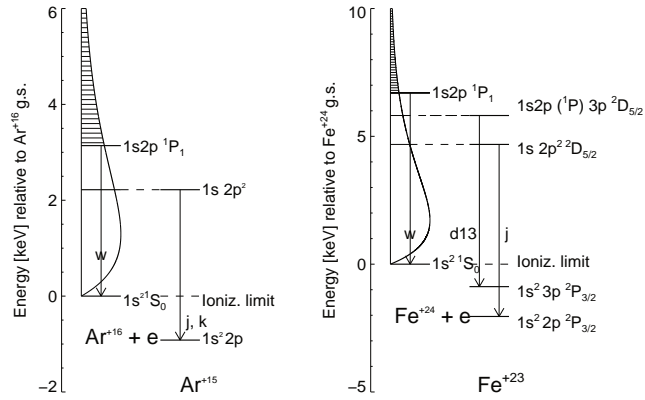}
	\caption{The diagram of energy levels for the argon and iron ions: Ar$^{+15}$ (left) and Fe$^{+23}$. The free electron exciting satellite lines j and \emph{d}13+\emph{d}15 is assumed to have Maxwell-Boltzmann distribution with $T_\mathrm{e}~=~15$~MK (Ar) and $T_\mathrm{e}~=~20$~MK (Fe). The line \emph{w} is excited by electrons with energies above 3.14~keV (Ar) and 6.7~keV (thatched areas) respectively, the Fe{\sc~xxiv} satellite line \emph{j} is excited by mono-energetic electrons with E~=~4.69~keV, and the Fe{\sc~xxiv} satellite \emph{d}13 is excited by mono-energetic electrons with E~=~5.82~keV.}
	\label{Fig:bjs2}
\end{figure}
%
%
\subsubsection{Non-Maxwellian analysis of X-ray line spectra}
\label{Sect:4.3.2}

\citet{Gabriel79} developed a technique for determining the departure of the electron velocity distribution from a Maxwellian. A departure from a Maxwellian distribution can be detected because dielectronic recombination lines are produced only by the electrons with a given energy, \textit{i.e.}, by electrons whose energy is exactly equal to the energy difference between the initial and excited states of the dielectronic capture process (equal to within the energy width of the excited state, determined by its natural lifetime). In contrast to this, allowed lines formed by electron impact excitation can be produced by any electron with an energy greater than the threshold energy of the excited state.

An energy diagram showing which parts of the assumed Maxwellian electron population contribute to the excitation of the parent allowed \emph{w} lines as well as dielectronic satellite lines (\emph{j}, \emph{k}, \emph{d13}) is displayed in Figure \ref{Fig:bjs2} for the He-like \ion{Ar}{XVII} and \ion{Fe}{XXV} ions. It is clearly seen that by increasing the temperature, the number of electrons contributing to the parent line intensity changes relative to the number of electrons at the resonance, contributing to the satellite line intensity. The dependence of the satellite to resonance line intensity ratio scales approximately as $1/T$. Similarly, the intensities of allowed lines will be increased if power-law tails (\textit{e.g.}, $\kappa$-distributions) are present. During flares, such power-law tails are observed by RHESSI at energies of $\gtrsim$\,4\,keV (see Section \ref{Sect:4.1}). The increase of the intensities of the allowed lines arise because the enhanced number of high-energy particles at these energies increase the corresponding excitation rate. In contrast to that, the dielectronic lines formed at lower energies will not be affected by this high-energy tail. Therefore, the ratio of intensities of the dielectronic to satellite lines will decrease \citep{Gabriel79}.

In order to apply this technique very high-resolution spectrometers resolving the dielectronic satellite lines are needed. The technique was successfully applied to laboratory plasmas \citep[see, \textit{e.g.}][]{Bartiromo85}. In solar spectra, these satellite lines are hard to resolve. However, \citet{Seely87} provided a non-thermal analysis of X-ray line spectra of three flares based on this technique, finding that the dielectronic satellite lines were increased during the early phase of a flare. This is contrary to the prediction of what happens in the presence of a high-energy tail or HXR burst, which were observed independently. The increase of the dielectronic satellites could also not be explained by multi-thermal plasma.

Instead, the observations were explained by the presence of a non-Maxwellian electron energy distribution more peaked than the Maxwellian one, if the peak occurs at energies corresponding to formation of dielectronic satellite lines. It was found that such distributions occur in addition to the high-energy tails (see below) and that the occurrence of this distribution coincided with the occurrence of hard X-ray bursts.

%
\begin{figure} 
	\centering
	\includegraphics[width=0.7\textwidth,clip=]{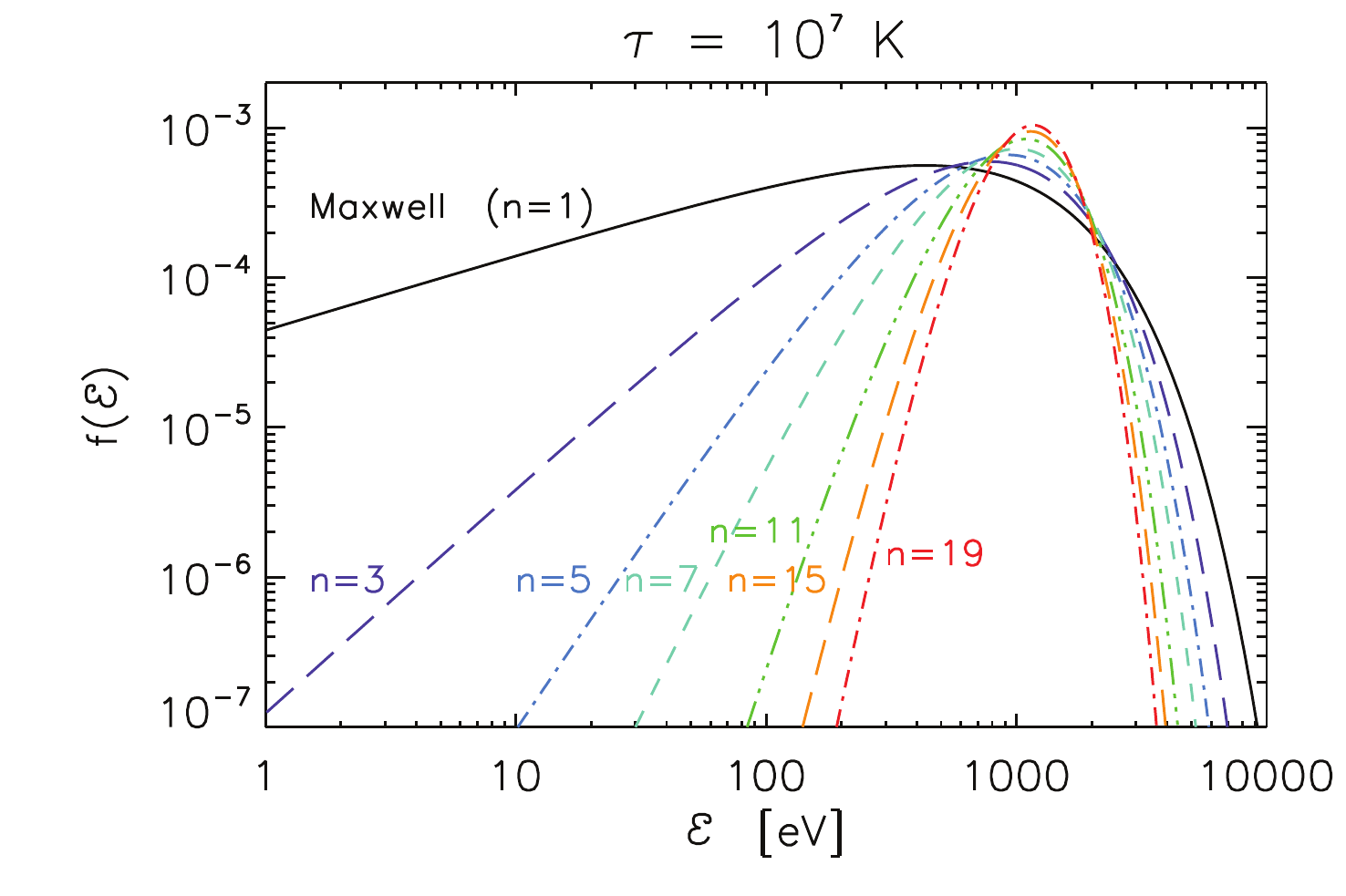}
	\caption{The shape of the Maxwellian ($n$ = 1) and $n$-distributions with $n$ = 3, 5, 7, 11, 15, and 19. The distributions have the same mean energy corresponding to $\tau$\,=\,10$^7$ K. The deviation from the Maxwellian distribution rises with value of $n$. }
	\label{Fig:n_dst}
\end{figure}

These peaked electron distributions (Figure \ref{Fig:n_dst}) have a different shape than a Maxwellian or a $\kappa$-distribu\-tion, and no high-energy tail. \citet{Dzifcakova98} modified the expression for the non-Maxwellian distribution used by \citet{Seely87} to the $n$-distribution
\begin{equation}
	f_{n}(E)dE = B_{n}\frac{2}{k_\mathrm{B}T\sqrt{\pi}}\left(\frac{E}{k_\mathrm{B}T}\right)^{n/2} \mathrm{exp}\left(-\frac{E}{k_\mathrm{B}T}\right)dE\,,
	\label{Eq:n_distribution}
\end{equation}
where $B_{n}$ is the normalization constant, $n$ and $T$ are free parameters. The mean energy of the $n$-distribution is $\left<E\right>=\frac{n+2}{2}k_\mathrm{B}T=\frac{3}{2}k\tau$ where $\tau$ is the pseudo-temperature defined from the mean kinetic energy, with the same physical meaning as temperature for the Maxwellian distribution (Figure \ref{Fig:n_dst}).

The physical conditions for the origin of the $n$-distribution were studied by \citet{Karlicky12b} and \citet{Karlicky12a}. \citet{Karlicky12b} investigated the possibility that the $n$-distribution corresponds to the so-called ``moving Maxwellian'', \textit{i.e.}, a Maxwellian with a velocity drift. This distribution matches well the $n$-distribution; however, for the observed values of $n$, drift velocities several times higher than the thermal velocity are required. Of course, such a ``moving Maxwellian'' is unstable due to the Buneman instability. Nevertheless, the $n$-distribution in solar flares persist for several minutes or more \citep{Dzifcakova08a,Kulinova11}. \citet{Karlicky12a} found that a distribution similar to a $n$-distribution can also originate in the double layers formed along flare loops. A strong gradient of the electric field in narrow double layers is able to form a stable non-Maxwellian distribution similar to a $n$-distribution with a Maxwellian background. As shown by \citet{Dzifcakova13b}, a strongly peaked electron distribution formed on double layers together with a Maxwellian background distribution can produce spectra with enhanced intensities of satellite lines observed in solar flares.

The ionization equilibrium for $n$-distributions was calculated by \citet{Dzifcakova98} and updated by \citet{DzifcakovaDudik15}. Calculations of the synthetic spectra \citep{Dzifcakova07} showed that the ionization equilibrium and subsequently the spectra are affected by the distribution function, with the ionization peaks being more peaked. The synthesis of the Fe, Ca and Si helium-like line spectra during flares were performed for $n$-distributions by \citet{Dzifcakova06} and \citet{Kulinova11}. The influence on the total radiative losses and continua was subsequently studied by \citet{Dudik11,Dudik12}.

%
\begin{figure}
	\centering
	\includegraphics[width=0.49\textwidth,clip=]{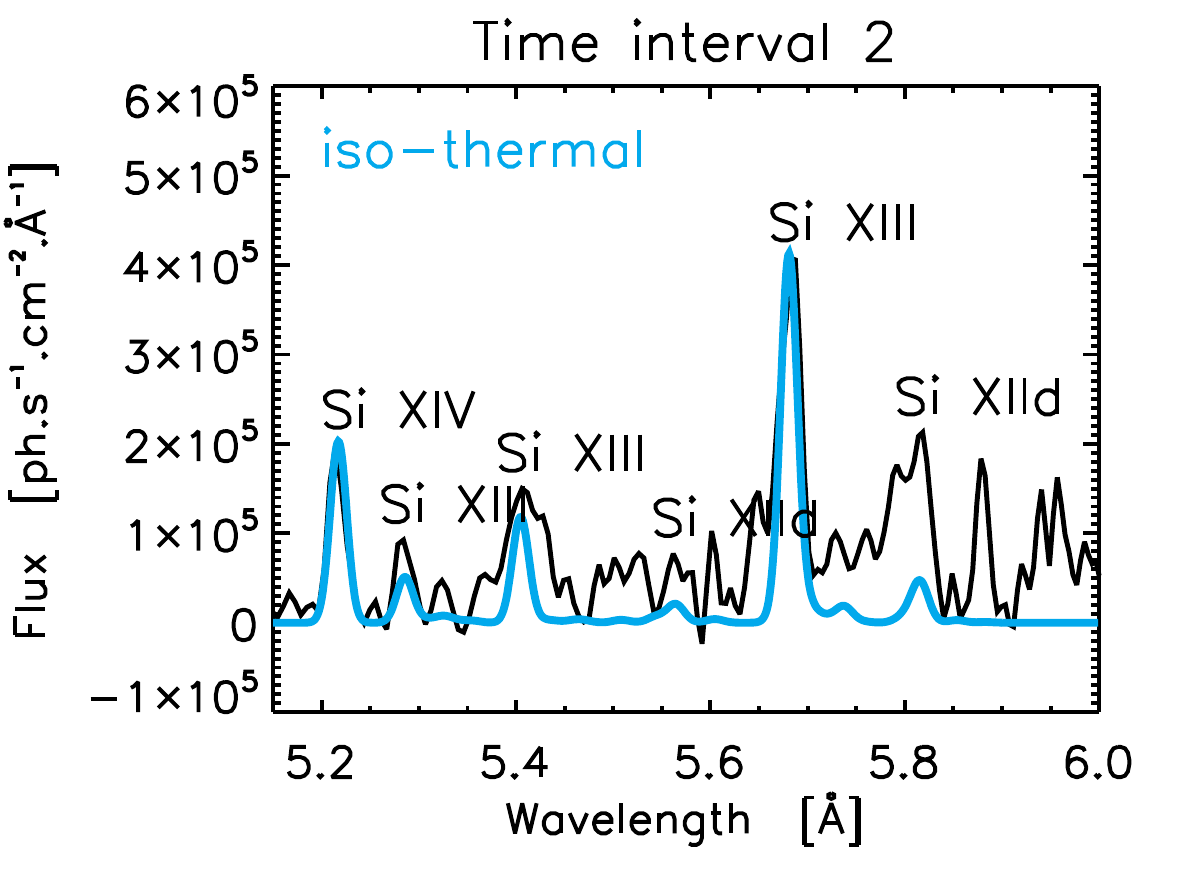}
	\includegraphics[width=0.49\textwidth,clip=]{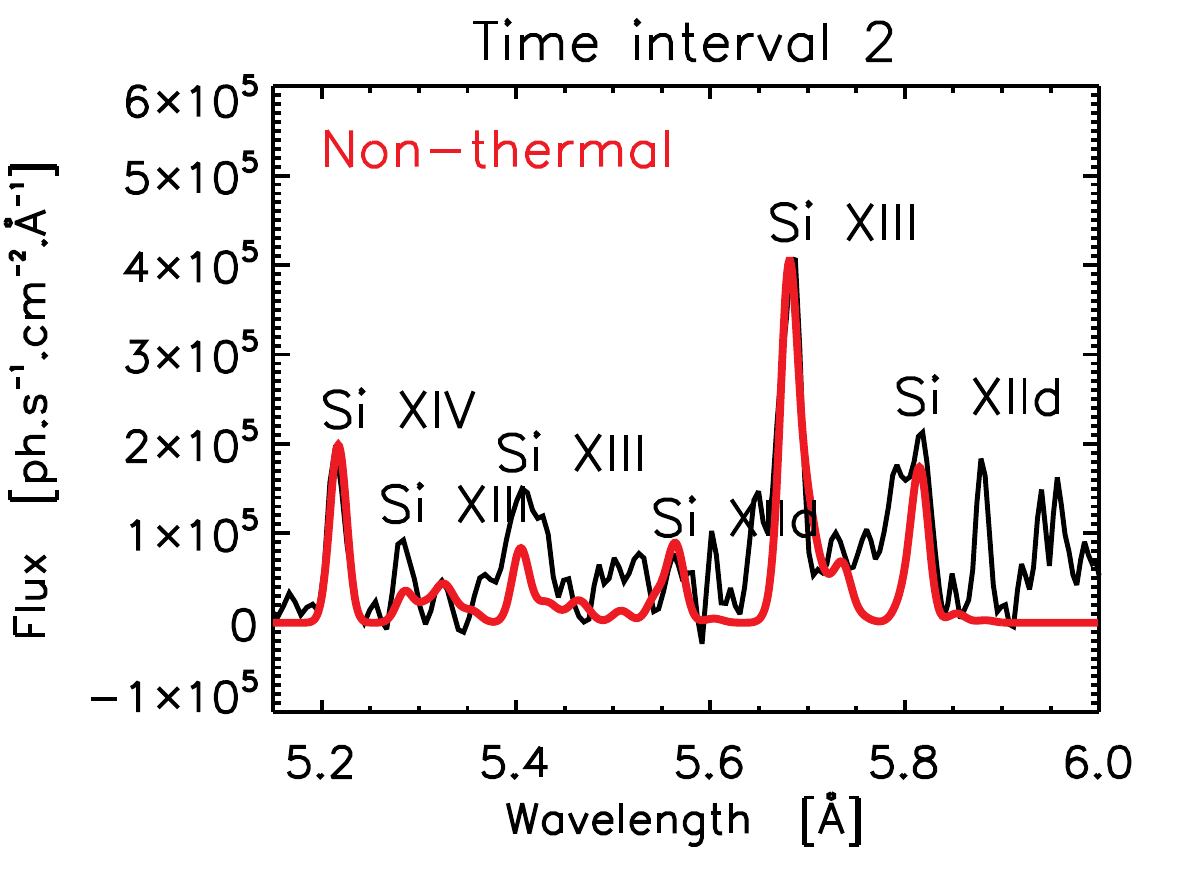}
	\caption{The comparison of the observed spectrum in the RESIK channel 4 with synthetic spectra for the Maxwellian distribution (left) and $n$-distribution with $n$\,=\,11 (right).
	Credit: \citet{Dzifcakova08a}, reproduced with permission \textcircled{c} ESO.
	\label{Fig:resik_si}}
\end{figure}

\citet{Dzifcakova08a} proposed the use of changes in the ionization and excitation equilibria with $n$-distribu\-tions for diagnostics from the flare X-ray line spectra observed at 5--6 \AA~by the RESIK spectrometer. These diagnostics utilized the ratio of the allowed lines of \ion{Si}{XIV} to \ion{Si}{XIII}, which are sensitive to the ionization state, together with the ratio of the allowed \ion{Si}{XIII} line to satellite lines \ion{Si}{XII}d, which is sensitive to the $n$-distribution. The $n$-distributions with strong deviation from the Maxwellian, $n$\,$\gtrsim$\,11, were able to explain high intensities of the satellite lines during the impulsive flare phase (Figure \ref{Fig:resik_si}). This increase was about a factor two in comparison to Maxwellian spectra. The spectra could not be interpreted as multithermal although the DEM was calculated from the whole RESIK X-ray line spectrum. Furthermore, the presence of the $n$-distribution correlated well with the occurrence of type-III radio bursts. 

%
\begin{figure}
	\centering
	\includegraphics[height=12cm,angle=90,clip=]{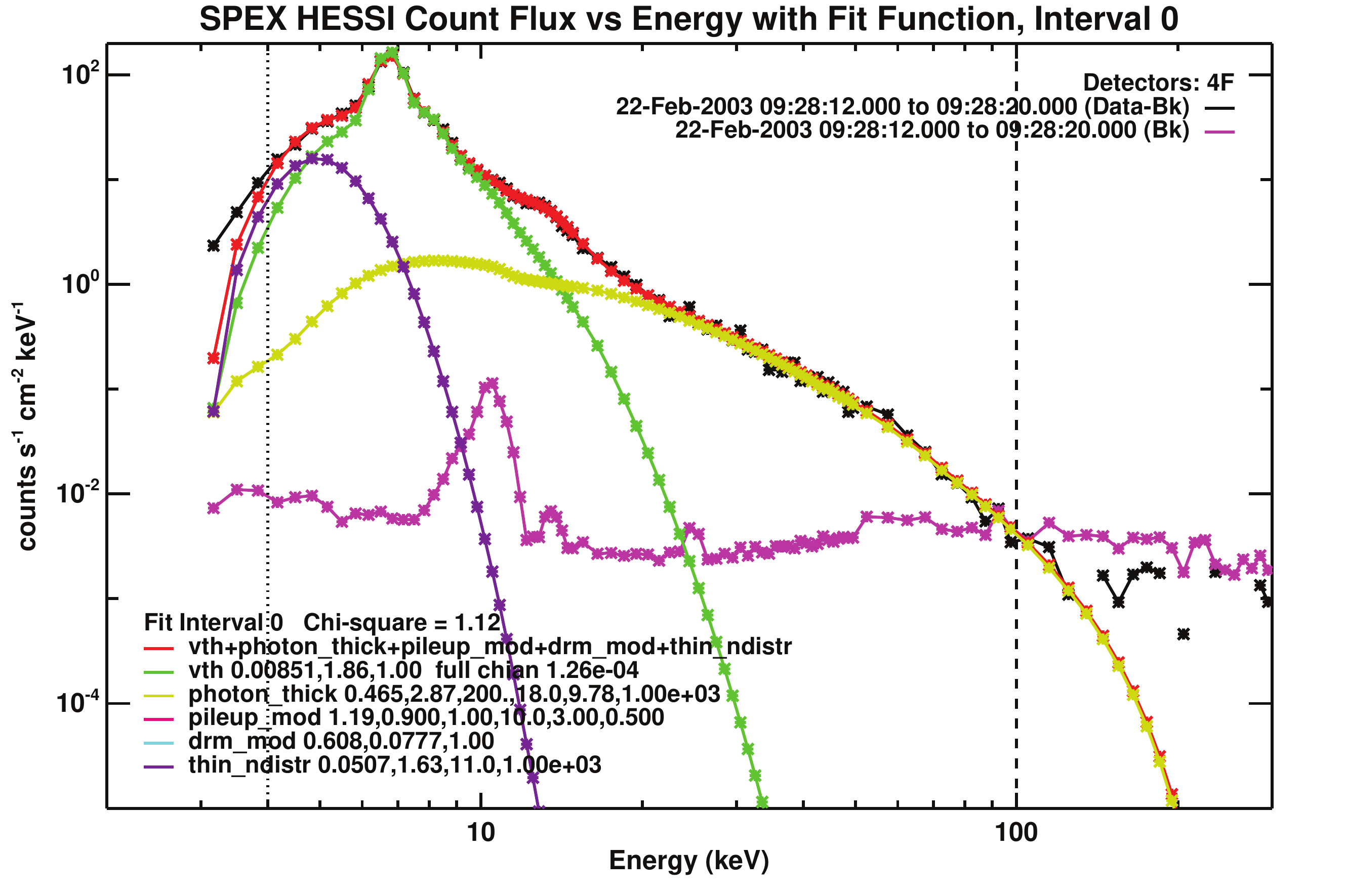}
	\caption{RHESSI spectrum fits of \citet{Kulinova11} including an $n$-distribution with $n$\,=\,11 (purple color). The thermal and high-energy tail components are shown in green and yellow, respectively. Red color denotes the sum of all components; black is the observed spectrum.
	Credit: \citet{Kulinova11}, reproduced with permission \textcircled{c} ESO.
	\label{Fig:rhessi_n_distr}}
\end{figure}

\citet{Kulinova11} used the non-thermal diagnostics proposed by \citet{Dzifcakova08a} for the analysis of three RESIK flares. An analysis of RHESSI data with energies under 10\,keV (corresponding to RESIK energy range) was also performed. Although RHESSI is strongly attenuated for these low energies, these authors successfully diagnosed the $n$-distribution from RESIK data as well for two flares where the RHESSI data were available (Figure \ref{Fig:rhessi_n_distr}). Besides this non-thermal component corresponding to $n$-distribution, the RHESSI spectra also contained a thermal component and a high-energy power-law tail. The parameters of the $n$-distribution derived from RESIK and RHESSI spectra agreed within their respective errors. The emission measure calculated from RHESSI for the $n$-distribution was about one order higher than for the thermal component of the plasma that explained the non-Maxwellian character of RESIK line spectra \citep{Dzifcakova11p}. As in previous cases, the intensities of the satellite lines were enhanced by up to a factor two during the impulsive phase and flare maximum. This enhancement coincided with the presence of type-III radio bursts observed at radio frequency regions of 2.7--15.4\,GHz.

\citet{Dzifcakova11a} calculated theoretical spectra for a composite $np$-distribution characterized by a $n$-distribution bulk and a power-law tail. They showed that such a composite distribution is also able to produce enhanced dielectronic satellites, while the Maxwellian distribution with power-law tails alone cannot produce this effect. 

The dielectronic satellite lines from other elements in RESIK X-ray spectra usually have low intensities and are blended with allowed lines. This makes the application of a method analogous to \citet{Dzifcakova08a} difficult. However, the enhancement of \ion{Ar}{XVII} line intensities were reported using the RESIK spectra for some flares. This enhancement can also potentially be explained by the overabundance of electrons at several keV similar to the $n$-distribution \citep{Sylwester08}.

Independent theoretical diagnostics of the $n$-distribution were also proposed. These methods rely on the predicted disappearance of recombination edges in the free-bound continuum \citep{Dudik12} or the polarization of gyro-resonance emission in micro and radio wave region \citep{Fleishman14}.

%
\begin{figure}[ht]
	\centering
	\includegraphics[width=0.49\textwidth,angle=0]{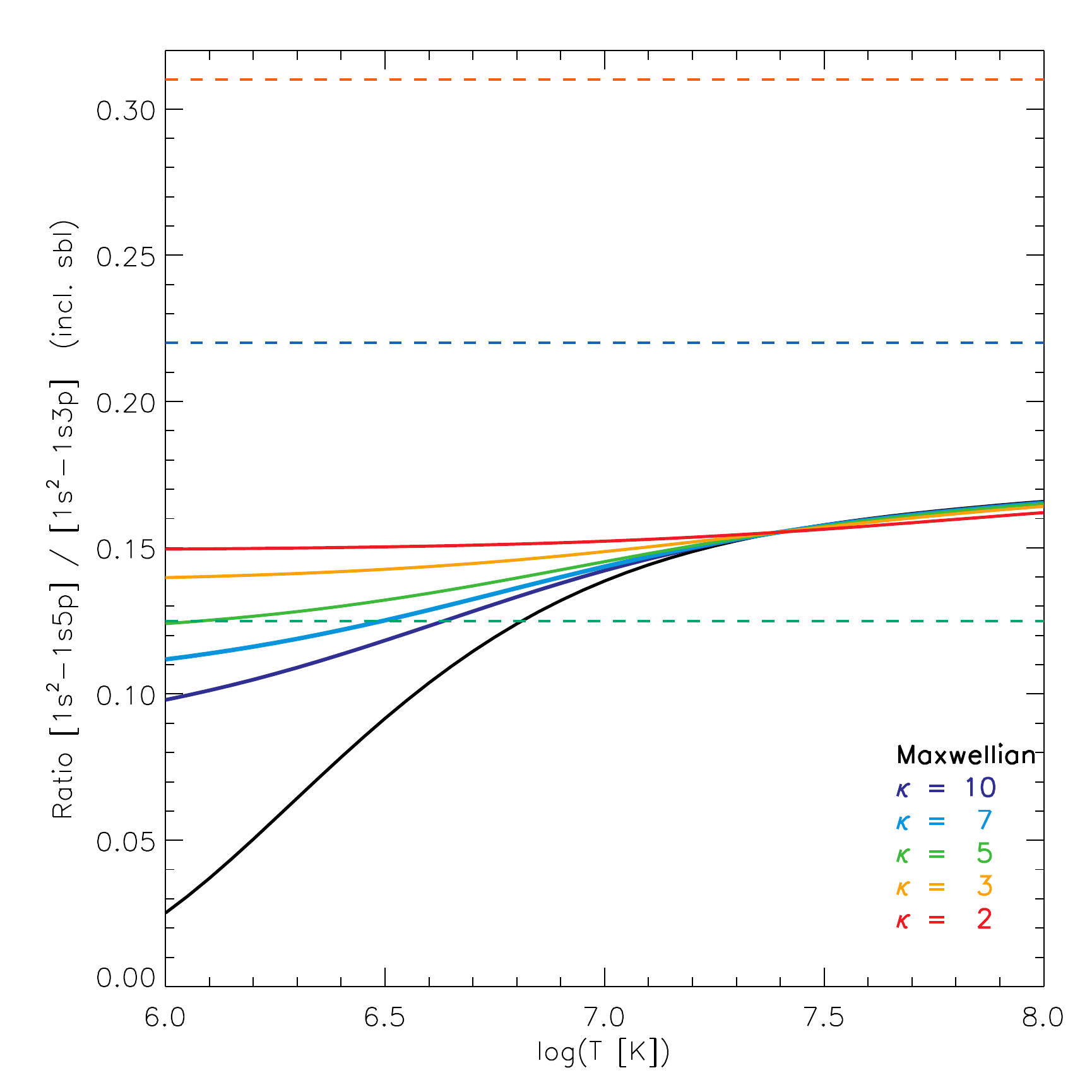}
	\includegraphics[width=0.49\textwidth,angle=0]{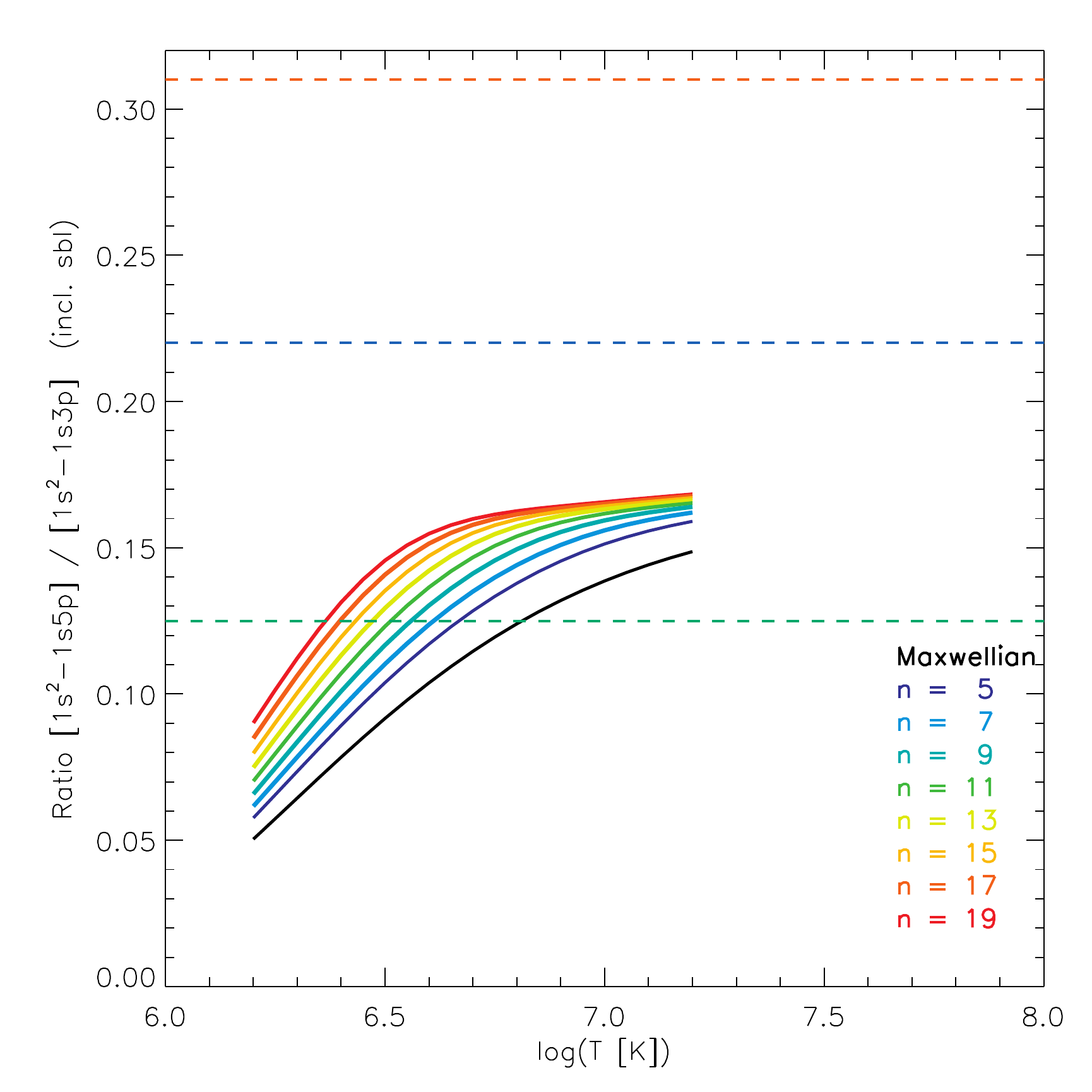}
	\caption{\ion{Si}{XIII} $1s^2-1s5p$ to $1s^2-1s3p$ line intensity ratio as a function of temperature and the non-Maxwellian $\kappa$-distributions (left) and $n$-distributions (right panel). The values of individual non-Maxwellian parameters are indicated. Observed line ratios are indicated by dashed horizontal lines, corresponding to the rise (red), maximum (blue), and decay (green) phases of C5.8 SOL2003-02-22T09:29 flare.}
	\label{Fig:high-n-kappa}
\end{figure}

\subsubsection{H- and He-like high-$n$ transitions}
\label{Sect:4.3.3}

During the impulsive phase, RESIK observed emission at wavelengths corresponding with allowed transitions from very high levels characterized by the principal quantum number $n$ up to 6--7 \citep{Kepa06}. These intensities were strongly enhanced, by a factor of more than 2 when compared to the transitions from the lower levels ($n$\,=\,3, see dashed lines in Figure \ref{Fig:high-n-kappa}). In contrast to that, during the soft X-ray peak phase the intensities of $n$\,=\,6--7 lines became close to the expected values. Since the atomic data for these high-$n$ levels were unavailable at the time, \citet{Kepa06} used approximate methods to calculate the synthetic line intensities. These higher-$n$ transitions are sensitive to local plasma conditions and are used in laboratory plasma diagnostics \citep{Chen07}. The observed intensity ratios could not be explained with an isothermal or multi-temperature approach.

In principle, non-thermal electrons that are present during the impulsive phase of flares could cause the observed increase. To verify this, we have used the cross-sections from the recent UK APAP-network calculations \citep{Fernandez16} to predict the effects of a $\kappa$- or $n$-distribution on the line ratios (Figure~\ref{Fig:high-n-kappa}). By considering a $\kappa$ or $n$-distribution, it is possible to account for measurements collected for some flares during the decay phase only, but not during the rise phase (Figure~\ref{Fig:high-n-kappa}). The high-energy tail of a $\kappa$-distribution do enhance the ratios at lower temperatures, but not at higher temperatures. Only a moderate increase is present for the $n$-distributions.

Therefore, even assuming the presence of non-Maxwellian electron distributions, this does not explain the strongly enhanced intensities of lines formed from high-$n$ levels. Another possibility is that transient ionization conditions in recombining plasma prevail in the emitting source region. These may result in the overpopulation of higher-$n$ levels \citep{Dupree68}. Such persistent presence of recombining plasmas at the temperature of formation of the observed lines is however unlikely. An alternative mechanism contributing to overpopulation of higher-$n$ levels could be charge exchange \citep{Wargelin08}. Despite these possibilities, the mechanism increasing the high-$n$ line intensities has not yet been identified.

%
%
%
%
%
\section{Departures from ionization equilibrium}
\label{Sect:5}
When studying the observed emission, it is common to assume that the excitation and de-excitation processes occur independently of ionization and recombination, because the timescales of these processes are extremely different, by orders of magnitude. It is also common to assume that the plasma is in ionization equilibrium, \textit{i.e.} the ion charge distribution is in a steady state. Such situations were assumed in Sections \ref{Sect:3}--\ref{Sect:4}. However, is has long been recognised that solar events exist for which this latter assumption is likely to break down, mostly because of the large spectral variability observed on timescales shorter than the ion equilibration times. We describe the timescales below, in a separate subsection. We then summarise our current knowledge, based on some older and more recent observations and models. 

%
%
\subsection{Atomic timescales}
\label{Sect:5.1}
A plasma contains atoms, ions, free electrons and radiation and may be permeated by a magnetic field. Interaction between these species collectively through collisional and radiative processes establishes the population structure of the emitting atom or ion. In turn, the population structure reflects key parameters of the plasma, such as electron and ion temperature, electron density, the thermal structures of the plasma, its chemical composition and ionization state and its dynamics.

The lifetimes of the various states of ions vary enormously and determine the relaxation times of the atomic populations. The order of these timescales, together with the plasma development times and their values relative to observation times, determines the modelling approach.

The key lifetimes consist of two groups. The first group just includes the {\it atomic parameters} (ground, metastable and excited state radiative decays and auto-ionizing decay). The second group depends on {\it plasma conditions}, especially particle density involved in processes such as free particle thermalization, charge-state change including ionization and recombination, and redistribution of population amongst excited states. Mass flows are also of importance (Section \ref{Sect:5.2} and \ref{Sect:5.3}). These two groups need to be compared with each other and with the {\it plasma timescales}, representing relaxation times of transient phenomena, plasma ion diffusion across temperature and density gradients and observation times.

It is often assumed that the ion populations respond instantaneously to changes in the plasma temperature and density and the ionization balance is calculated in {\it ionization equilibrium}. However, whenever there is an extremely rapid change in the plasma conditions, the ionization and recombination timescales can be greater than the plasma timescales. This implies that the local ionization balance is no longer determined exclusively by the local temperature and density conditions, but depends on the past history of the temperature, density and ionization state of the plasma. Therefore, the {\it time-dependent ion populations} need to be determined:
\begin{equation}
	 \left(\frac{\partial}{\partial t} + \vec{u} . \vec{\nabla} \right)N^{(z)} = n_\mathrm{e} [q^{(z-1)}N^{(z-1)} + (q^{(z)}+\alpha^{(z)})N^{(+z)} + \alpha^{(z+1)}N^{(z+1)}]
	\label{Eq:Ion_time_evol}
\end{equation}
where $\vec{u}$ represents bulk flow velocity and the left-hand side is generally no longer zero as in Equation (\ref{Eq:Ion_eq}). We remind the reader that $q$ and $\alpha$ are the {\it effective} ionization and recombination coefficients. They give the contribution to the growth rates
for the ground state population, due to the effective ionization, which includes direct and excitation/auto-ionization contributions, and the effective recombination, which includes radiative, dielectronic and three-body contributions (see Section \ref{Sect:3.3}). 

\begin{figure}[!ht]
	\centering
	\includegraphics[width=0.8\textwidth,angle=0]{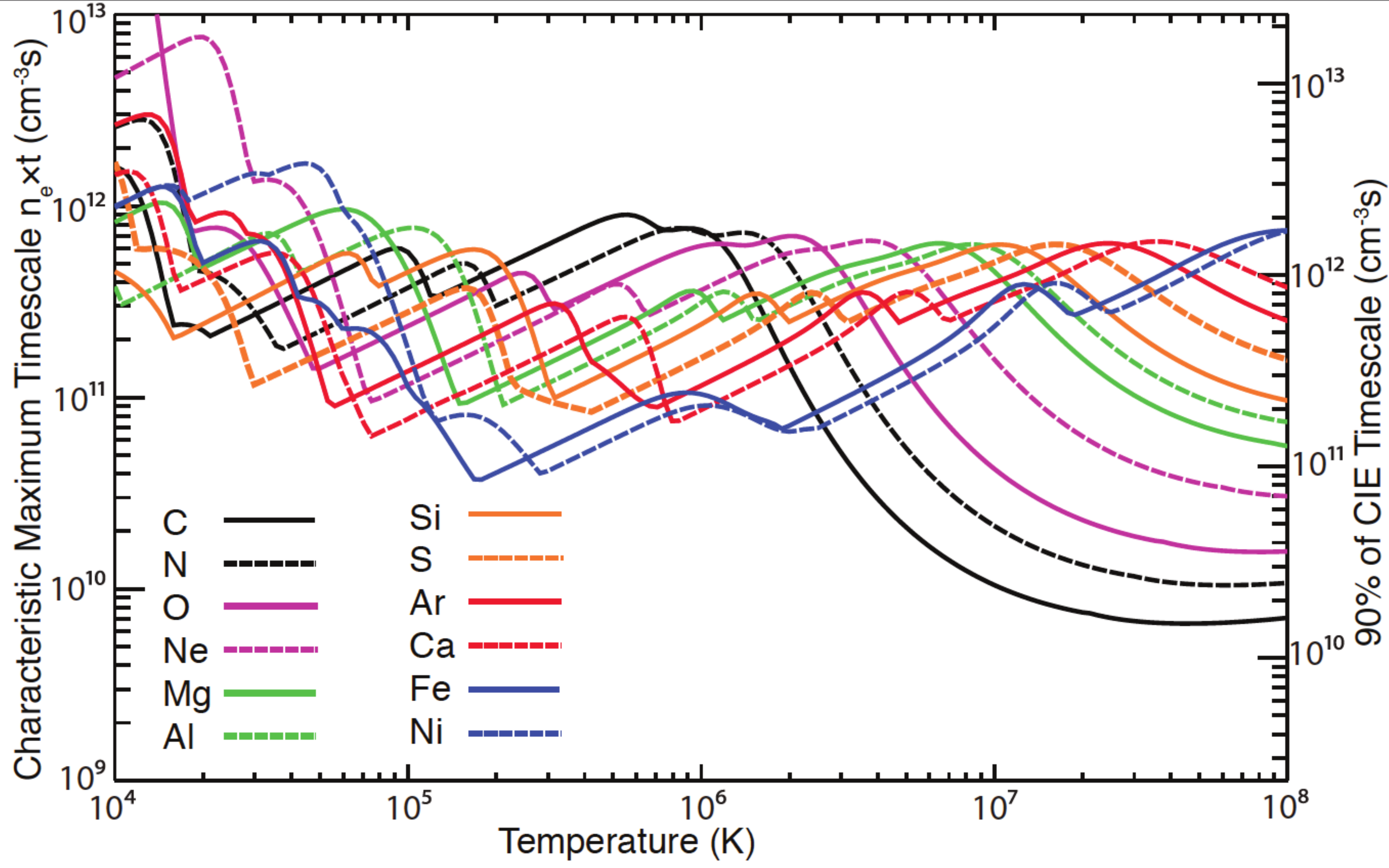}
	\caption{Characteristic maximum timescales for ion equilibration as a function of $T$ and $Z$. Individual elements are depicted by different colors and linestyles. The axis at the right-hand side gives the density-weighted timescale for equilibration of 90\% of the ions. From \citet{Smith10}, \textcircled{c} AAS. Reproduced with permission. 
\label{Fig:NEI_timescales}}
\end{figure}

Calculations of the ionization equilibration timescales have been performed by several authors, \textit{e.g.}, \citet{Golub89}, \citet{Reale08}, and \citet{Smith10}. In \citet{Smith10}, it was shown that, for the case of no flows ($\vec{u}$\,=\,\textbf{0}), the characteristic product of the electron density and maximum timescale for equilibration (see Figure \ref{Fig:NEI_timescales}) is of the order of $n_\mathrm{e} t$\,$\approx$\,10$^{11}$--10$^{12}$ cm$^{-3}$\,s and that the most abundant ion is often the slowest to reach equilibrium. This means that at densities of few 10$^{9}$--10$^{11}$ cm$^{-3}$ typical of the active corona and the transition region, the solar plasma can be out of ionization equilibrium if intensity changes occur on timescales less than about 10--100\,s. Such timescales for intensity changes are indeed often observed. Non-equilibrium ionization may be generated by a variety of processes in the solar corona and transition region, ranging from impulsive plasma heating and cooling to mass flows through a steep temperature gradient as in the transition region. Since the line intensity (Equation \ref{Eq:line_intensity}) depends directly on the relative ion abundance given by Equation (\ref{Eq:Ion_time_evol}), the presence of non-equilibrium ionization will necessarily lead to transient, time-dependent phenomena that can be characterized by fast intensity changes. 

%

%
%
\subsection{Non-equilibrium ionization in the transition region}
\label{Sect:5.2}

It has been known for a long time that the transition region (TR) is highly dynamic. Even in the transition region of the quiet Sun, downflows are ubiquitous, so the ions are flowing into regions with lower temperatures, given the steep gradient. Non-equilibrium ionization effects are therefore important in the transition region. The IRIS \citep{DePontieu14} observations with high temporal cadence and spatial resolution (compared to previous missions) have now unambiguously confirmed this. Hi-C \citep{Kobayashi14} high-resolution and high-cadence observations have also shown the dynamical nature of the low-temperature emission. Intensity changes on timescales of the order of seconds have now been observed in a multitude of phenomena outside of flares by IRIS and Hi-C \citep[\textit{e.g.},][]{Testa13,Regnier14,Hansteen14,Peter14Sci,Tian14,Vissers15,Tajfirouze16,Martinez16}.

\subsubsection{1-D hydrodynamic modelling}
\label{Sect:5.2.1}

Earlier 1-D hydrodynamical modelling showed that non-equilibrium ionization has significant effects on line intensities, electron densities and even relative abundances in the transitions region. The literature is extensive \citep[see, \textit{e.g.},][]{Raymond78,Dupree79,Borrini82,Noci89,Raymond90,Hansteen93,Spadaro94,Edgar00}.

More elaborate 1-D hydrodynamical modelling with an adaptive grid and a self-consistent calculation of the total radiative losses was introduced with the HYDRAD code \citep{Bradshaw03a}. HYDRAD calculations confirmed strong enhancements in some TR ions such as C IV \citep{Bradshaw04}. We note that ions such as C IV are known to be 
{\it anomalous}, in the sense that their intensities are much stronger than those from ions formed at similar 
temperatures, assuming ionization equilibrium in the low-density limit \citep[see, \textit{e.g.}][and references therein]{DelZanna02}.

The highly dynamic nature of the transition-region emission as observed by IRIS was recently studied in terms of 
 non-equilibrium ionization by a number of authors. \citet{Doyle12} and \citet{Doyle13} applied the GCR technique 
to predict the response of TR lines to short heating bursts occurring at loop footpoints. In \citet{Doyle13}, the \ion{Si}{IV} 1393.76~\AA\ and \ion{O}{IV} 1401.16~\AA\ lines were studied. It was found that about 0.2\,s after the occurrence of the burst, \ion{Si}{IV} responds with an intensity increase by a factor of more than three, while the intensity of \ion{O}{IV} is increased only weakly. With time, the \ion{Si}{IV} intensity returns closer to the equilibrium values. The authors concluded that a rapid increase of \ion{Si}{IV} over \ion{O}{IV} is a signature of non-equilibrium ionization.

\citet{Judge12} studied the non-equilibrium ionization effects in H, He, C, N, O, and Si during and after the heating of chromospheric plasma described by 1-D hydrodynamic equations. In their model, the chromospheric plasma at heights of about 2\,Mm was first heated and accelerated upward for about 30\,s (model A therein), and then heated for another 30\,s, until coronal temperatures of about 1\,MK were reached. The evolution of the heated plasma in a calculation assuming ionization equilibrium was found to be more variable than the case involving full non-equilibrium treatment. 

%
%
\begin{figure}[!ht]
	\centering
	\includegraphics[width=6cm]{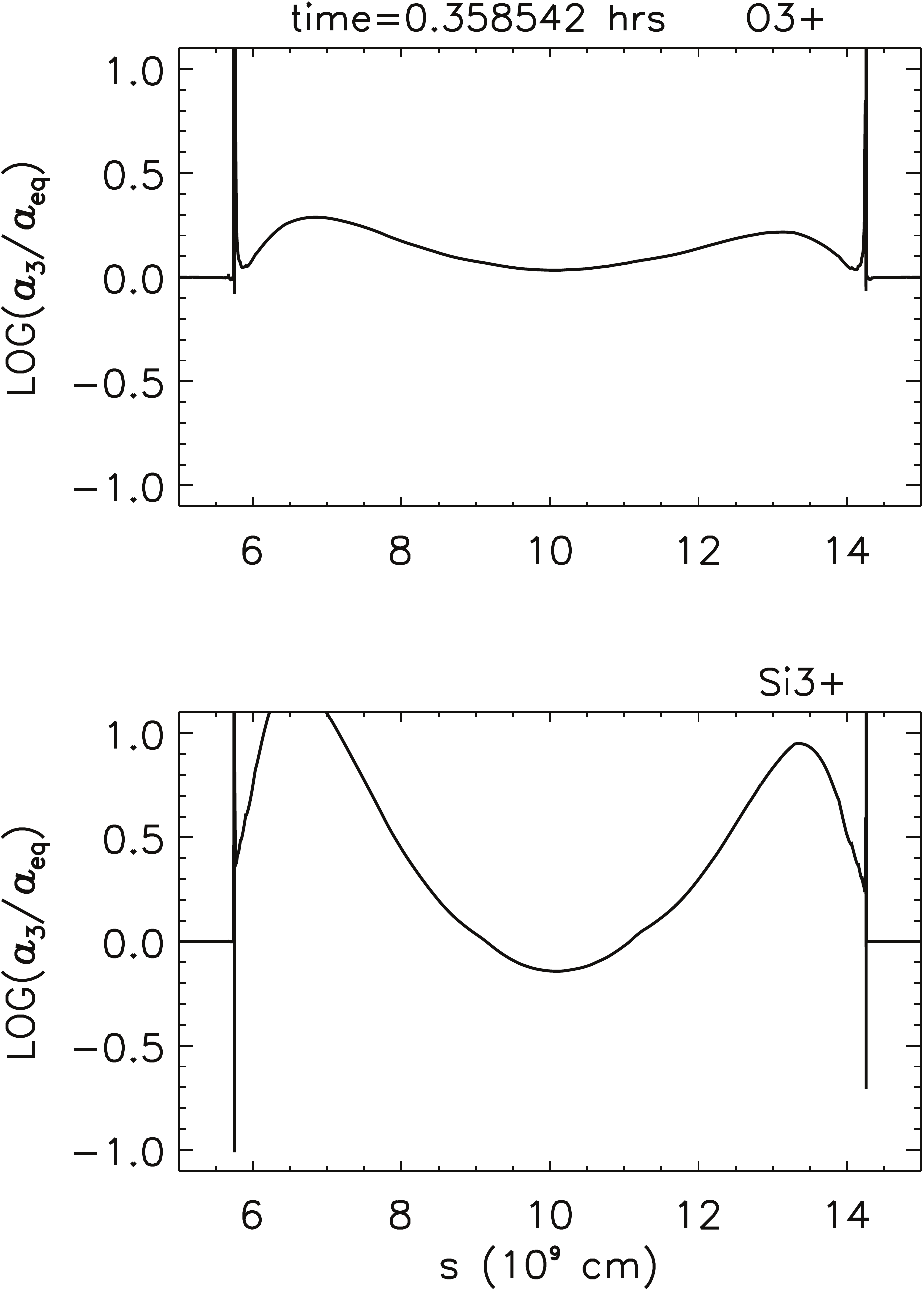}
	\caption{Non-equilibrium/equilibrium population ratios as a function of the loop length for \ion{O}{IV} (top panel) and \ion{Si}{IV} (bottom panel). From \citet{Giunta16}. \textcircled{c} SISSA. Reproduced with permission.
\label{Fig:Hydr_simulation}}
\end{figure} 
%
\subsubsection{An example of non-equilibrium ionization in IRIS TR lines}
\label{Sect:5.2.2}

To show how the non-equilibrium ionization influences the IRIS TR lines, we present an example hydrodynamic simulation including the non-equilibrium ionization. The emission of two lines, \ion{Si}{IV} 1393.76\AA~and \ion{O}{IV} 1401.16\AA~and their response to non-equilibrium ionization is investigated using numerical simulations of coronal loop hydrodynamics, which reproduce a flaring small-scale event, and are compared with IRIS observations of April 2, 2014.

The standard set of hydrodynamic equations for the conservation of mass, momentum and energy are solved, in one-dimensional magnetically confined loop with a 80\,Mm length, using the Adaptively Refined Godunov Solver \citep[ARGOS,][]{Antiochos99, MacNiece00} code following the approach of \citet{Susino10}. To simulate the behavior of a small flare, the case of an impulsive heating event with a pulse cadence of 1/4 of the loop cooling time, \textit{i.e.} $t$=250~s~=~$\tau_{cool}/4$ \citep{Serio91}, has been chosen \citep[Run 9 in][]{Susino10}. An energy of 10$^{24}$~erg for each pulse \citep{Parker83,Parker88} is supplied to an initial static loop with apex temperature of 7.5$\times$10$^5$~K.

The system of equations for the ionization balance (Equation \ref{Eq:Ion_time_evol}) are then solved providing the variations of the ion populations with the time for \ion{Si}{IV} and \ion{O}{IV} \citep{Lanza01}. The initial state at $t$=0~s is defined as the static equilibrium loop solution before the heating is supplied. 

Figure \ref{Fig:Hydr_simulation} shows the ratio between the transient ($a_3$) and equilibrium ($a_{eq}$) ionization balance calculations as a function of the loop length $s$ when the heating is turned on for \ion{O}{IV} and \ion{Si}{IV}. During a pulse, \ion{Si}{IV} is enhanced at loop footpoints by a factor $\approx$~10 compared to equilibrium, while \ion{O}{IV} is enhanced only by a factor of $\approx$3. This shows that \ion{Si}{IV} depends strongly on non-equilibrium ionization and \ion{O}{IV}, in contrast, is less affected.

This different response to transient ionization for the two ions may be directly compared with IRIS observations investigating the behavior of the two lines \ion{Si}{IV} 1349 \AA\, and the \ion{O}{IV} 1401 \AA\, and the changes of their intensities during a flaring event with respect to a quiet region \citep{Giunta16}. Assuming that in the quiet Sun region both ions are in ionization equilibrium, whereas the rapid variations of plasma conditions due to a transient event, such as a flare, lead to departure from ionization equilibrium, then the following relation is derived for each line \citep{Giunta16}:
\begin{equation}
{I_{AR}\over I_{QS}}={\int_{\Delta T}G(T_\mathrm{e},n_\mathrm{e},t)dT\over \int_{\Delta T}G(T_\mathrm{e},n_\mathrm{e})dT}={a_3 \over a_{eq}}
\label{obs_comparison}
\end{equation}
where, $I_{AR}$ and $I_{QS}$ are the line intensities of a given line respectively in a flaring region and in a quiet region, $a_3$ and $a_{eq}$ stand for the relative ion abundance of the \ion{Si}{iv} and \ion{O}{iv} out of and in equilibrium, respectively, and $a_3/a_{eq}$ is the ion's population ratio as shown in Figure \ref{Fig:Hydr_simulation}.

%
\begin{figure}
	\centering
	\includegraphics[width=7cm]{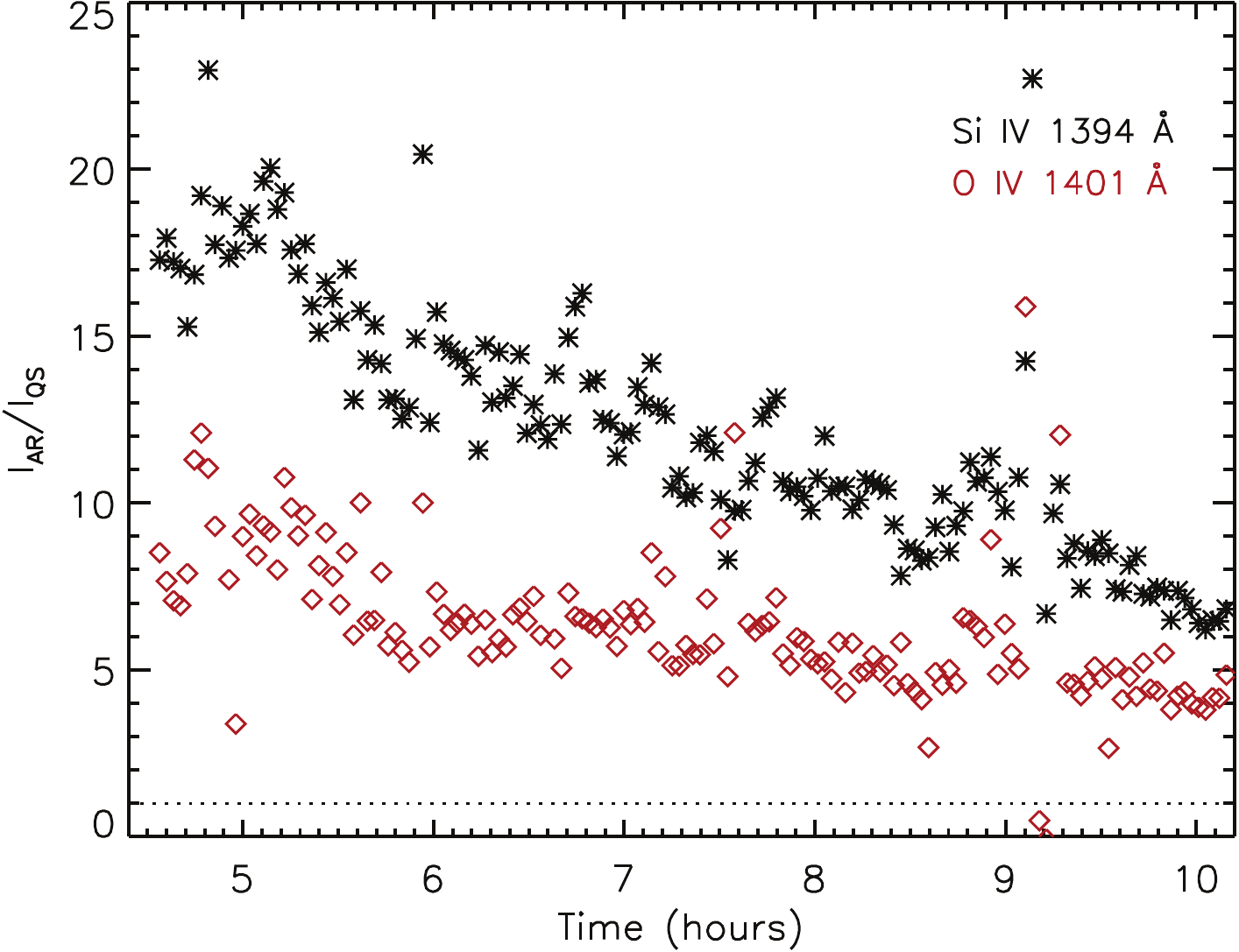}
	\caption{Observed ratios between the intensities of the flaring region, $I_\mathrm{AR}$, and the quiet region, $I_\mathrm{QS}$, as a function of time for \ion{Si}{IV} 1394 \AA\,(black points) and \ion{O}{IV} 1401 \AA\, (red points). From \citet{Giunta16}. \textcircled{c} SISSA. Reproduced with permission.
\label{Fig:TR_Observations}}
\end{figure} 

These results can be compared directly to observations. To do that, the data taken on April 2, 2014 from 04:33 to 10:11 UT are used. IRIS was pointed to an active region close to the limb. The observation consists of a large coarse 8-step raster, with steps 8$\times$2'', field of view was 14$'' \times$119$''$, and step cadence of 16.4~s. The raster cadence was 131~s with 155 raster positions. To increase the signal, each raster is summed along the $X$ direction and two regions have been selected along the $Y$ direction (200 pixels each): active region (AR) and a quiet Sun (QS). Although \ion{O}{IV} 1401.16\AA~is much weaker than \ion{Si}{IV} 1393.76\AA, the counts are sufficient if binned along $X$ for each raster. Background-subtracted intensity ratios $I_\mathrm{AR}/I_\mathrm{QS}$ are shown in Figure \ref{Fig:TR_Observations} as a function of time for the two lines. We see that \ion{Si}{IV} is enhanced by a factor $\approx$7--23, while \ion{O}{IV} intensity shows a less defined increase, $\approx$5 on average, in broad agreement with the enhancement due to non-equilibrium ionization of the hydrodynamic simulations shown in Figure \ref{Fig:Hydr_simulation}.

\subsubsection{Non-equilibrium ionization in 3-D modeling with Bifrost and consequences for IRIS spectra}
\label{Sect:5.2.3}

While 1-D modeling is able to capture both the importance and the details of non-equilibrium ionization along a single magnetic structure, solar atmosphere and especially the TR consists of many different structures. Realistic modeling of the TR involving many different structures is possible only using 3-D codes. In this section, we discuss the setup as well as the results obtained by the Bifrost code \citep{Gudiksen11}. Bifrost is an MHD code for stellar atmospheres, solving the standard MHD equations by utilizing a staggered mesh explicit code with 6th order spatial differential operator and 3rd order time accuracy achieved by the Hyman predictor-corrector scheme. Boundary conditions are non-periodic and the reflection at boundaries is minimized. Since the code is explicit, it includes a diffusive operator split into a small global diffusive term and a local hyper diffusion term.

The implementation of the non-equilibrium ionization in the Bifrost simulation is discussed in detail by \citet{Olluri13a}. The Bifrost simulation includes wavelength-dependent non-LTE radiative transport in the photosphere and lower chromosphere, optically thin losses in the corona, and field-aligned heat conduction. Heating in the simulation occurs due to many dissipation events relaxing the stressed magnetic fields (see \citet{Hansteen15} and \citet{Carlsson16} for more details). The non-equilibrium ionization is implemented via the DIPER package \citep{Judge94}. This approach considers collisional and radiative transitions (ionization, recombination, excitation, and spontaneous emission) between several tens of energy levels for each element. The total number of energy levels has to be restricted in order to make the simulation tractable on current supercomputers. Equations for populations of each level $N_i$ of the form
\begin{equation}
	\frac{\partial N_i}{\partial t} + \vec{\nabla} . (N_i \vec{u}) = \sum\limits_{j \ne i}^{N_k} N_j P_{ji} - N_i \sum\limits_{j \ne i}^{N_k} P_{ij}\,
	\label{Eq:NonEq_Ioniz_Bifrost}
\end{equation}
are solved. Here, $\vec{u}$ is the plasma bulk velocity, $N_k$ is the total number of levels in the model ion, which can be either excitation or ionization levels, and $P_{ij}$ is the rate coefficient between levels $i$ and $j$. The Equation (\ref{Eq:NonEq_Ioniz_Bifrost}) is solved for 12 levels for Si, 14 for O, and 20 for \ion{Fe}{X}--\ion{Fe}{XV}.

Following this implementation, \citet{Olluri13b} found significant departures from the ionization equilibrium in the Bifrost simulation. In particular, individual oxygen ions existed over much wider ranges of temperatures than in the ionization equilibrium \citep[see Figure 3 in][]{Olluri13b}. For example, \ion{O}{IV} was found to exist over the range of log($T$\,[K])\,$\approx$\,4.0--5.8, and high ion fractions of \ion{O}{VI} was found at log($T$\,[K])\,$\approx$\,6.0 instead of 5.5 as would be expected from equilibrium. These departures from equilibrium arose as a combination of advection bringing plasma from different temperature regions together with \textit{e.g.} long recombination timescales for \ion{O}{III}--\ion{O}{IV} as well as long ionization timescales for \ion{O}{V}. The occurrence of significant non-equilibrium ionization was found to have strong consequences for plasma diagnostics. In particular, the density diagnostics from \ion{O}{IV} intercombination multiplet around 1400\,\AA~was found to yield results different by up to an order of magnitude if the non-equilibrium ionization was taken into account. This occurred since the \ion{O}{IV} emission in the simulation was formed at much lower temperatures than those in the ionization equilibrium. By taking the non-equilibrium ionization into account, \citet{Olluri15} found that the spatially averaged \ion{Si}{IV} and \ion{O}{IV} intensities compared well with the observed ones.

The influence of non-equilibrium ionization on IRIS spectra has been discussed by \citet{DePontieu15} and \citet{Martinez16}. \citet{DePontieu15} found that the IRIS resolution of 0.3$''$ does not reduce the significant non-thermal broadening previously observed at lower resolution of about 1$''$ in the \ion{Si}{IV} 1402.77\,\AA~line. This broadening is of the order of 20 km\,s$^{-1}$ and is correlated with line intensity. Using the Bifrost code, \citet{DePontieu15} found that the observed correlation between the line broadening and intensity can be reproduced only if non-equilibrium ionization is taken into account. Furthermore, the non-equilibrium ionization produced the non-thermal broadening of the order of 2--10 km\,s$^{-1}$, a significant contribution to the observed one. \citet{Martinez16} reported that the \ion{Si}{IV}\,/\,\ion{O}{IV} ratio is positively correlated with the \ion{Si}{IV} intensity, and that the nature of the correlation depends on the feature observed. The strongest increase of \ion{Si}{IV} relative to \ion{O}{IV} was found in active region spectra. Using the Bifrost MHD simulations, the authors have found that the non-equilibrium ionization significantly (but not fully) reduces the discrepancy between the observed and predicted \ion{Si}{IV}\,/\,\ion{O}{IV} ratios and that these ratios become dependent on the \ion{Si}{IV} intensities similarly as observed. The authors explained this correlation as a consequence of the interplay between the thermal stratification of the atmosphere, its dynamics, and the non-equilibrium ionization.

%
%
%
\subsection{Non-equilibrium ionization in the solar lower corona}
\label{Sect:5.3}
Although apparently the solar corona often seems quasi-static, there are several cases where non-equilibrium ionization can be important. For example, it is quite possible that short-lived releases of energy occur at relatively low densities, where the plasma is likely out of ionization equilibrium. Here we briefly summarize some recent studies. 

\citet{Bradshaw03a} studied cooling loops and found that advection flows of the order of 10--20 km\,s$^{-1}$ can carry \ion{C}{VII} into cooler regions where none should exist in equilibrium. This was a consequence of the long recombination timescales of \ion{C}{VII}, which were about 2$\times 10^3$ s in the simulation due to densities being lower than about 10$^{10}$ cm$^{-3}$. As a result, the loop cooled more slowly than it would have done in the ionization equilibrium. \citet{Bradshaw03b} investigated the non-equilibrium ionization in a coronal loop transiently heated at its apex. Significant departures from ionization equilibrium were again found. The non-equilibrium ionization of Fe was of particular importance, leading to nearly 5 times higher emissivities compared to the equilibrium calculations. Furthermore, the emissivity profile along the loop was more uniform than in the equilibrium case.

Short-lived heating events in the solar corona were also considered by \citet{Bradshaw06}. In their simulation, a diffuse, low-density loop was heated by a 30 s nanoflare. During this time, the apex density increased from 10$^7$ to 10$^9$ cm$^{-3}$, while the electron temperature reached 25\,MK during the first 0.5\,s and was kept above 10\,MK for more than 20\,s. As a result of this rapid heating, explosive chromospheric evaporation set in, characterized by strong blueshifts in \ion{Fe}{VIII} during the first 15 seconds, followed by blue-shifted emission from higher charged states of Fe. The ionic composition was found to be strongly affected by the heating and evaporation. In particular, the lowest charge states were found at highest temperatures. E.g., at 20\,s, \ion{Fe}{VIII} was found to be formed at log($T$\,[K])\,$\approx$\,6.8, while \ion{Fe}{IX}--\ion{Fe}{XIV} were found at progressively lower temperatures. Higher ionic stages lagged even further behind, suggesting that the high-temperature lines, originating above 10\,MK in ionization equilibrium, would likely not be observed due their emission measure being very low. These results were verified by \citet{Reale08} using a range of models with heating durations between 5 and 180\,s. These authors concluded that hot plasma above $\approx$5\,MK would not be detectable if the nanoflare durations were shorter than about 1 minute. We note that such results present a considerable obstacle in coronal heating studies, especially considering that some of the current instrumentation exhibits ``blind spots'' in terms of reliable DEM reconstruction at temperatures above log($T$\,[K])\,$\approx$\,6.8 \citep{Winebarger12b}.

Nanoflare heating occuring in storms that heat different loop threads were considered by \citet{Bradshaw11}. These authors found that hot plasma above 5\,MK is present, but the individual AIA channels would be dominated by warm emission formed close to ionization equilibrium at temperatures of 1--2\,MK. For the AIA 131\AA~channel, which is sensitive to temperatures above 10\,MK, it was found that the hot plasma contributes only up to 6\% of the observed signal at most, and that only then in cases characterized by strong total heating. Nevertheless, this hot emission was found again to be mostly out of equilibrium in short-duration heating models.

\citet{Dzifcakova16} investigated the occurrence of non-equilibrium ionization arising from periodic high-energy electron beams. The beams were simulated by changing the distribution function from a Maxwellian to a $\kappa$-one with the same bulk of the distribution but with highly different $T$. The periods considered were of the order of several tens of seconds, similar to the exposure times of coronal spectrometers such as EIS. The plasma was found to be out of equilibrium for all cases with densities lower than $10^{11}$\,cm$^{-3}$. Instantaneous spectra showed fast intensity changes in several lines. However, the spectra averaged over one period appeared multithermal, with DEMs similar to the ones observed for coronal loops. 

Finally, significant out of equilibrium effects were found when studying the intensities of coronal lines at the base of newly-reconnected field lines in the hot cores of active regions by \cite{Bradshaw11a}. Such modelling was developed to provide a plausible explanation, within an interchange reconnection scenario \citep{DelZanna11a}, for the coronal outflows, regions where hot ($T > $ 1 MK) coronal lines show progressively stronger outflows above sunspots and plage areas \citep{DelZanna08}. The reconnection process between a higher-pressure (higher temperature and density) hot loop and a surrounding field would initiate a rarefaction wave travelling towards the chromosphere. After about 300\,s in the simulation, the wave produces outflows in the coronal lines which are similar to the observed ones and are stronger in the higher temperature ones. The lower-temperature lines around 1-2 MK are still formed near ionization equilibrium, while the higher-temperature ones around 3 MK are enhanced by a factor of two when time-dependent ionization is considered. The rapid expansion cools the plasma that is not recombining fast enough. 

To summarize, non-equilibrium ionization effects can be significant at all electron temperatures in the solar corona, with details depending on the nature of the heating and cooling processes, and the associated flows. These non-equilibrium effects are crucial for interpretation of hot plasma above 5\,MK in temperature. Such plasma contains important information on the nature of coronal heating, but non-equilibrium effects may prevent detection with present instrumentation.

%
%
\begin{figure}
	\centering
	\includegraphics[width=0.9\textwidth,clip=]{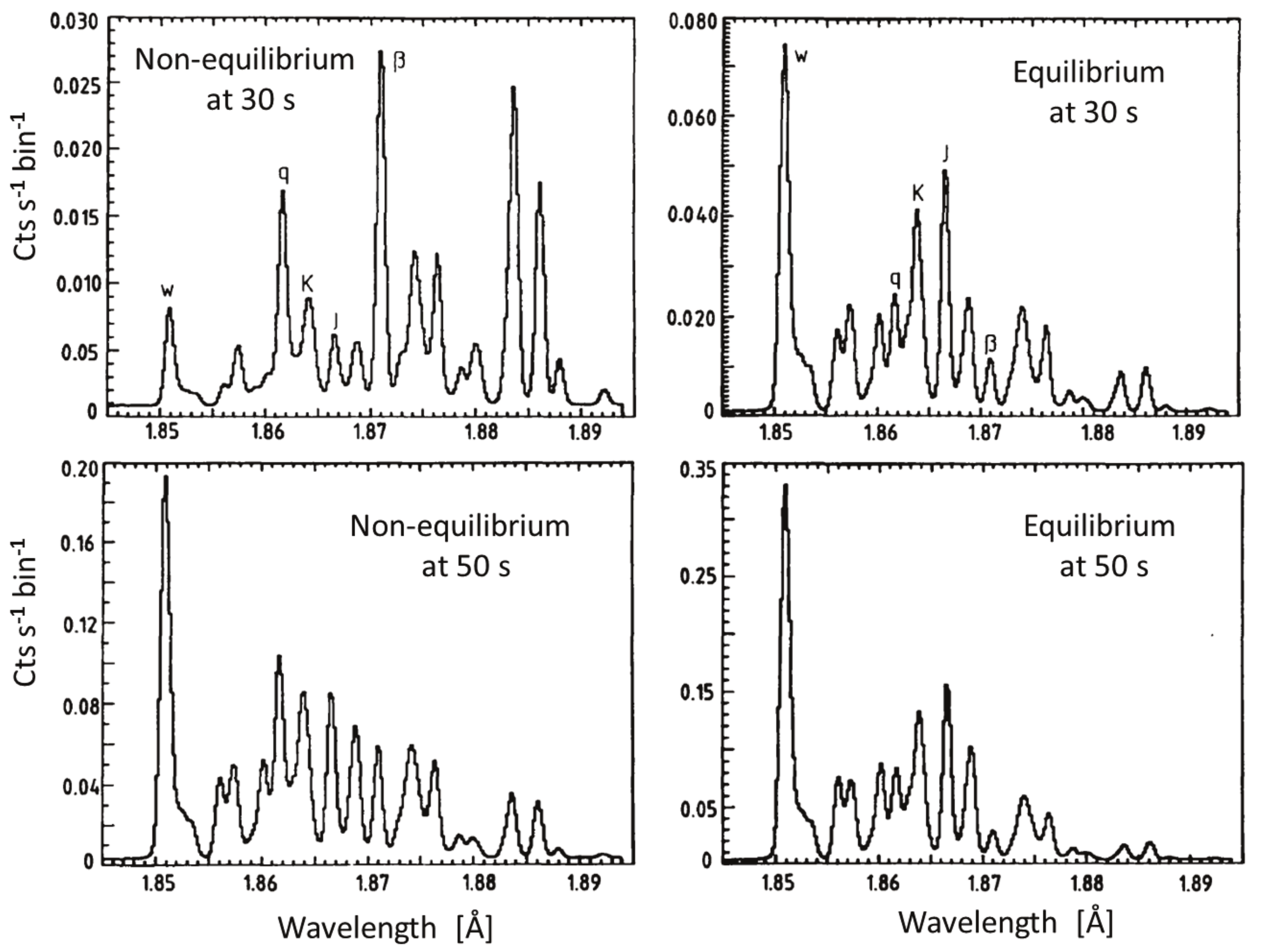}
	\caption{Simulated SMM BCS \ion{Fe}{xxv--xxii} spectra for the flare SOL1980-11-12T17:05 30\,s and 50\,s after the start of transient heating. The left panels show results from a model with non-equilibrium ionization, and the right panels shows results from an equilibrium ionization calculation.
	Credit: \citet{Mewe85}, reproduced with permission \textcircled{c} ESO.
\label{Fig:NEI_bjs5}}
\end{figure}
%
%
\subsection{Non-equilibrium ionization in solar flare spectra}
\label{Sect:5.4}
The high densities (10$^{11}$--10$^{12}$\,cm$^{-3}$) typical of solar flare plasma \citep[for recent measurements see, \textit{e.g.},][]{Graham11,DelZanna11,Milligan12,Young13,Graham15,Doschek15} indicate less favourable conditions for the occurrence of non-equilibrium ionization (see Figure \ref{Fig:NEI_timescales}) than in the non-flaring solar corona. Nevertheless, non-equilibrium ionization could still be present, especially during the early phase when the electron density is still low.

The best measurements of non-equilibrium ionization effects in flare plasma have been obtained in the past by several instruments which observed the \ion{Fe}{XX}--\ion{Fe}{XXVI} X-ray emission lines around 1.84--1.92\AA. The ratios of inner-shell excitation lines such as the \ion{Fe}{XXIV} $q$ vs. the resonance lines of the He-like \ion{Fe}{XXV} depends strongly on the ratio of the Li-like to the He-like relative abundance \citep{Gabriel72}. Another diagnostic is the \ion{Fe}{XXIII} inner-shell $\beta$ line. As an example, we show in Figure~\ref{Fig:NEI_bjs5} simulated spectra of this complex obtained by \citet{Mewe85a}, who studied a flare observed by SMM/BCS. The authors considered heating of a 40\,Mm long loop of initial temperature of 3.2\,MK and density of 6\,$\times$10$^9$\,cm$^{-3}$. Energy deposition occurred either at the apex, at footpoints, or in the form of a chromospheric heating by fast electron beams. It was found that the Ca equilibration times are about 30\,s, while for Fe they are about a minute. In the apex heating case, the peak temperature and density was 21\,MK and 2\,$\times$10$^{10}$\,cm$^{-3}$. In the footpoint heating case, the equilibration times were shorter as a result of the lower temperatures and higher densities; 15\,MK and 5\,$\times$10$^{10}$\,cm$^{-3}$, respectively. In the synthetic spectra, it was found that the inner-shell lines such as $q$ and $\beta$ were enhanced compared to the directly excited resonance lines of He-like \ion{Fe}{XXV} and \ion{Ca}{XIX}. Furthermore, the ratios of these high-ionization resonance lines relative to the continuum were decreased. This arose since the continuum is dominated by H and He, and is therefore unaffected by the non-equilibrium ionization effects in highly ionized metals.

The technique was also applied by \citet{Doschek79} to NRL SOLFLEX observations of the Fe lines in flares. They found strong departures from equilibrium, as the plasma can be always under-ionized. However, \citet{Doyle81} found the data were consistent with ionization equilibrium when they interpreted the observations with different atomic data.

\citet{Doschek81} applied the same technique as \citet{Doschek79} to NRL SOLFLEX observations of the Ca lines in flares. A similar analysis using the Ca lines observed by SMM/XRP was carried out by several authors, see, \textit{e.g.} \cite{Gabriel83,Antonucci84}. Significant uncertainties were present, but no obvious indications of departures from ionization equilibrium were found. \citet{Doschek87} reviewed the SMM and \textit{Hinotori} observations but found no clear evidence of transient ionization effects, although some marginal indications were present during the impulsive phase. With some modelling they found that these effects can persist for times of up to several 10$^{11}/n_\mathrm{e}$ seconds. This means that, at higher densities, the non-equilibrium ionization effects last shorter than at lower densities. If the flare plasma has densities lower than 10$^{11}$\,cm$^{-3}$, the non-equilibrium ionization can last several seconds or more; however detection is then hampered by the correspondingly low emission measures in combination with the detector sensitivities.

Time-dependent ionization is nowadays often included in hydrodynamical models of flare loops, although the results are often close to those obtained assuming equilibrium. Differences have however been found. For example, \citet{Bradshaw04} applied the HYDRAD code to modelling of emission of a small compact flare, finding significant differences when non-equilibrium ionization was included. An 80\,Mm loop was heated by a 300\,s heating pulse located at the apex, raising its temperature from an initial value of 1.6\,MK to about 7\,MK, accompanied by a density increase from 4.7$\times 10^8$\,cm$^{-3}$ by a factor of about three. Flows of up to 200 km\,s$^{-1}$ were produced. Strong enhancements of \ion{He}{I}, \ion{He}{II}, and \ion{C}{IV} emissivities were found in the impulsive phase, later followed by a decrease. At the position of 20\,Mm along the loop and 60 seconds into the simulation, \ion{Fe}{XIV} and \ion{Fe}{XV} were found to be enhanced over their equilibrium values by a factor of 28 and 6, respectively, but were depleted at the position of 10\,Mm along a loop leg. In contrast to this, \ion{Fe}{XVIII} was found to be depleted at all positions along the loop. Later on, at 210\,s, the \ion{Fe}{XIV}--\ion{Fe}{XVI} emissions were close to their equilibrium value, while \ion{Fe}{XVII} was found to be enhanced in the upper parts of the loop. However, \ion{Fe}{XIX} was not present at all in the non-equilibrium ionization simulation.

Simulations using the HYDRAD code have also been carried out recently for a variety of chromospheric evaporation studies in solar flares, including chromospheric heating by electron beams and Alfv\'{e}n waves \citep{Reep13,Reep15,Reep16,Polito16a}. These simulations include the presence of non-equilibrium ionization by default, often without an accompanying equilibrium calculation. For example, the simulation of \citet{Polito16a} predicts the largest \ion{Fe}{XXI} and \ion{Fe}{XXIII} evaporation velocities of more than 250\,km\,s$^{-1}$ occur at the start of the chromospheric evaporation process, and decrease later on. This predicted behavior is in agreement with that observed. We note that the apex densities reached in such simulations are typically of the order of 10$^{11}$\,cm$^{-3}$, so that the non-equilibrium effects cannot be excluded especially in smaller flares.

Occurrence of non-equilibrium ionization in a reconnecting current sheet below an erupting flux rope has been studied by \citet{Shen13}. These authors found that the lower part of the current sheet is under-ionized, while the upper part at heights of about 2\,$R_\odot$ is over-ionized. This is because in its lower part, the current sheet is heated on timescales shorter than the ionization timescales, while in its upper part, the heated plasma cools and expands more quickly than the recombination timescales. Flows related to the reconnection also played a role. Consequently, temperatures derived from observations would be under- and over-estimated at low and large heights, respectively, if equilibrium conditions were assumed. Furthermore, AIA count rates predicted from the non-equilibrium simulation differed significantly from the one obtained assuming ionization equilibrium. In particular, the discernible thin bright region visible in 131\,\AA~and 94\,\AA~AIA images, corresponding to the centre of the current sheet, extends in the non-equilibrium simulation to larger heights than in the equilibrium one. In the regions outside of the current sheet, \textit{i.e.}, in the ambient corona, non-equilibrium effects were significant (larger than 35\% changes in line intensities) only at heights above $\approx$2\,$R_\odot$, where the electron density were low.


%
\subsection{Non-equilibrium ionization and solar wind}
\label{Sect:5.5}
Since the electron temperatures in the solar wind do not correspond to the local charge state of the solar wind, the solar wind is also not in ionization equilibrium. The ionic state in the solar wind is commonly inferred using Equation (\ref{Eq:Ion_time_evol}) applied to cases of solar wind expansion \citep[\textit{e.g.},][]{Ko97,Edgar00}, which explicitly includes $\vec{u}$\,$\ne$\,0. In a steady state characterized only by radial outflows, \citet{Ko97} showed that whether the ionic fractions correspond to equilibrium, or are ``frozen-in'', depends not only on the local electron temperature, but also on the competition between the ion equilibration timescale (see Section \ref{Sect:5.1}) and the expansion timescale $t_\mathrm{exp}$, which is given by
\begin{equation}
	t_\mathrm{exp} = \left| \frac{u}{n_\mathrm{e}}\frac{\partial n_\mathrm{e}}{\partial r} \right|^{-1}\,.
	\label{Eq:wind_exp_timescale}
\end{equation}
At large radii $r$, the density $n_\mathrm{e}$ is small while the wind velocities $u$ are large, meaning that $\partial N^{(+z)} / \partial r \to 0$, \textit{i.e.}, the ion charge state is ``frozen-in''. \citet{Ko97} showed that this almost always occurs within 5\,$R_\odot$ for all the $T(r)$ and $n_\mathrm{e}(r)$ studied. For some elements, such as C, O, and Mg, the ionic composition is frozen-in (and thus out of equilibrium) at distances larger than 2\,$R_\odot$, while for Si and Fe, the freezing-in occurs at radii larger than about 2.5\,$R_\odot$ and 4\,$R_\odot$, respectively (see Figure 5 therein).

Since the frozen-in charge state depends on the local conditions in the solar wind source region, it can be used to obtain constraints on these source regions \citep{Ko97,Esser00}. It is possible to do this with a variety of models including different Poynting flux values or magnetic field geometries (open or closed) in different regions such as coronal holes, and even eruptions constituting transients in the solar wind \citep[\textit{e.g.},][]{Lynch11,Gruesbeck11,Landi12c,Landi12b,Landi14b,Lepri12,Oran15,Rodkin16}.

Evolution of charge state along an open magnetic field line was studied in a series of papers by \citet{Landi12c,Landi12b,Landi14b}. It was found that wind-induced departures from ionization equilibrium existed in the solar atmosphere, producing up to a factor of 3 differences in total radiative losses of the upper chromosphere and transition region when compared to equilibrium ion composition \citep{Landi12b}. Carbon and oxygen were the elements showing the most pronounced departures from equilibrium. \citet{Landi12c} showed that the ions emitting the brightest lines (such as \ion{C}{IV}, \ion{N}{V}, \ion{O}{V}--\ion{O}{VI}) especially at low heights in coronal holes were out of equilibrium. Most ions except the highest ionization states were found to be over-abundant. However, at larger heights, some ions such as \ion{O}{V}--\ion{O}{VI} and \ion{Fe}{VIII}--\ion{Fe}{XII} were close to equilibrium, with departures being smaller than 25\% in the case of Fe ions. \citet{Landi14b} used the models of \citet{Hansteen95}, \citet{Cranmer07}, and \citet{Oran13} to predict the intensities of emission lines and the fast solar wind charge state distributions, and found that while transition-region lines are predicted well, coronal line intensities and the fast wind charge state distribution are underpredicted, possibly due to the fast wind encountering difficulties to reach the high ionization stages corresponding to local conditions.

\citet{Oran15} calculated the charge state evolution covering all latitudes in a realistic open magnetic field, and obtained similar results. In particular, the \ion{O}{VIII}\,/\,\ion{O}{VII} and \ion{C}{VII}\,/\,\ion{C}{VI} charge state ratios are underpredicted when compared to \textit{Ulysses} observations. Similarly, the intensities of the S, Si, and Fe lines were underpredicted when compared to \textit{Hinode}/EIS observations of a coronal hole. These authors suggested that the ionization rates in the low corona, which they assumed to be Maxwellian, are likely underpredicted. It was found that adding a second Maxwellian having 2\% of particles and temperature of 3\,MK produced much better agreement with both the observed charge state composition and the EIS intensites. We note that \citet{Cranmer14} suggested that mild suprathermal tails characterized by a $\kappa$-distribution with $\kappa$\,=\,10--25 can be sufficiently energetic to enhance the O charge state in the solar wind.

%
%
%
%
%
\section{Future Instrumentation for Detection of Non-Equilibrium Phenomena}
\label{Sect:6}

As we have shown, detection of the non-equilibrium effects in the solar atmosphere through remote-sensing observations has proven difficult due to limitations of the instrumentation or the small magnitude of the expected effects. A plasma may only be out of ionization equilibrium for tens of seconds in some cases \citep[or a few seconds in the case of][]{Doyle13}, while the emission measure of the transient plasma may be very low (Section \ref{Sect:5.4}). Simultaneous coverage of consecutive ion stages of an element would allow transient ionization to be tracked through the atmosphere, while high sensitivity observations of strong lines are needed to capture the evolution at cadences of order seconds. To study transient ionization during flares, new instruments with much improved sensitivity would be required. 

Remote-sensing detection of non-Maxwellian electron distributions provides a further challenge: to be sensitive to the distribution, a pair of emission lines from an ion should have significantly different excitation thresholds, yet such lines are often widely separated in wavelength. This also puts strong demands on the radiometric calibration of the instrument \citep[cf.,][]{BenMoussa13,DelZanna13a,Warren14}, which should have accuracies of 10\% or better over this wavelength range. In addition, the lines are often weak compared to the strongest lines from the ion \citep[\textit{e.g.},][]{Mackovjak13,Dudik15}, so high sensitivity is again critical. Additionally, in the EUV and especially the X-ray range high spectral resolution is needed to resolve blended lines (Section \ref{Sect:4.3}).


In this section, we discuss the capabilities of future remote-sensing instruments for detecting signatures of non-equilibrium processes in the solar atmosphere. We include both instruments under construction (Section \ref{Sect:6.1}) and future design concepts (Section \ref{Sect:6.2}). In situ instruments were discussed already in Sections \ref{Sect:2.5} and \ref{Sect:2.6}, which highlighted future capabilities for improved measurements that will reach closer to the Sun than previously achieved.

%
%
\subsection{Instruments under construction}
\label{Sect:6.1}

At present, several instruments are under construction. These will be flown on \textit{Solar Orbiter}, \textit{Interhelioprobe}, \textit{Proba-3}, the \textit{International Space Station} (ISS), as well as sounding rockets.

\subsubsection{Solar Orbiter}
\label{Sect:6.1.1}
The \textit{Solar Orbiter} will be the first mission to fully explore the interface region where solar wind and heliospheric structures originate. It will carry 6 remote-sensing and 4 in situ instruments that will link the solar wind to its source regions in the solar atmosphere. \textit{Solar Orbiter} has been selected as first medium-class mission of ESA's Cosmic Vision 2015--2025 programme, implemented with NASA, and is due to be launched in 2018. 
The \textit{Solar Orbiter} will operate in coordination with the \textit{Solar Probe Plus} \footnote{\url{http://solarprobe.jhuapl.edu}}, 
a NASA mission also due to be launched in 2018 which will fly into the low solar corona, providing a unique opportunity for a deep investigation of the processes by which the Sun produces and accelerates the solar wind and energetic particles, and also probe how the solar corona is heated.

The Spectral Investigation of the Coronal Environment \citep[SPICE,][]{Fludra13} on board the \textit{Solar Orbiter} operates at extreme ultraviolet wavelength in two bands, 704--790\,\AA~and 973--1049\,\AA, which are dominated by emission lines from a wide range of ions of H, C, O, N, Ne, S, Mg, Si and Fe. These lines are formed at temperatures from $10^4$ to $10^7$ K, allowing the study of the solar atmosphere from the chromosphere in the Lyman-$\beta$ line up to the flaring corona (\textit{e.g.} \ion{Fe}{XX} 721\,\AA). SPICE will mainly observe cool lines, so in principle could provide useful measurements to study non-equilibrium ionization in the TR. Telemetry limitations will however put some constraints on the number of lines and the field of view. Also, the sensitivity will allow exposure times of the order of seconds only for the stronger lines. 

The \textit{Spectrometer Telescope for Imaging X-rays} (STIX) will perform imaging spectroscopy in the X-ray region 4 to 150 keV using a Fourier-transform imaging technique similar to the one used by the previous \textit{Yohkoh}/HXT and RHESSI spacecraft. A set of cadmium telluride (CdTe) X-ray detectors provide spectral resolution of $\sim1$~keV at 6\,keV, with unprecedented spatial resolution and sensitivity (near perihelion). A detailed description of the imaging technique is given in \citet{Benz12}. 

The set of in-situ instruments on \textit{Solar Orbiter} will allow plasma composition and energy distribution functions of individual particle species to be measured in the vicinity of the probe. On the other hand, STIX will play a key role by measuring non-thermal electron distribution functions in the solar atmosphere. A connection from the solar surface to the heliosphere can thus be made. In particular, how the distribution functions are modified by transport processes operating in the heliosphere.

\subsubsection{Interhelioprobe}
\label{Sect:6.1.2}

The Russian Interhelioprobe \citep[IHP,][]{Oraevsky01} program includes sending two interplanetary probes on orbits similar to \textit{Solar Orbiter}, reaching as close to the Sun as $\approx$60 solar radii. The probes are planned for launch in 2025 and 2026, and they will enable stereoscopic observations of solar activity. Every one of the gravitational assist maneuvers will increase the inclinations of the orbital plane to the ecliptic, reaching 30$^\circ$ towards later phase of the missions. At perihelion, the structures of the solar atmosphere will be observed with a spatial resolution $\approx$3 times better than from Earth distance, with the solar radiative fluxes up to 13 times greater. Both IHP spacecraft will be equipped with the identical $\approx$150\,kg science payloads consisting of 20 instruments for remote and in-situ plasma diagnostics \citep{Kuznetsov15}.

On both IHP spacecraft, Polish-led Bragg bent crystal spectrometer ChemiX \citep{Siarkowski16} will be placed. This spectrometer will observe active region and flaring plasma in the spectral range 1.5--9\,\AA~using 10 monocrystal wafers bent cylindrically, which allow a factor 10 improvement in the spectral resolution over RESIK. Careful construction and appropriate selection of materials will cause the X-ray fluorescence produced by irradiance of solar X-rays on the instrument's material, which commonly forms a background emission, to be minimized allowing for reliable continuum measurements. Four crystals will cover slightly overlapping spectral ranges 1.500\,--\,2.713\,\AA, 2.700\,--\,4.304\,\AA, 4.290\,--\,5.228\,\AA, and 5.200\,--\,8.800\,\AA. The overlaps will allow for cross-calibration. The use of cooled CCD detectors to record the spectra will result in an order of magnitude increase of the continuum to background signal compared to RESIK. The other three pairs of identical crystal-detector units will form three dopplerometers. Analysis of their high spectral resolution measurements will allow for precise determinations of the line-of-sight component of plasma motions.

%
\begin{figure}[!ht]
	\centering
	\includegraphics[width=10cm]{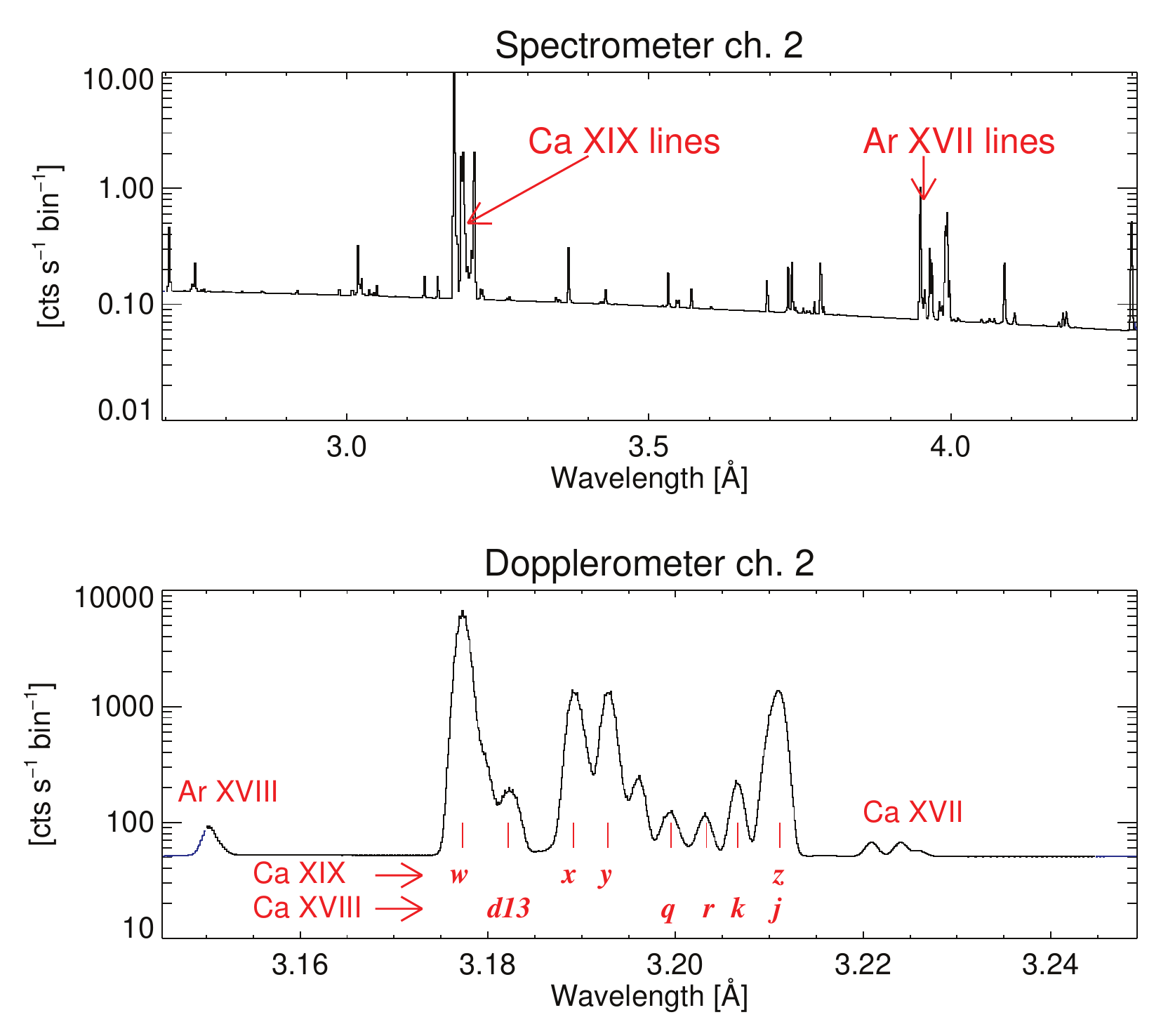}
	\caption{Simulated ChemiX spectra as seen on the CCD at 1~AU predicted using the {\sc chianti} database. Top: Spectrum in channel 2 containing \ion{Ca}{XIX} and \ion{Ar}{XVIII} line groups. Bottom: Dopplerometer channel 2 spectrum that contains mostly the Ca He-like ion spectra in high resolution. A number of satellite lines are well resolved.}
	\label{Fig:Chemix_simulated_spectra}
\end{figure}

We note that the ChemiX spectral range contains dielectronic satellite lines that can be used to discern the non-Maxwellian effects (Section \ref{Sect:4.3.2}) from a range of elements and at a range of energies of 1.4--8.2\,keV; this will enable sampling of the distribution function. Figure~\ref{Fig:Chemix_simulated_spectra} shows examples of simulated spectra from channels 1 and 4. This simulation assumes plasma with temperature $T_{e} \approx 18.7$\,MK and emission measure EM $\approx 5.6~\times$\,10$^{49}$\,cm$^{-3}$; \textit{i.e.}, conditions corresponding to a typical GOES M5 class solar flare. Collimator and CCD efficiency are assumed to be unity.

\subsubsection{SolpeX on ISS}
\label{Sect:6.1.3}
Another instrument capable of collecting high-resolution soft X-ray spectra and detect polarisation is \textit{Solar Polarimeter in X-rays} (SolpeX). This instrument consists of three functionally independent blocks which are included within the Russian instrument KORTES, to be mounted on the Sun-pointed platform aboard the \textit{International Space Station} (ISS). The three SolpeX units are: a simple pin-hole soft X-ray imager and spectrophotometer with $\approx$20$''$ spatial resolution, spectral resolution of 0.5\,keV, and high-cadence of 0.1\,s; Bragg Polarimeter B-POL observing at 3.9--4.6\,\AA~ with a low linear polarization detection threshold of 1--2\%, and finally a fast Rotating Drum X-ray Spectrometer (RDS) with high time resolution of 0.1\,s.
Such a combination of measuring blocks offers an opportunity to reliably measure possible X-ray polarization and high-cadence spectra of solar flares, in particular during the impulsive phase. Polarized bremsstrahlung and line emission due to the presence of directed particle beams can be detected, and measurements of the velocities of evaporated hot plasma will be made. 


Lines of only a few elements suitable for polarization detections are present in the B-POL spectral range. To extend the spectral coverage over the entire soft X-ray range (1--23\,\AA), the RDS spectrometer was included. A novelty of this instrument is the wavelength assignation to each photon Bragg-reflected from the crystal. This will be achieved by using the timing information on the photon arrival on the silicon drift detector. The symmetrical location of detectors on both sides of the drum with respect to the direction towards the Sun represents the configuration necessary to fulfill the tasks of a Dopplerometer \citep{Sylwester15}. Eight flat crystals are planned to be used, two of them identical. Two identical crystals are placed in the classical dopplerometer configuration. The combination of all the rotating crystals will enable a continuous spectral coverage, extending from 0.3\,\AA~to~22.8\,\AA, recorded every 0.1\,s. 

%
\subsubsection{Marshall Grazing-Incidence X-ray Spectrometer}
\label{Sect:6.1.4}

The \textit{Marshall Grazing-Incidence X-ray Spectrometer} (MAGIXs) is a NASA spectrometer designed to observe the Sun at energies of 6--24\,\AA (0.5--2\,keV) with a spectral resolution of 22\,m\AA~and spatial resolution of $\approx$5$''$. The instrument will be flown on a sounding rocket, allowing for $\approx$5\,min observations of a portion of the solar corona. The spectral range of MAGIXs contains many \ion{Fe}{XVII}--\ion{Fe}{XX} spectral lines \citep{DelZannaMason14,Schmelz15}. A goal of this mission is to constrain the amount of hot plasma in the solar active region cores. As discussed in Section \ref{Sect:5.3}, the emission measure of this plasma may be low due to the non-equilibrium ionization effects. Additionally, the multiple \ion{Fe}{XVII}--\ion{Fe}{XVIII} lines observed within the MAGIXs wavelength window may offer opportunities of detecting high-energy tails, possibly in combination with the AIA 94\,\AA~filter observations.

\subsubsection{Proba-3}
\label{Sect:6.1.5}

\textit{Proba-3} is the first ESA mission to test new technologies to fly two spacecraft with tight constraints on their alignment. Proba-3 will perform solar coronographic observations. The first spacecraft is the occulter. The second one, at about 150 meters away, has the coronograph. The instrument will obtain measurements of the polarized brightness in the visible and of the forbidden \ion{Fe}{XIV} coronal line. As we have discussed in Section~\ref{Sect:3.5.2}, observations of the forbidden visible line, in combination with those of the EUV \ion{Fe}{XIV} lines (\textit{e.g.} from \textit{Hinode}/EIS if the instrument will still be operational), are potentially very useful to indicate if non-Maxwellian distributions are present in the solar corona. 

%
%
\subsection{Proposed future instruments}
\label{Sect:6.2}


%
\subsubsection{The EUVST instrument proposed for Solar-C}
\label{Sect:6.2.1}

The Extreme-Ultraviolet Spectroscopic Telescope \citep[EUVST,][]{Teriaca12}, previously also named LEMUR (Large European Module for solar Ultraviolet Research) is an image-stabilized imaging spectrometer with a 0.14$''$ pixel size and 0.3$''$ spatial resolution that should observe in several spectral ranges, 170--215\,\AA, 690--850\,\AA, 925--1085\,\AA, and 1115--1275\,\AA~in the first order and 463--542\,\AA~and 557--637\,\AA~in the second order with a spectral resolution $\lambda / \Delta \lambda$ of about 1.6--3\,$\times$10$^4$. The design enables very short exposure times, up to 0.1\,s for the brightest spectral lines. The EUVST has been proposed as an instrument for the JAXA-led Solar-C mission during the ESA's M4 mission call in 2014, but was not approved at the time.

Multiple, widely-spaced spectral bands will enable non-Maxwellian diagnostics from lines formed at widely different energies, for example \ion{O}{IV}. In addition, the high spatial resolution will allow the occurrence of non-equilibrium phenomena to be pin-pointed to individual structures in the atmosphere.

\subsubsection{Microcalorimeters}
\label{Sect:6.2.2}

In the early 1980s, it was realized that microcalorimeters have the potential to measure the energy of an X-ray keV photon with an accuracy of a few eV. This is comparable with the spectral resolution of crystal Bragg spectrometers, however the effective area of a microcalorimeter is 4--5 orders of magnitude better. A microcalorimeter is a unique instrument as it allows to simultaneously determine the energy and the position of each X-ray incoming photon. At present, the best energy resolution (1.8\,eV at 6\,keV) is obtained with microcalorimeters with superconducting transition-edge sensors \citep[TESs,][]{Benz12}. In the future, metallic magnetic microcalorimeters (MMCs) should enable sub-eV resolution. Thanks to their enormous sensitivity, microcalorimeters can revolutionize understanding of coronal plasma processess, especially as concerns the non-equilibrium plasma dynamics of solar flares and active regions \citep{Laming10}. Using appropriate imagers direct observations of the spectra of microflares and possibly nanoflares will be possible. With the spectral resolution of the order of 1\,eV, the spectral line profiles can be studied in detail including thermal, Doppler, microturbulence and multitemperature components. The presence of such components was seen in SMM BCS spectra of the main phase of flares (\textit{cf.}, Figure\,\ref{Fig:XRP}).

Microcalorimeters could enable the detection of departures from ionization equilibrium in the impulsive phase, particularly in the 6.7\,keV Fe line complex (Section \ref{Sect:5.4}). Intensity ratios of dielectronic satellites to the resonance line can also be used to determine the electron temperature, while the inner-shell excitation lines can be used to determine the abundance ratio of the Li-like/ He-like Fe ions. In ionization equilibrium, the temperatures from these two ratios should be equal but non-equality would indicate either an ionizing or recombining plasma. 

Furthermore, the \ion{Fe}{XXI}\,/\,\ion{Fe}{XXII} ratios observed with crystal spectrometers on SMM were used to detect high electron densities in solar flares, approaching $10^{12}$--$10^{13}$ cm$^{-3}$ \citep{Phillips96}. The microcalorimeter would provide similar spectral resolution, but with significantly wider bandpass, throughput and spatial resolution. 

%
%
%
%
%
%
\section{Concluding remarks}
\label{Sect:7}
We have shown that non-equilibrium processes, including the presence of non-Maxwellian distributions and non-equilibrium ionization, can be important in a range of observed phenomena in the solar atmosphere. These include the transition region, which is highly variable, solar active regions, and solar flares, and finally the solar wind. Not including these non-equilibrium processes into the observational analysis can lead to severe misinterpretation of observations, while omitting these in time-dependent models can have significant consequences for the modelled dynamics.

Regarding the non-Maxwellian distributions, we note that strong constraints have been obtained for the presence of high-energy particles in the quiet solar atmosphere. This indicates that a range of structures in the quiet corona are probably close to being Maxwellian. In other coronal and transition region structures however, signatures of non-Maxwellian distributions were detected using EIS and IRIS. The non-Maxwellians are also routinely observed in the solar wind. It is currently unknown how the origin of the non-Maxwellian solar wind relates to structures in the low solar corona observed in UV, EUV, and X-rays. Although theoretical diagnostics for detection of these non-Maxwellians in the wind source regions are being developed, they often rely on weaker spectral lines and/or the use of large spectral ranges to capture lines formed at different excitation thresholds. This brings additional constraints in terms of instrument calibration. In particular, the relative calibration uncertainties should be lowered in future missions, and careful accounting of the in-flight instrument degradation needs to be performed. Finally, to understand coronal heating and the formation of the solar wind, we need a set of remote-sensing and in-situ observations that cover the solar atmosphere seamlessly from the chromosphere to the upper corona and then to the wind. The set of spacecraft under construction, such as \textit{Solar Orbiter}, \textit{Solar Probe Plus}, and \textit{Interhelioprobe}, will for the first time help with this task, but should be supplemented with next-generation remote-sensing instruments operating in UV and EUV, such as the proposed EUVST instrument.

Solar flares represent a class of dynamic phenomena for which non-Maxwe\-llian distributions are routinely detected. It is however important to constrain the non-Maxwellian properties not only at high energies where the bremsstrahlung emission is observed with spectrometers such as RHESSI, but also at intermediate and low energies. The combination of RHESSI and AIA observations have shown that simple power-law fits to the high-energy bremsstrahlung emission are often incompatible with the emission detected at lower energies. Finally, the situation is further complicated by the presence of the peaked $n$-distributions especially during the impulsive phase of solar flares. To disentangle the various contributions to the observed emission, high spatial resolution will be needed to understand the formation and propagation of these non-Maxwellians in the various structures involved in the flare. Additionally, we are currently in the curious situation of having better flare line spectra from stellar observations made by instruments such as \textit{Chandra} and \textit{XMM-Newton} than we do of the Sun \citep{Laming10}. A set of instruments under construction will resolve this mismatch, enabling direct comparisons with stellar observations in addition to addressing the issues discussed above. Furthermore, we note also that the observations of X-ray lines formed at lower temperatures in active region cores may lead to strong constraints on the properties of the electron distribution function there. The MAGIXs instrument will address this issue.

Dynamic phenomena in the solar atmosphere characterized by transient heating or cooling lead to the occurrence of non-equilibrium ionization. Although non-equilibrium ionization is more pronounced at lower densities, it can be important for nearly all layers of the outer solar atmosphere, from the transition region to corona and even flares. Recent theoretical work has shown that non-equilibrium ionization can dramatically decrease the discrepancies between the transition region line intensities observed by IRIS. In addition, the transient ionization decreases significantly the emission measure from the high-temperature emission expected to arise as a result of nanoflare heating of the corona. Diagnosing non-equilibrium ionization from observations is not straightforward as it requires observing ions with very different ionization and/or recombination rates, such as the Si IV and O IV ions observed with IRIS, or consecutive ionization stages of a single element. This further emphasizes the need to simultaneously observe numerous spectral lines from many ions. Such observations must also go hand-in-hand with time-dependent numerical modeling.

In summary, the presence and importance of non-equilibrium phenomena in the solar atmosphere and wind is now beyond question. Careful evaluation of these processes should be carried out for each emitting structure observed; a task that complicates the observational analysis. For such undertaking to be routinely performed in the future, high-quality spectroscopic observations are indispensable. Publicly available spectroscopic databases including the necessary atomic data, such as CHIANTI, OPEN-ADAS, and KAPPA, are a cornerstone of such analysis. The accessibility of modeling codes such as HYDRAD and BIFROST is also invaluable.

%
%
%
%
\begin{acks}
The authors thank the anonymous referee for numerous improvements to the manuscript. The authors also acknowledge useful discussions with M. Battaglia, P. Heinzel and A. Zemanov\'{a}. J.D. and E.Dz. authors acknowledge support by Grant Agency of the Czech Republic, Grant No. 17-16447S, and institutional support RVO:67985815 from the Czech Academy of Sciences. G.D.Z. and H.E.M. acknowledge funding from STFC. J.D., G.D.Z., and H.E.M. acknowledge funding from Royal Society via the Newton Alumni programme. P.R.Y. acknowledges support from NASA grant NNX15AF25G. A.G. acknowledges the in house research support provided by the STFC. B.S. and J.S. acknowledge support from the Polish National Science Centre grant 2013/11/B/ST9/00234. M.O. was supported by NASA grant NNX08AO83G at UC Berkeley. L.M. was supported by the UK Science and Technology Facilities Council grant ST/K001051/1. The authors benefitted greatly from participation in the International Team 276 funded by the International Space Science Institute (ISSI) in Bern, Switzerland. CHIANTI is a collaborative project involving the University of Cambridge (UK), the George Mason University (USA), and the University of Michigan (USA). ADAS is a project managed at the University of Strathclyde (UK) and funded through memberships universities and astrophysics and fusion laboratories in Europe and worldwide.
\end{acks}

%
\bibliographystyle{spr-mp-sola}
\bibliography{review-R2}

\end{article} 
\end{document}